\documentclass[a4paper, amsfonts, amssymb, amsmath, reprint, showkeys, nofootinbib, twoside, superscriptaddress, longbibliography]{revtex4-2}
\usepackage[table]{xcolor} 
\usepackage[english]{babel}
\usepackage[utf8]{inputenc}
\usepackage{booktabs}
\usepackage[colorinlistoftodos, color=green!40, prependcaption]{todonotes}
\usepackage[pdftex, pdftitle={Article}, pdfauthor={Author}]{hyperref} % For hyperlinks in the PDF

\usepackage{amsthm}
\usepackage{mathtools}
\usepackage{physics}
\usepackage{graphicx}
\usepackage[left=18mm,right=18mm,top=25mm,columnsep=15pt, bottom=20mm]{geometry}
\usepackage{adjustbox}
\usepackage{placeins}
\usepackage[T1]{fontenc}
\usepackage{lipsum}
\usepackage{csquotes}
\usepackage{xfrac}
\usepackage{enumitem}
\usepackage{array}
\usepackage{multirow}
\usepackage{svg}

\renewcommand{\arraystretch}{1.6}

\newcommand{\bsg}{\ln \mathcal{B}_{\mathcal{S}, \mathcal{G}}}
\newcommand{\etaC}{\eta_{\mathrm c}}
\newcommand{\etaR}{\eta_{\mathrm r}}
\newcommand{\pyr}{\,\mathrm{yr}^{-1}}

\newcommand{\Hz}{\,\mathrm{Hz}}
\newcommand{\kHz}{\,\mathrm{kHz}}
\newcommand{\ms}{\,\mathrm{ms}}
\newcommand{\s}{\,\mathrm{s}}
\newcommand{\apjl}{Astrophys. J. Lett.}
\newcommand{\mnras}{Mon. Not. R. Astron. Soc.}
\newcommand{\apjs}{Astrophys. J. Suppl. Ser.}

\begin{document}

\title{Dedicated-frequency analysis of gravitational-wave bursts from core-collapse supernovae with minimal assumptions}

\author{Yi Shuen C. Lee}
    \email[]{ylee9@student.unimelb.edu.au}% Your name
    \affiliation{School of Physics, University of Melbourne, Parkville, Victoria 3010, Australia.}
    \affiliation{Australian Research Council Centre of Excellence for Gravitational Wave Discovery (OzGrav),
University of Melbourne, Parkville, Victoria 3010, Australia.} 
\author{Marek J. Szczepa\'{n}czyk}
    \email[]{marek.szczepanczyk@fuw.edu.pl}% Your name
    \affiliation{Faculty of Physics, University of Warsaw, Ludwika Pasteura 5, 02-093 Warszawa, Poland.}
    \author{Tanmaya~Mishra}% Your name
    \affiliation{Department of Physics, University of Florida, PO Box 118440, Gainesville, Florida 32611-8440, USA.}
\author{Margaret~Millhouse}% Your name
    \affiliation{School of Physics, University of Melbourne, Parkville, Victoria 3010, Australia.}
    \affiliation{Center for Relativistic Astrophysics, Georgia Institute of Technology, Atlanta, Georgia 30332, USA.}
\author{Andrew Melatos}% Your name
    \affiliation{School of Physics, University of Melbourne, Parkville, Victoria 3010, Australia.}
    \affiliation{Australian Research Council Centre of Excellence for Gravitational Wave Discovery (OzGrav),
University of Melbourne, Parkville, Victoria 3010, Australia.}

\begin{abstract}
Gravitational-wave (GW) emissions from core-collapse supernovae (CCSNe) provide insights into the internal processes leading up to their explosions. Theory predicts that CCSN explosions are driven by hydrodynamical instabilities like the standing accretion shock instability (SASI) or neutrino-driven convection, and simulations show that these mechanisms emit GWs at low frequencies ($\lesssim 0.25 \kHz$). Thus the detection of low-frequency GWs, or lack thereof, is useful for constraining explosion mechanisms in CCSNe. This paper introduces the dedicated-frequency framework, which is designed to follow-up GW burst detections using bandpass analyses. The primary aim is to study whether low-frequency (LF) follow-up analyses, limited to $\leq 256 \Hz$, constrain CCSN explosion models in practical observing scenarios. The analysis dataset comprises waveforms from five CCSN models with different strengths of low-frequency GW emissions induced by SASI and/or neutrino-driven convection, injected into the Advanced LIGO data from the Third Observing Run (O3). Eligible candidates for the LF follow-up must satisfy a benchmark detection significance and are identified using the coherent WaveBurst (cWB) algorithm. The LF follow-up analyses are performed using the \textit{BayesWave} algorithm. Both cWB and \textit{BayesWave} make minimal assumptions about the signal's morphology. The results suggest that the successful detection of a CCSN in the LF follow-up analysis constrains its explosion mechanism. The dedicated-frequency framework also has other applications. As a demonstration, the loudest trigger from the SN 2019fcn supernova search is followed-up using a high-frequency (HF) analysis, limited to $\geq 256 \Hz$. The trigger has negligible power below 256 Hz, and the HF analysis successfully enhances its detection significance. 

\end{abstract}

\maketitle

\section{Introduction}
\label{sec:intro}

Core-collapse supernovae (CCSNe) are energetic explosions of massive stars ($\gtrsim 8 M_\odot$) at the end of their lifetimes. CCSNe synthesize heavy elements during the explosion process, which are subsequently dispersed into the interstellar medium and inherited by the next generation of stars~\cite{Woosley_2002, Janka_2012}. CCSNe are also known to produce compact objects remnants like neutron stars and black holes~\cite{Burrows_1995, Woosley_2005, Kotake_2006}; these remnants have been observed by the the Advanced Laser Interferometer Gravitational-Wave Observatory (LIGO) \cite{LIGOmain} and Advanced Virgo \cite{VIRGOmain} detectors. CCSNe play a vital role in stellar formation and evolution, which has led to extensive theoretical \cite{Bethe_1985} and numerical \cite{Janka_2017, Muller_2020, Abdikamalov_2020} studies of their explosion mechanisms over the last few decades. It is believed that the initial collapse of the iron core produces an outward-bound hydrodynamic shock. Left to its own devices, however, the shock stalls within milliseconds due to energy dissipation, and fails to eject stellar material \cite{Bethe_1990, Muller_2020}. This suggests the existence of a secondary process that revives the outward progress of the shock \cite{Janka_2012}.

Constraining the explosion mechanisms of CCSNe requires accurate probing of the stellar interior, before and during the explosion. Electromagnetic signatures of the pre-explosion dynamics cannot penetrate the atmosphere of the progenitor. Neutrinos and gravitational-waves (GWs), on the other hand, propagate unscattered and unobstructed through the stellar atmosphere. Thus they carry information on the physical processes that drive the explosions. SN 1987A is the only CCSN observed to date accompanied by neutrino emissions \cite{SN1987A_1, SN1987A_2, SN1987A_3}, providing the first empirical evidence for a neutrino-driven explosion \cite{Burrows_1986, Woosley_SN1987A}. Transient GW (burst) emissions due to asymmetrical motions in and around the compact stellar core are also expected during CCSNe \cite{Ott_2009, Abdikamalov_2020}. From the first observing run (O1) to the second segment of the fourth observing run (O4b), the LIGO-Virgo-KAGRA (LVK) collaboration reported nearly 100 published GW detections~\cite{GWTC1, GWTC2, GWTC2.1, GWTC3} and over 200 candidates\footnote{The list of O4 GW candidates is available at: \url{https://gracedb.ligo.org/superevents/public/O4/}.}, but none are associated with CCSNe. Given the current detector sensitivity, it is unlikely that GWs associated with CCSNe can be detected through independent, all-sky searches. However, a CCSN occurring within our galaxy would likely be accompanied by neutrino and electromagnetic signals, which could help localize the source and enable a targeted GW search. The highly anticipated detection of GWs from CCSNe in the ongoing and future observing runs incentivizes studies to improve GW burst analysis methods.

% The LVK detectors commenced the third segment of the fourth observing run (O4c) in June 2025, and are expected to detect several dozen more GW events by the end of the run.

Physical processes occurring within CCSNe can be characterized by the frequency of their GW signatures~\cite{Abdikamalov_2020}. Simulations have shown that the standing accretion shock instability (SASI)~\cite{Blondin_2002} and neutrino-driven convection~\cite{Bethe_1990} contribute to reviving stalled shock waves, ultimately driving CCSN explosions, and these processes generate low-frequency signals ($\lesssim 0.25 \kHz$). Low-frequency GWs from CCSNe are therefore useful for constraining the explosion mechanisms. By dividing the frequency range into two bins above and below $0.25~\kHz$, Ref.~\cite{Szczepanczyk_2022} compares the low- and high-frequency reconstructions of CCSN GWs. The study finds that for CCSN models with stronger low-frequency emissions, such as SFHx~\cite{SFHx} and mesa20~\cite{mesa20_pert}, the low-frequency reconstructions are more accurate than the high-frequency and full-band reconstructions. Since bandpass analyses can improve reconstruction, they may likewise be useful for detecting, or enhancing sensitivity to, frequency-specific GW signatures.

However, the potential benefits of bandpass analyses have so far been explored only under idealized noise assumptions; the analysis in Ref.~\cite{Szczepanczyk_2022} assumes Gaussian detector noise colored by simulated power spectral densities (PSDs). In real observing scenarios, reconstruction accuracy may differ because the noise PSDs are not precisely known, and the data contain non-Gaussian noise transients (glitches), which are known to reduce detector sensitivity of GW bursts like CCSNe \cite{Buikema_2020, GWTC3, allskyO3}. Moreover, Ref.~\cite{Szczepanczyk_2022} considers only the reconstruction accuracy using bandpass analysis and does not address detection sensitivity. Therefore, the central goal of this paper is to develop a robust analysis framework for exploring broader applications of bandpass analyses in realistic observing conditions. Standard GW burst analyses for CCSNe consider the frequency range $32-2048\Hz$, and different classes of glitches occupy different regions of the time-frequency plane~\cite{GravitySpy_O3}. Hence one should ask whether excluding high frequencies from burst analyses can be used to detect, or enhance sensitivity to, low-frequency gravitational wave (GW) signatures, and vice versa. To address this question, we propose the \textit{dedicated-frequency framework} which follows-up eligible detection candidates with bandpass analyses, e.g. the low-frequency (LF; $\leq 256 \Hz$) and high-frequency (HF; $\geq 256 \Hz$) follow-up analyses. By limiting the band, the outcomes of the LF (HF) analyses are not influenced by irrelevant HF (LF) glitches. We assess the astrophysical significance of detection candidates using empirical noise backgrounds drawn from real detector data. 

The dedicated-frequency framework uses a hierarchical analysis pipeline comprising two independent algorithms: coherent WaveBurst (cWB) \cite{cWB1, cWB2, cWB3, cWB_library, Mishra_2024} and \textit{BayesWave} \cite{BayesWave, BayesWave2, BayesWave3}, which are used by the LVK community to characterize generic GW bursts with minimal assumptions about the source and signal morphology~\cite{GWTC1, GWTC2, GWTC2.1, GWTC3}. The performances of cWB and \textit{BayesWave} with CCSNe have been studied separately in Refs. \cite{Szczepanczyk_2021} and \cite{Raza_2022} respectively. However, \textit{BayesWave} is computationally intensive; it uses a reversible jump Markov Chain Monte Carlo (RJMCMC) algorithm to marginalize over its model dimensions. Therefore, \textit{BayesWave} is typically used to follow up cWB candidate events, serving as a semi-independent validation and reassessment of the candidates. One demonstrated benefit of the hierarchical cWB and \textit{BayesWave} pipeline is its ability to improve detection significance~\cite{Kanner_2016}, and its implementation is also a recognized practice in burst searches~\cite{allskyO1, allskyO2, allskyO3}. Accordingly, we present the methodology of the dedicated-frequency framework in the context of the hierarchical pipeline: eligible candidates identified through a full-band cWB search are followed up by \textit{BayesWave} using LF and HF bandpass analyses.

As noted above, the detection of low-frequency GW signatures from the SASI and neutrino-driven convection, or lack thereof, places constraints on the explosion mechanism of a CCSN. Hence the primary objective of this paper is to study whether the detection of low-frequency GW signatures using the LF follow-up analyses can be used to constrain CCSN models. This study uses GW signals extracted from five distinct three-dimensional CCSN simulations, each with varying strengths of low-frequency GW emissions attributed to SASI and neutrino-driven convection. Although targeted CCSNe searches~\cite{CCSN_optical_O1_O2, CCSN_optical_O3, SN2023ixf} are more feasible with current detector sensitivity, many analyses such as Refs.~\cite{Szczepanczyk_2022, allskyO1, allskyO2, allskyO3} adopt an all-sky approach that does not rely on prior knowledge of the source or its location, allowing for unbiased sensitivity estimations across the entire sky. Therefore, for the LF analyses in this paper, we adopt the all-sky approach to minimize bias in our results toward any specific sky location. The secondary objective of this paper is to explore whether the HF follow-up analysis within the dedicated-frequency framework can enhance the detection significance of targeted CCSNe search candidates with minimal low-frequency power. This study is conducted using the loudest event from SN 2019fcn~\cite{CCSN_optical_O3}. To ensure relevance to practical observing scenarios, both the LF and HF studies are carried out using real O3 data~\cite{OpenDataII}.

The rest of this paper is organized as follows. Section \ref{sec:GW_CCSN} discusses the physical processes within CCSNe, which produce GW signatures at different frequencies. Section \ref{sec:dedicated_frequency} presents the dedicated-frequency framework in two parts: (i) an overview of the burst analysis pipelines in Section \ref{sec:analysis_algorithms}, and (ii) the applications and workflow in Section \ref{sec:df_workflow}. Section \ref{sec:LF_CCSN} presents the methods and results of the LF study, demonstrating its application in constraining CCSN explosion mechanisms. Section \ref{sec:HF_SN2019fcn} demonstrates an application of the HF follow-up using the loudest event of SN 2019fcn. We conclude in Section \ref{sec:conclusion}, with discussions of future work and broader applications of the dedicated-frequency framework.

\section{Gravitational-wave signatures of core-collapse supernovae}
\label{sec:GW_CCSN}

CCSNe occur when stars with mass $8\,M_\odot \lesssim M \lesssim 100\,M_\odot$ enter the final stages of exoergic nuclear fusion~\cite{Woosley_2005}. If the iron core of a star exceeds the effective Chandrasekhar mass (${\sim}1.5\,M_\odot$) \cite{Woosley_2002}, the gravitational instability triggers core-collapse. The infall of stellar material compresses the core, forming a protoneutron star (PNS)~\cite{Bethe_1979, Baron_1985}. Once the core density exceeds nuclear density, further compression is no longer possible and a rebound occurs. The rebound process, otherwise known as core bounce, launches an outbound hydrodynamic shock wave~\cite{Colgate_1966}. When the shock encounters the still-collapsing outer core, it loses energy through the dissociation of heavy nuclei into nucleons and neutrinos \cite{Bethe_1990, Burrows_2013}. The outbound shock stalls and fails to produce an explosion. Therefore, a secondary shock-revival mechanism is required to trigger the explosion and this remains an active area of research~\cite{Janka_2012, Janka_2017, Burrows_2021}. 

One way to study the physical mechanisms within CCSNe leading up to their explosion is to analyze the post-bounce GW emissions. The explosion mechanism is strongly influenced by progenitor rotation (see Ref.~\cite{Janka_2012} and references therein). In this paper, we focus on slowly-rotating progenitors as they are expected to be the most common CCSN sources~\cite{Heger_2005}. Their GW emissions exhibit two distinct spectral signatures: (i) a high-frequency component starting at ${\sim} 0.4\kHz$, which increases in frequency over time, and (ii) a low-frequency component confined to $\lesssim 0.25\kHz$. The two components are discussed in Sections~\ref{sec:GW_HF} and \ref{sec:GW_LF} respectively.

\subsection{High frequencies}
\label{sec:GW_HF}

Multi-dimensional numerical models have shown that high-frequency ($\gtrsim0.4\kHz$) GW emissions predominantly originate within the PNS~\cite{Morozova_2018, Mezzacappa_2024}. The processes that give rise to such emissions include sustained Ledoux convection and convective overshoot (see Ref.~\cite{Andresen_2016} and references therein). Sustained Ledoux convection arises from persistent lepton gradients within the PNS~\cite{Wilson_1988}. Convective overshoot occurs when Ledoux convection extends into the stable outer layer~\cite{Murphy_2009}. Quadrupolar ($\ell=2$) oscillations driven by intermittent aspherical accretion onto the PNS \cite{Andresen_2016} also emit high-frequency GWs. The emission of high-frequency GWs typically commences $0.1{-}0.2\s$ post-bounce. In the early stages, the high-frequency emissions are driven by low-order ($n=1,2$) $g$-modes\footnote{Restoring forces of $g$-mode oscillations are exerted by buoyancy i.e. gravity.}. As the PNS contracts and the equation of state (EOS) stiffens over time, the frequency of the GW signal increases. In the later stages, the high-frequency GW emissions are associated with the fundamental ($n=0$) $f$-modes on the surface of the PNS \cite{Morozova_2018, s25}. This evolutionary timeline varies based on the physical properties of the progenitor star (e.g. mass and rotation), as well as the PNS EOS. Altogether, the high-frequency GW emissions probe the evolution of the PNS structure and hence the EOS, but they do not explicitly reveal details about shock revival or the explosion itself.

\subsection{Low frequencies}
\label{sec:GW_LF}
 
Two hydrodynamical instabilities have been identified as plausible shock-revival mechanisms: (i) the standing-accretion-shock instability (SASI)~\cite{Blondin_2002} and (ii) neutrino-driven convection~\cite{Bethe_1990, Mezzacappa_1998}. The SASI arises due to asymmetric radial density and velocity fluctuations, causing the initially spherical shock front to oscillate. The oscillation modulates the accretion flow, and the aspherical movement of matter results in the emission of low-frequency ($\lesssim 0.25\kHz$) GW~\cite{Andresen_2016, Hayama_2018, Mezzacappa_2024}. GW emissions from SASI diminish upon the revival of the shock~\cite{s25, Powell_2021}. The convection scenario, on the other hand, suggests the development of an entropy gradient as the outward shock weakens. The entropy gradient arises due to neutrino heating in the gain region\footnote{The gain region is where the energy gain through neutrino absorption exceeds the energy loss through neutrino emission \cite{Bethe_1985}.}, and the outflow of hot neutrinos revives the shock. The convective instabilities also modulate the accretion flow~\cite{Foglizzo_2006}, but they produce weaker low-frequency GWs~\cite{Andresen_2016}. While the physical mechanisms behind the low-frequency emissions cannot be fully disentangled, the observation of low-frequency GW signatures offers insights into the pre-explosion processes and the timing of shock revival, which are useful for constraining the CCSN explosion mechanism.

 \begin{figure}[t]
     \centering
     \includegraphics[width=\linewidth]{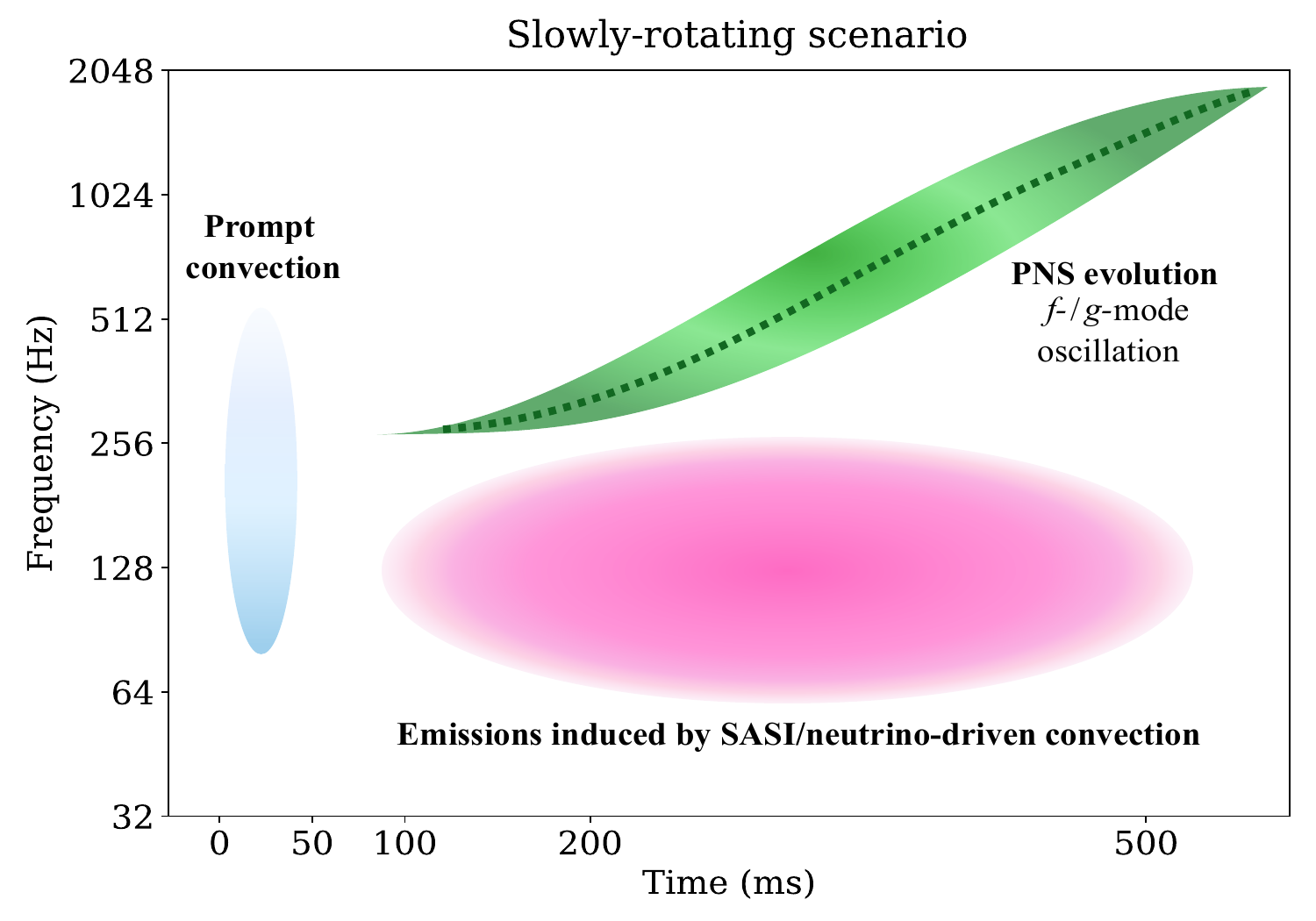}
     \caption{Schematic of GW emissions from CCSNe with slowly-rotating progenitors in the time-frequency plane. The scales on the axes are representative only. The actual durations and spectra of the GW signatures vary depending on the physical properties of the CCSN progenitor, and not all CCSNe will exhibit every signature shown.}
     \label{fig:CCSN_GWs}
 \end{figure}

 In some slowly-rotating progenitors, Ledoux convection occurring shortly (${\lesssim}10\ms$) after core bounce also emits GWs. This short-lived convection, lasting $\lesssim50 \ms$, is otherwise known as prompt convection and is typically followed by a quiescent phase lasting ${\sim}0.1\s$, with minimal GW emissions before the onset of the low- and high-frequency emissions described above. Figure~\ref{fig:CCSN_GWs} shows where the different GW signatures discussed above are located in the time-frequency plane. Simulations have shown that the peak GW frequency of prompt convection varies based on the EOS and numerical perturbation seeds, with values ranging across $0.1{-}0.7\kHz$~\cite{s25, Marek_2009, Ott_2009, Mezzacappa_2024}. Indeed, the GW emissions from prompt convection may overlap in frequency with the low-frequency emissions from SASI and neutrino-driven convection. However, as depicted in Figure~\ref{fig:CCSN_GWs}, the emission from prompt convection does not overlap in time with other low-frequency emissions and hence can be distinguished. Moreover, due to its brief duration, prompt convection contributes only a small fraction of the overall signal power~\cite{Andresen_2016, s25}.

\section{Dedicated-frequency framework}
\label{sec:dedicated_frequency}

As discussed in Section \ref{sec:GW_CCSN}, GW signatures at different frequencies probe different interior dynamics of CCSNe. Can one improve the characterization of frequency-specific GW signatures by ignoring contributions from irrelevant frequency ranges? Here, we introduce the dedicated-frequency framework, which is designed to follow-up GW burst detections with bandpass analyses. In Section \ref{sec:analysis_algorithms}, we overview the relevant burst analysis pipelines used in the framework. In Section \ref{sec:df_workflow} we discuss the applications and workflow of the dedicated-frequency analyses. 

\subsection{Hierarchical burst analysis pipeline}
\label{sec:analysis_algorithms}

The LVK all-sky burst searches \cite{allskyO1, allskyO2, allskyO3} implement multiple minimally-modelled burst analysis algorithms to ensure consistency. Two algorithms are coherent WaveBurst (cWB) and \textit{BayesWave}. The hierarchical pipeline, in which \textit{BayesWave} is used as a follow-up to cWB, enhances GW burst detection confidence \cite{Kanner_2016}, especially for complex waveforms (e.g. binary black holes and generic white noise bursts) with low signal-to-noise ratio (SNR). %Since the dedicated-frequency framework targets CCSN GW signals with nontrivial time-frequency spectrograms, the hierarchical pipeline is an appropriate option. 

The cWB algorithm uses the multi-resolution \textit{WaveScan} transform \cite{Wavescan} to compute excess power and cross-power statistics in the time-frequency data, and employs the constrained maximum likelihood formalism \cite{cWB2, cWB4} to reconstruct the signal waveforms and sky location. Each identified event is tagged with estimated summary statistics, describing the time-frequency structure, signal strength, and coherence across the multi-detector network. In order to limit the trigger production due to non-stationary detector noise, cWB implement vetoes related to some of the summary statistics to reduce excess background and false alarms. For example, one may discard triggers with network cross-correlation coefficient, $c_\mathrm{c}=E_{\rm c}/(E_{\rm c}+E_{\rm n})$, below a nominal threshold ($=0.5$ in this study), where $E_{\rm c}$ and $E_{\rm n}$ denote the coherent signal energy and residual noise energy respectively~\cite{cWB2}. Events passing the summary-statistic vetoes are ranked by cWB's main detection statistic $\etaC$, which is effectively the coherent network SNR, defined by \cite{Mishra_2021, Mishra_2024}
\begin{equation}
    \etaC= \sqrt{\frac{E_{\rm c}}{1+\tilde{\chi}^2(\max \{1,\tilde{\chi}^2\}-1)}}.
    \label{eq:cWB_DS}
\end{equation}
The reduced chi-squared statistic $\tilde{\chi}^2=E_{\rm n}/N_{\rm DoF}$ in Equation \ref{eq:cWB_DS} serves as an auxiliary glitch veto. For coherent signals, $\tilde{\chi}^2$ is approximately unity because the residual noise energy $E_{\rm n}$ is expected to obey the $\chi^2$-distribution, with degrees of freedom $N_{\rm DoF}$ proportional to the number of \textit{WaveScan} pixels used in the event reconstruction. Conventionally, events with $\tilde{\chi}^2>2.5$ are vetoed~\cite{CCSN_optical_O3}. 

On top of the standard cWB vetoes, we use a machine-learning classification algorithm called eXtreme-Gradient Boost (XGBoost) to further distinguish GW signals and noise transients, i.e. to reduce false-alarm triggers \cite{Mishra_2021, Mishra_2022, Mishra_2024}. The XGBoost model for generic burst searches is trained on a representative set of background noise events and stochastic white-noise-burst (WNB) signals\footnote{WNBs are band-limited and temporally localized signals with flat spectra that resemble white noise, but with power exceeding the average noise floor.} that do not correspond to any known GW sources~\cite{Szczepanczyk2022_XGBoost}. Signal and noise classification with XGBoost is achieved by applying additional post-production vetoes on the cWB summary statistics, and the detection statistic reduces to
\begin{equation}
\label{eq:reduced_eta}
    \etaR = \etaC W_{\rm XGB}.
\end{equation}
In Equation~\ref{eq:reduced_eta}, $W_{\rm XGB}$ denotes the XGBoost penalty factor, which has a value between 0 (noise) and 1 (signal) \cite{Mishra_2021}. Unless stated otherwise, all cWB analyses in this paper incorporate the XGBoost optimization and use $\eta_r$ as the detection statistic.

\textit{BayesWave}, on the other hand, takes a Bayesian approach when characterizing non-Gaussian features in the data. The algorithm reconstructs the data $\vb*{d}$ using three independent models, namely the coherent signal plus Gaussian noise ($\mathcal{S}$) model, the incoherent glitch plus Gaussian noise ($\mathcal{G}$) model and the pure Gaussian noise model ($\mathcal{N}$). The $\mathcal{S}$ and $\mathcal{G}$ \textit{BayesWave} models are constructed using sine-Gaussian wavelet frames \cite{BayesWave}. A parallel-tempered Reversible Jump Markov Chain Monte Carlo (RJMCMC) algorithm is used to sample the number of wavelets $N$ and wavelet parameters for model $\mathcal{M} \in \{\mathcal{S}, \mathcal{G}, \mathcal{N}\}$. The associated Bayesian evidence $p(\vb*{d}|\mathcal{M})$ is then computed using thermodynamic integration \cite{BW_TI}. Model selection in \textit{BayesWave} is conducted by comparing the Bayes factor between the models. The detection statistic of \textit{BayesWave} is the log Bayes factor between $\mathcal{S}$ and $\mathcal{G}$,
\begin{equation}
    \bsg = \ln p(\vb*{d}|\mathcal{S}) - \ln p(\vb*{d}|\mathcal{G}),
    \label{eq:CCSN_bsg}
\end{equation}
which scales not only with the network SNR, but also the model complexity quantified by $N$, and the number of detectors $\mathcal{I}$, viz. $\bsg\sim \mathcal{O}(\mathcal{I}N\ln\mathrm{SNR}_\mathrm{net})$ \cite{BayesWave2, Lee_2021}. 

Since the RJMCMC algorithm in \textit{BayesWave} is computationally intensive, \textit{BayesWave} is not used to analyze extended data segments, e.g. all-sky searches. Instead, \textit{BayesWave} serves as a follow-up tool for analyzing targeted data segments flagged by other burst searches. In this work, we first run cWB to analyze the full dataset, and then use \textit{BayesWave} to follow up on triggers that satisfy a nominal cWB detection threshold. The $\etaR$ in cWB compares signals against Gaussian noise, whereas the log Bayes factor $\bsg$ in \textit{BayesWave} compares signals against glitches. This sequence of analyses is called the hierarchical pipeline and is commonly used in the LVK all-sky burst searches~\cite{allskyO1, allskyO2, allskyO3}. %the \textit{BayesWave} follow-up provides a partly independent reassessment of astrophysical significance for cWB triggers. 

\subsection{Applications and workflow}
\label{sec:df_workflow}

The dedicated-frequency framework is intended as a follow-up tool, used exclusively for events that satisfy a benchmark detection criterion. For example, an event qualifies for dedicated-frequency follow-up only if its False Alarm Rate (FAR) is less than or equal to one per year (FAR$\leq 1 \pyr$); we describe how FAR is calculated from the event's detection statistic in Section~\ref{sec:O3_background_ccsn}. In other words, the dedicated-frequency framework does not redefine the standard LVK detection criteria, but instead follows up standard detections. In this framework, eligible candidates are identified using the standard (full-band) burst analysis which covers the band $32{-}2048 \Hz$. Once identified, they can be followed up with low-frequency (LF) and high-frequency (HF) analyses, below and above $256\Hz$, respectively. The LF–HF boundary can be customized to study different sources and scientific questions. The frequency range may also be divided into more than two bands if needed. However, for the purposes of the CCSNe study presented in this paper, we limit ourselves to two bands, LF and HF, to distinguish SASI and neutrino-driven convection ($\lesssim 0.25 \kHz$) from PNS oscillations ($\gtrsim 0.4 \kHz$); additional bands lack clear physical motivation and would incur significant computational cost.

The dedicated-frequency framework implements the hierarchical pipeline as follows. cWB identifies signals by clustering contiguous time-frequency pixels with coherent power exceeding the Gaussian noise floor; a candidate GW event is recorded when the accumulated SNR ($\etaC$) of nearby pixel clusters surpasses a nominal threshold, typically $\etaC=7$. A dedicated low-frequency search using cWB in its current configuration may not effectively detect SASI and neutrino-driven convection emissions, especially in broadband signals, due to insufficient power clustering in the low-frequency range. However, cWB has a low runtime and well-suited for processing extensive datasets. Therefore, we use the full-band cWB analysis to identify triggers, and then use \textit{BayesWave} to follow-up on eligible (i.e. astrophysically-relevant) cWB triggers with the dedicated-frequency (LF/HF) analyses. Eligible cWB triggers are defined based on an arbitrary significance threshold. In this paper, a cWB trigger must satisfy FAR$\leq 1 \pyr$ to qualify for dedicated-frequency follow-up, unless stated otherwise; we discuss this choice in detail in Section~\ref{sec:O3_background_ccsn}. \textit{BayesWave} is used for the dedicated-frequency follow-ups because the successive application of cWB followed by \textit{BayesWave} improves detection significance for signals with non-trivial time-frequency spectrograms~\cite{Kanner_2016}, and is a established practice in burst analyses~\cite{allskyO1, allskyO2, allskyO3}.  

In \textit{BayesWave}, the coherent signal ($\mathcal{S}$) and incoherent glitch ($\mathcal{G}$) models are constructed by summing a set of continuous sine-Gaussian wavelets. A sine-Gaussian wavelet is intrinsically parameterized by its central frequency. Therefore, when restricting the band to LF or HF, we restrict the central frequency prior, not the overall wavelet spectrum. Consequently, some wavelet power during an LF analysis may leak into the HF band, and vice versa. For a model $\mathcal{M} \in \{\mathcal{S}, \mathcal{G}, \mathcal{N}\}$ parameterized by $\vb*{\theta}^\mathcal{M}$, the Bayesian evidence is given by 
\begin{equation}
    p(\vb*{d}|\mathcal{M}) = \int d\vb*{\theta}^\mathcal{M} p(\vb*{\theta}^\mathcal{M}|\mathcal{M})p(\vb*{d}|\vb*{\theta}^\mathcal{M}, \mathcal{M}),
    \label{eq:CCSN_modelevidence}
\end{equation}
where $p(\vb*{\theta}^\mathcal{M}|\mathcal{M})$ is the prior and $p(\vb*{d}|\vb*{\theta}^\mathcal{M}, \mathcal{M})$ is the likelihood. In \textit{BayesWave}, the likelihood is calculated in the frequency domain (see Equation~4 in Ref.~\cite{BayesWave2}), which is restricted to match the dedicated-frequency (LF/HF) band. %This ensures that only the relevant frequencies are included in the likelihood and, thus the $\bsg$ calculation.

For real data containing glitches, the \textit{BayesWave} dedicated-frequency follow-up analyses must be applied to both the background (instrumental noise) and foreground (potential GW) triggers. The background analysis provides false-alarm rate measurements for assessing the astrophysical significance of foreground detections. In addition to the dedicated-frequency follow-ups, \textit{BayesWave} also performs the full-band analysis on eligible events to provide significance estimates independent of cWB.

We demonstrate applications of the LF and HF follow-up analyses in Sections \ref{sec:LF_CCSN} and \ref{sec:HF_SN2019fcn} respectively.

\section{Constraining CCSN explosion mechanisms}
\label{sec:LF_CCSN}

The primary motivation of the dedicated-frequency framework is to constrain CCSN models for eligible detection candidates, in real observing scenarios. The detection of low-frequency ($\lesssim 0.25 \kHz$) GW signatures, or lack thereof, can help select between CCSN models and explosion mechanisms. Furthermore, low-frequency GW emissions fall within the most sensitive frequency bands of existing interferometric GW detectors like LIGO and Virgo\footnote{See representative O3 noise curves in Figure 2 of Ref. \cite{GWTC3}}. In this section, we apply the LF follow-up to simulated GW signals from five distinct CCSN models, featuring different amplitudes of low-frequency GWs induced by the SASI or neutrino-driven convection. We then assess whether CCSN models with prominent low-frequency GW signatures achieve higher detection efficiency with the LF follow-up, compared to those with little to no low-frequency emissions.

In Section \ref{sec:CCSN_models}, we detail the key features of the five CCSN models used in this study. The CCSN signals are injected into real O3 data. Noise background measurements are required to evaluate the significance of burst triggers. We present the background measurements in Section \ref{sec:O3_background_ccsn}. In Section \ref{sec:ccsn_injections}, we discuss the properties of the injected CCSN signals and how we ensure that they are detectable up to a nominal significance threshold in O3 data. The analysis results are presented and interpreted in Section \ref{sec:ccsn_results}.

\subsection{CCSN models}
\label{sec:CCSN_models}

We use the predicted GW waveforms of five three-dimensional CCSN models. The selected models are a subset of those used to test the LVK search sensitivity for GWs associated with SN 2023ixf \cite{SN2023ixf}, one of the closest CCSNe observed in the last decade. For this proof-of-principle study, we choose models that represent typical CCSNe, where the progenitors have solar metallicity and do not rotate~\cite{Janka_2017, Abdikamalov_2020, allskyO3}.
\begin{itemize}
    \item The SFHx model \cite{SFHx}, otherwise known as s15, has a progenitor with zero-age main sequence (ZAMS) mass equal to $15\,M_\odot$ and a soft, i.e. low-pressure, EOS as described in Ref. \cite{SFHx_eos}. The soft EOS results in more vigorous SASI activity, and hence strong low-frequency GW emissions ($0.05{-}0.2\kHz$). The high-frequency component of the signal reaches up to ${\sim}1\kHz$ and is attributed to the PNS surface $g$-mode oscillations. This simulation is truncated ${\sim}0.35\s$ after core bounce.
    \item The s25 model \cite{s25} has a progenitor with ZAMS mass equal to $25 \, M_\odot$ and assumes the SFHo EOS \cite{SFHx_eos}, which differs slightly from SFHx in terms of the mass-radius relationship. The GW signal starts with distinct low-frequency GW emissions at ${\sim}0.1\kHz$ associated with prompt convection\footnote{Figure 4 in Ref.~\cite{s25} shows that the GW energy emitted $\lesssim0.1\s$ after core bounce constitutes $\lesssim2$\% of the total GW energy. That is, prompt convection contributes only a small fraction of the total signal power. Our analysis of s25 waveforms injected into O3 data also shows that the recovered power within $\lesssim 0.1 \s$ after core bounce constitutes $< 0.1$\% of the total recovered power, suggesting that prompt convection is generally not detectable in the presence of a colored PSD.}. This is followed by distinct low-frequency emissions ($0.05{-}0.2\kHz$) associated with SASI, and high-frequency emissions ($0.4{-}1\kHz$) associated with the surface $f$- and $g$-modes of the PNS. The simulation is truncated ${\sim}0.6 \s$ after core bounce.
    \item The D15 model \cite{D15} has a progenitor with ZAMS mass equal to $15 \, M_\odot$ and is simulated using the D-series \textsc{Chimera} code \cite{Chimera}. The GW signal is dominated by high-frequency emissions peaking at ${\sim}1\kHz$, largely due to Ledoux convection in the PNS. There are also secondary emissions below ${\sim}$0.25$\kHz$ associated with the SASI and neutrino-driven turbulent convection. The explosion occurs ${\sim}0.5 \s$ after core bounce and the simulation is truncated at ${\sim}0.75\s$.
    \item The mesa20\_pert model~\cite{mesa20_pert} has a progenitor with ZAMS mass equal to $20 \, M_\odot$ and assumes the SFHo EOS \cite{SFHx_eos}. Precollapse perturbations are introduced through an aspherical matter velocity field, leading to increased turbulence in the gain region. The GW signal is dominated by high-frequency components ($0.3{-}1.2\kHz$) associated with the PNS contraction, and is accompanied by low-frequency components ($0.05{-}0.2\kHz$) due to convection and the SASI. The simulation is truncated ${\sim} 0.52 \s$ after core bounce.
    \item The s18 model~\cite{s18} has a progenitor with ZAMS mass equal to $18\, M_\odot$ and is simulated using the neutrino hydrodynamics \textsc{coconut-fmt} code \cite{coconut_fmt}. The GW signal peaks between $0.8{-}1\kHz$ due to $g$-mode PNS oscillations. There is minimal low-frequency emission associated with the SASI. The shock revival driving the explosion occurs $\sim 0.25 \s$ after core bounce, and the simulation is truncated at $0.89\s$.
\end{itemize}
The extraction of GW signals from CCSN simulations is computationally expensive, so the simulations of SFHx~\cite{SFHx}, s25~\cite{s25} and mesa20\_pert~\cite{mesa20_pert} are truncated before the GW signal is fully developed, i.e. they exclude signals from the shock revival phase. %\ysl{We discuss how GWs are extracted from CCSN simulations in Appendix \ref{app:GW_CCSN}.}

\begin{figure*}[t]
    \centering
    \includegraphics[width=\textwidth]{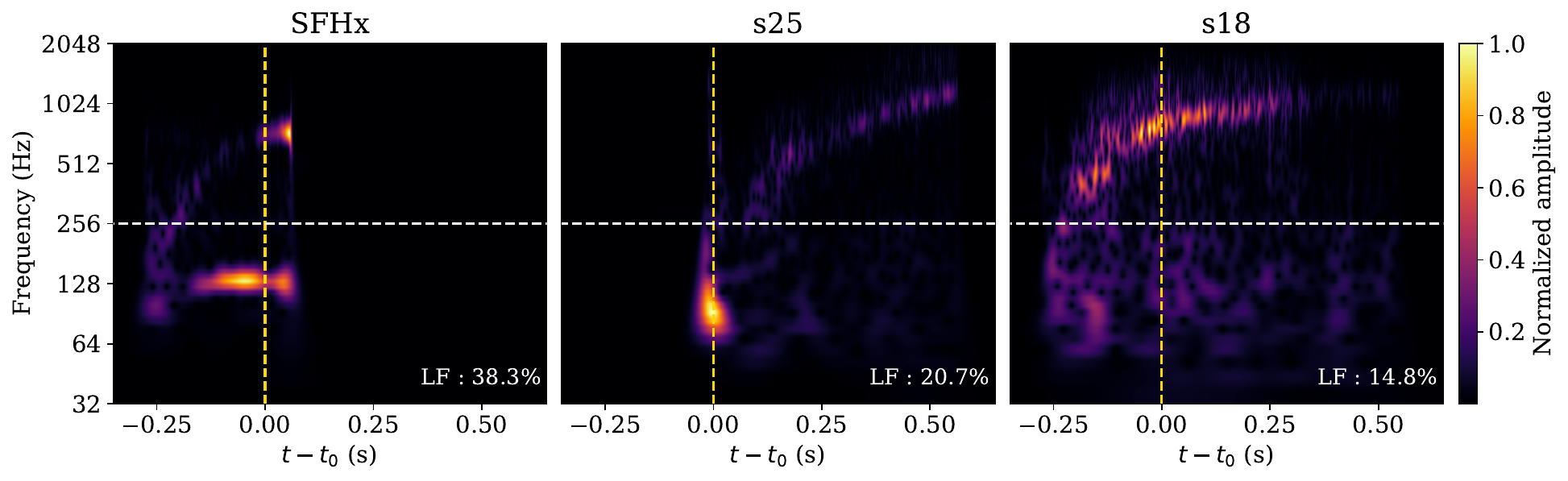}
    \caption{CWT time-frequency spectrograms of a sample GW signal for SFHx (left), s25 (middle) and s18 (right). The vertical axis shows the full-band analysis frequency range (32$-$2048$\Hz$) and the white horizontal lines at $256\Hz$ indicate the boundary between LF and HF components, as per the dedicated-frequency framework. The horizontal axis shows the time $t$ relative to the central time $t_0$ of the signal, and the orange vertical lines indicate $t=t_0$. The horizontal scales are the same for all three plots, and the approximate duration of the SFHx, s25 and s18 signals are 350 ms, 625 ms and 900 ms respectively. The color bar represents the linearly scaled amplitude of the signals; the minimum and maximum amplitudes in each panel correspond to the values 0 and 1 respectively. The LF energy divided by the total signal energy is quoted in the bottom right corner of each plot.}
    \label{fig:CCSN_spectrograms}
\end{figure*}

Different SASI signatures are observed for the selected models. The models are listed above in descending order of low-frequency power, i.e. SFHx waveforms generally have the strongest LF GW emissions, followed by s25 and so on. We use the continuous wavelet transform (CWT)~\cite{Henshaw_2024} to quantify the LF power in each model. To visualize the process, we show the CWT time-frequency spectrograms of arbitrary selected sample waveforms from SFHx (left), s25 (middle) and s18 (right) in Figure \ref{fig:CCSN_spectrograms}. The white horizontal lines at $256\Hz$ divide the LF and HF components below and above the lines. The horizontal axis in Figure \ref{fig:CCSN_spectrograms} shows the time, $t$, relative to the squared-strain-weighted central time,
\begin{equation}
    t_0 = \frac{1}{h_{\rm{rss}}^2}\int_{-\infty}^{\infty} dt \, h(t)^2t.
    \label{eq:central_time}
\end{equation} 
 The root-sum-squared (rss) strain amplitude of the $+$ and $\times$ polarizations,
\begin{equation}
    h_{\rm{rss}}^2=\int_{-\infty}^{\infty} dt \, [h^2_+(t) + h^2_\times(t)],
    \label{eq:hrss}
\end{equation}
is proportional to the total energy of the GW signal, and serves as a normalization factor in Equation \ref{eq:central_time}. The orange vertical lines in Figure~\ref{fig:CCSN_spectrograms} indicates $t=t_0$ of each waveform. By the definition in Equation \ref{eq:central_time}, $t_0$ is the time at which the majority of the signal energy is concentrated. Visual inspection of Figure \ref{fig:CCSN_spectrograms} reveals that the LF features of SFHx (left) and s25 (middle) are close to $t_0$, suggesting LF emissions for SFHx and s25 contribute considerably to the total signal power. For s18, on the other hand, $t_0$ roughly aligns with the peak emission at ${\sim}1\kHz$, suggesting HF emissions have a stronger influence. To quantify the extent of LF emissions, we compute the signal energy for frequency $f\leq256 \Hz$ (i.e. below the white horizontal lines) divided by the total signal energy in the range $32 \Hz \leq f \leq 2048 \Hz$. For the particular waveforms plotted in Figure \ref{fig:CCSN_spectrograms} for the models SFHx, s25 and s18, the LF contributions are 38.3\%, 20.7\% and 14.8\% respectively. We repeat this calculation for 175 randomly realized waveforms per CCSN model. We find that, on average, the LF emissions contribute 36.6\%, 19.4\%, 18.4\%, 16.2\% and 14.9\% to the overall signal energy of the SFHx, s25, D15, mesa20\_pert and s18 models respectively. 

Note that Figure~\ref{fig:CCSN_spectrograms} and the LF contributions presented above are derived from the raw $ h(t)$ signal, without accounting for the detector response. When the signal is injected into real detector noise, the recovered $ t_0 $ and $ h_{\rm{rss}} $ are expected to differ in accordance with the noise PSD, which reflects the detector’s spectral sensitivity.

\subsection{Background measurements}
\label{sec:O3_background_ccsn}

To demonstrate the application of LF follow-ups in realistic observing scenarios, we inject GW signals from the five CCSN models into O3 data. Real detector data are susceptible to glitches and are therefore capable of producing false alarm triggers. In order to assess the significance of triggers produced by an analysis pipeline, one has to empirically measure the noise background, i.e. the rate of false alarms produced by the corresponding pipeline in the absence of GW signals. Trigger sensitivity varies unpredictably across different analysis pipelines. To address this, cWB first measures the background for all of O3, and \textit{BayesWave} independently follows up on the astrophysically-relevant cWB triggers to evaluate its own background. The independent background measurements with cWB and the \textit{BayesWave} follow-up account for the individual strengths and shortcomings of each algorithm, thereby improving the reliability of their respective significance estimates. The background measurements are used to evaluate the significance of detection candidates in terms of the FAR, which in turn is used to assess the eligibility of detection candidates for the dedicated-frequency follow-up.

We use the standard time-shift analysis to conduct the background measurements~\cite{timeslide_2010}. That is, we produce artificially extended detector background data by introducing temporal offsets, which are long enough to nullify any meaningful correlations between the outputs of two or more detectors. We choose to use the data from a two-detector configuration, comprising the LIGO Hanford (H) and LIGO Livingston (L) detectors. Previous works have shown that a three-detector configuration, comprising HL plus Virgo, does not outperform the HL-only network in terms of detection efficiency in O3 \cite{allskyO3, Szczepanczyk2022_XGBoost, Lee2024}. Altogether we accumulate 605 years of HL background by applying ${\sim} 2.4\times10^4$ time-shifts on 9.5 days of O3a data\footnote{We use O3a data because its overall glitch rate is lower than in O3b~\cite{GWTC3}. More specifically, we leverage the higher burst detection sensitivity in O3a to produce more detection candidates eligible for dedicated-frequency follow-ups.}. 

\subsubsection{cWB background}
\begin{figure}
     \centering
     \includegraphics[width=0.95\linewidth]{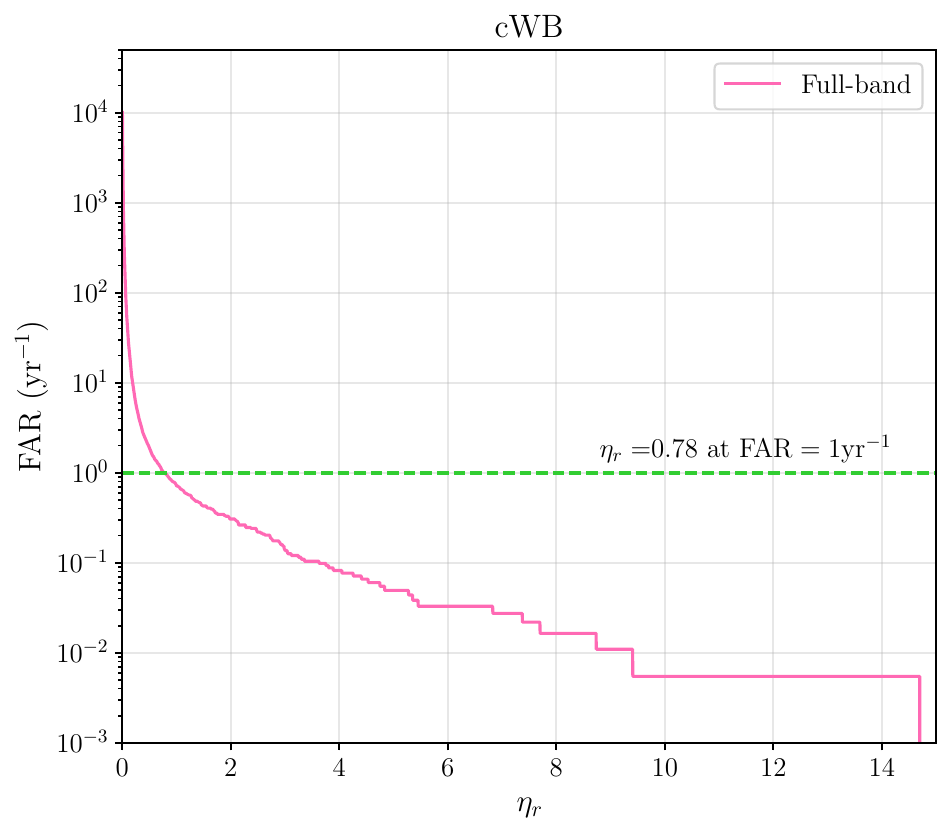}\\
     \includegraphics[width=0.95\linewidth]{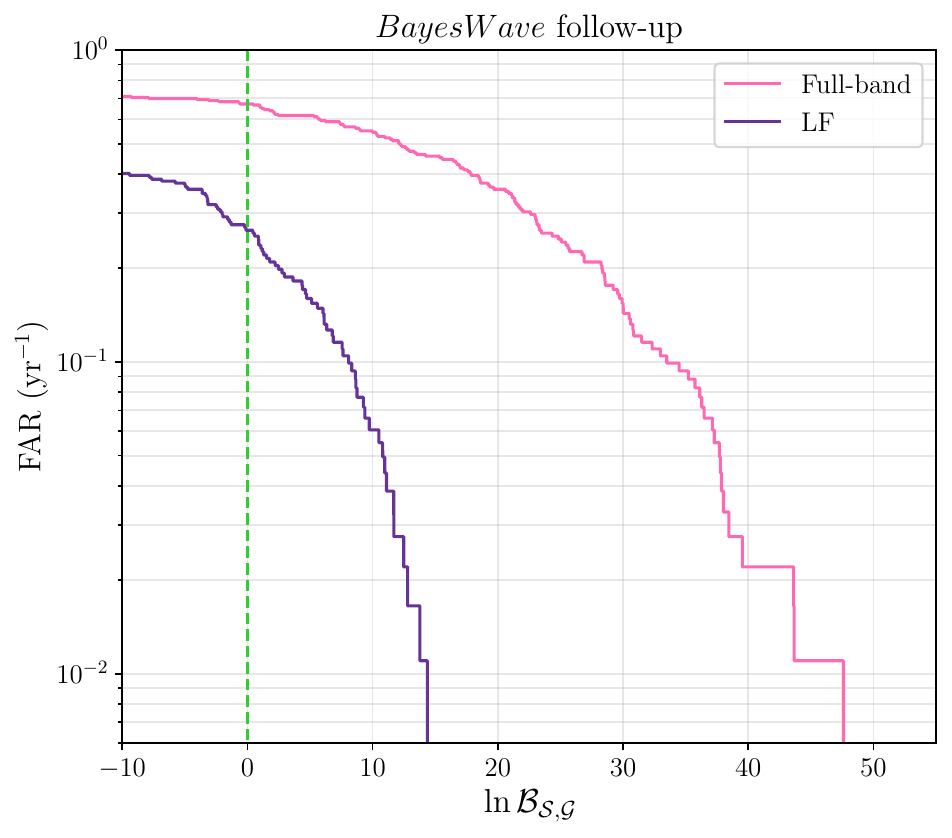}
     \caption{O3a background measurements. The top panel shows the FAR of the background triggers as a function of the detection statistic $\etaR$, for the full-band cWB analysis. The bottom panel shows the same but for the \textit{BayesWave} full-band (pink curve) and LF (purple curve) triggers, as a function of the detection statistic $\bsg$. The horizontal green line at FAR $=1\pyr$ (top panel) and the vertical green line at $\bsg = 0$ (bottom panel) indicate the detection thresholds for cWB and \textit{BayesWave} respectively.}
     \label{fig:CCSN_background}
 \end{figure}

The cWB background is measured by analyzing the entire time-shifted data set. The same XGBoost model is used for both the background and foreground cWB analyses. In this study, the model is trained using: (i) a randomly selected subset (70\%) of the background noise events and (ii) four sets of WNB signals with central frequency, bandwidth and duration uniformly sampled from overlapping subsets within the ranges $[24, 2048]\Hz$, $[10, 800]\Hz$ and $[0.1, 500]\ms$ respectively. The parameters of the WNB training sets are chosen based on the expected time-frequency volume of GW bursts (see Appendix A of Ref.~\cite{Szczepanczyk2022_XGBoost} for more details). The XGBoost training parameters for this study are similar to those in Ref.~\cite{SN2023ixf}. 

 cWB does not perform the dedicated-frequency follow-ups; it is only used to identify eligible candidates via the full-band analysis. Therefore, only the full-band background measurement is necessary. This measurement is conducted using the remaining 30\% (182 years) of data not used for the XGBoost model training, and is shown in the top panel of Figure~\ref{fig:CCSN_background}. The FAR is calculated as the number of background cWB triggers exceeding the corresponding $\etaR$, divided by the total background livetime of 182 years. The green horizontal line at FAR $=1\pyr$ indicates the nominal detection threshold, below which a cWB trigger qualifies as a detection candidate in this study, corresponding to $\etaR=0.78$. That is, a cWB trigger must satisfy $\etaR\geq0.78$ to qualify as a LF follow-up candidate. The reason for choosing the FAR $=1\pyr$ threshold is as follows. In the O3 all-sky burst search \cite{allskyO3}, an event is considered a significant detection for FAR $\leq 0.01 \pyr$. However, Ref. \cite{allskyO3} shows that typical non-rotating CCSNe with solar-metallicity, i.e. the models used in this study, are undetectable at such low FARs with O3 detector sensitivities. Therefore, to showcase the applications of the dedicated LF follow-up for anticipated CCSN detections with existing detector configurations, we arbitrarily increase the significance tolerance to FAR $\leq 1 \pyr$.

\subsubsection{\textit{BayesWave} background}
\label{sec:BW_background_CCSN}

 Eligible cWB triggers are followed up using the full-band and LF \textit{BayesWave} analyses. The full-band analysis checks that the cWB trigger is also detected by \textit{BayesWave}, while the LF analysis checks for low-frequency GW signatures, or the lack thereof. Therefore, we must conduct two separate background measurements for \textit{BayesWave}, one using the full-band analysis and the other using LF. \textit{BayesWave} is computationally expensive and therefore it is impractical to follow-up the entire time-shifted data (cf. cWB). Instead, \textit{BayesWave}'s background is measured by following-up cWB background triggers above a nominal significance threshold. This threshold is the same as the dedicated-frequency follow-up criterion, FAR $=1 \pyr$, discussed above. Since triggers with FAR $> 1 \pyr$ are not valid candidates for the LF follow-up, it is unnecessary to quantify their detection significance; they can be excluded from \textit{BayesWave}'s background measurements, to conserve computational resources. 

The \textit{BayesWave} analyses are configured as follows.  The analysis segment spans 4 seconds, centered on the event epoch, to provide a sufficiently long data segment for accurate noise PSD estimation. The event epoch corresponds to the central time of the trigger as recorded by cWB. However, burst signals are typically shorter than a second. Therefore, the time window in which \textit{BayesWave} is allowed to place signal and glitch wavelets is limited to 1 second, also centered on the event epoch. The sampling rate is twice the maximum frequency of the analysis band, e.g. the maximum frequency of the full-band analysis is 2048~Hz, so the sampling rate is 4096~Hz. This configuration applies to all \textit{BayesWave} analyses in this paper, and is similar to that used in Ref.~\cite{Raza_2022} for CCSNe analysis.

The bottom panel of Figure \ref{fig:CCSN_background} shows the backgrounds measured by \textit{BayesWave}; the pink and purple curves show the full-band and LF measurements respectively. The \textit{BayesWave} FAR is calculated the same as with cWB, except that $\etaR$ is replaced by $\bsg$. An event must be more consistent with the signal model than the glitch model to qualify as an astrophysically-relevant \textit{BayesWave} trigger, i.e. it must satisfy $\bsg > 0$. For reference, the green vertical line in the bottom panel of Figure~\ref{fig:CCSN_background} indicates where $\bsg = 0$. In the full-band and LF background, $\bsg = 0$ correspond to $\mathrm{FAR}=0.68\pyr$ and $0.26\pyr$ respectively. It is expected that $ \mathrm{FAR}<1\pyr$ for $\bsg>0$, because only a subset of cWB triggers with FAR $\leq1\pyr$ also qualify as \textit{BayesWave} triggers. Henceforth we focus our discussion on the background measurements for $\bsg>0$, as they represent the distribution of false alarms with astrophysically relevant detection statistics. 
 
The LF \textit{BayesWave} background (purple curve) is approximately an order of magnitude lower than the full-band background (pink curve) on average. There are two reasons for this. First, $\bsg$ scales with the SNR, as well as with the number of wavelets $N$, which reflects the model complexity. In the LF analysis, the HF contributions of the background triggers are disregarded, which reduces the trigger SNRs and the number of wavelets required for their \textit{BayesWave} reconstruction; these reductions collectively result in lower $\bsg$. Second, for events with minimal LF power, disregarding the HF contributions can cause the data to become more  consistent with the Gaussian noise model ($\mathcal{N}$) than with the non-Gaussian signal model ($\mathcal{S}$). That is, the log Bayes factor between the signal and Gaussian noise model ($\ln \mathcal{B}_{\mathcal{S}, \mathcal{N}}$) includes values below zero within its error bars $\Delta \ln\mathcal{B}_{\mathcal{S},\mathcal{N}}$, viz.
\begin{equation}
    \ln\mathcal{B}_{\mathcal{S},\mathcal{N}} \pm \Delta \ln\mathcal{B}_{\mathcal{S},\mathcal{N}} \leq 0.
    \label{eq:BW_nontriggers}
\end{equation} 
Events that satisfy Equation~\ref{eq:BW_nontriggers} are not astrophysically relevant and are therefore excluded from the FAR calculation, regardless of their $\bsg$. With this criterion in place, 17\% of the cWB background triggers are excluded from the \textit{BayesWave} full-band background measurements, cf. 49\% for the \textit{BayesWave} LF measurements. In other words, the LF analysis significantly reduces the noise background, which further explains why the LF background (purple curve) is lower compared to the full-band background (pink curve). 

Overall, the background measurements show that a LF-analysis \textit{BayesWave} trigger can achieve the same FAR as a full-band analysis trigger, with a lower $\bsg$.

\subsection{CCSN injection properties}
\label{sec:ccsn_injections}

As noted previously, only triggers with FAR $\leq 1 \pyr$ qualify for the \textit{BayesWave} LF follow-ups. Therefore we must ensure that the injected CCSNe also satisfy FAR $\leq 1 \pyr$ in the cWB full-band analysis, to qualify for the LF follow-up study. Here, we discuss how to use cWB to compute the appropriate signal amplitudes for the CCSN injections.

\begin{figure}[t]
    \centering
    \includegraphics[width=\linewidth]{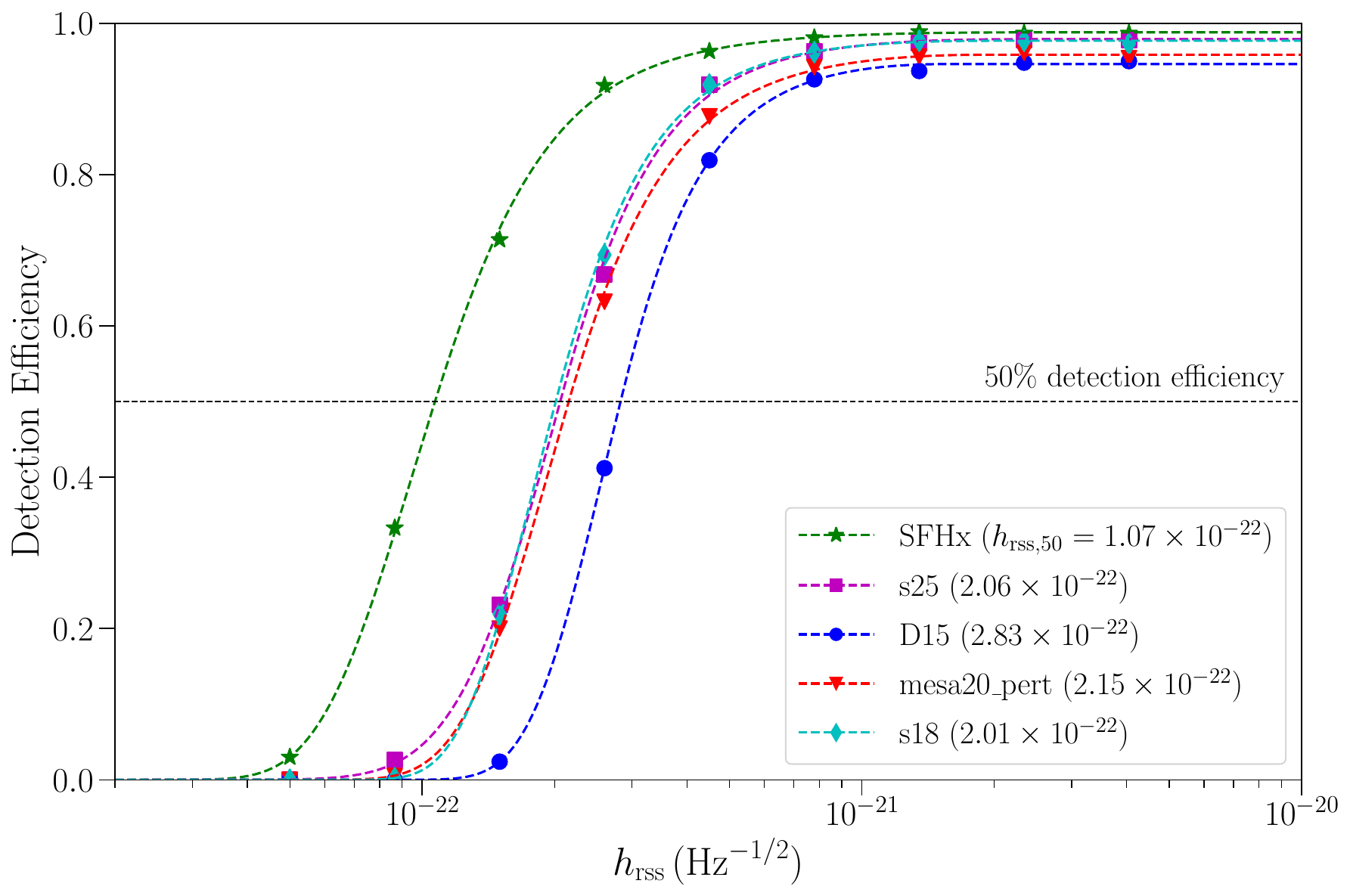}
    \caption{Detection efficiency for events with FAR $\leq 1 \pyr$ versus signal amplitude $h_{\rm{rss}}$. Each point represents the empirically measured detection efficiency with ${\sim}500$ injections. The colored points represent different CCSN models as indicated by the legend, and the dashed curves in corresponding colors show the least-square fit to a cumulative log-normal distribution. The numbers in parentheses are $h_{\rm{rss}, 50}$ (in units of $\Hz^{-1/2}$), i.e. the $h_{\rm{rss}}$ value that results in 50\% detection efficiency, as indicated by the horizontal dashed line.}
    \label{fig:hrss50}
\end{figure}

The detectability of GW bursts at a given FAR is typically quantified by their detection efficiency, as a function of $h_{\rm{rss}}$, defined in Equation \ref{eq:hrss}. The detection efficiencies are evaluated empirically by injecting the same set of signals into detector noise at different $h_{\rm{rss}}$, and then calculating the fraction of signals that are recovered by the full-band cWB analysis with FAR $\leq 1 \pyr$. In the LF analysis, the CCSN signals are injected at an amplitude $h_{\rm{rss}, 50}$, corresponding to 50\% detection efficiency. This choice aligns with the benchmark sensitivity used in standard burst searches and ensures that the CCSN signals are both detectable and eligible for LF follow-up. The model-specific $h_{\rm{rss}, 50}$ is evaluated as follows. First, we choose eight approximately uniform $h_{\rm{rss}}$ values from the range $5\times10^{-23} \, \mathrm{Hz}^{-1/2}\leq h_{\mathrm{rss}} \leq 4\times10^{-21}\, \mathrm{Hz}^{-1/2}$. Then, for each CCSN model, we inject ${\sim}500$ signals per $h_{\rm{rss}}$ value\footnote{The number of injections varies across the CCSN models, ranging from 526 to 585 injections, depending on the number of available waveforms for each model. However, for each CCSN model, the number of injections per $h_{\rm{rss}}$ value is the same.}, and compute the corresponding detection efficiency. The detection efficiencies as a function of $h_{\rm{rss}}$ are plotted as discrete data points in Figure \ref{fig:hrss50}. The different colors indicate different CCSN models. The $h_{\rm{rss}, 50}$ for each CCSN model is obtained by least-squares fitting a cumulative log-normal distribution function of $h_{\rm{rss}}$. Figure \ref{fig:hrss50} shows the cumulative log-normal fits as dashed curves in colors corresponding to the data points they are fitting. The $h_{\rm{rss}, 50}$ value for each model is enclosed within the parentheses in the legend. These are the $h_{\rm{rss}}$ values at which we inject the waveforms for the LF follow-up study.

For an isotropically emitting source, the source distance $r$ scales inversely with the $h_{\rm{rss}}$ as~\cite{Sutton_2013}:
\begin{equation}
    r^2 = \frac{GE_{\rm GW}}{\pi^2c^3h_{\rm{rss}}^2f_0^2}.
    \label{eq:hrss_vs_e}
\end{equation}
In Equation~\ref{eq:hrss_vs_e}, $G$ and $c$ respectively denote the gravitational constant and the speed of light, and $E_{\rm GW}$ is the fixed total energy emitted by the GW source. The central frequency $f_0$ is defined as
\begin{align}
    %f_0 &= \frac{1}{h_{\rm{rss}}^2}\int_{-\infty}^{\infty} df\, \abs{\tilde{h}(f)}^2 f\\
    f_0=\frac{2}{h_{\rm{rss}}^2}\int_{0}^{\infty} df\, \abs{\tilde{h}(f)}^2 f 
    \label{eq:central_freq}, 
\end{align}
where $\tilde{h}(f)=\tilde{h}(-f)$ denotes the Fourier transform of the real-valued time-domain amplitude $h(t)$. Table~\ref{table:hrss_dist} lists the $E_{\rm GW}$ and the peak frequency $f_{\rm peak}$ of the GW signal for the five CCSN models. Assuming that $f_{\rm peak}$ roughly approximates $f_0$, one can use Equation~\ref{eq:hrss_vs_e} to estimate the source distance $r_{50}$ corresponding to $h_{\rm{rss}, 50}$ for each CCSN model~\cite{Szczepanczyk2022_XGBoost}. The $h_{\rm{rss}, 50}$ from Figure~\ref{fig:hrss50} and the corresponding $r_{50}$ are also listed in Table~\ref{table:hrss_dist}. Using 50\% detection efficiency at FAR $\leq 1 \pyr$ as the detection benchmark, we find that all five CCSN models in this study are detectable at distances no greater than that of the Galactic Center (8.5 kpc). Among the five models, SFHx has the highest detectability, reaching distances up to 7.8 kpc; mesa20\_pert has the lowest detectability, reaching only 0.9 kpc.

\begin{table}[t]
\setlength{\tabcolsep}{0pt}
\rowcolors{4}{white}{white}
\renewcommand{\arraystretch}{1.4} 
\begin{tabular}{
    >{\centering\arraybackslash}p{2.2cm}
    >{\centering\arraybackslash}p{1.8cm}
    >{\centering\arraybackslash}p{1.3cm}
    >{\centering\arraybackslash}p{1.95cm}
    >{\centering\arraybackslash}p{1.2cm}
}
\hline

Model & $E_{\rm GW}$ [$M_\odot c^2] $ & $f_{\rm peak}$ [Hz] & $h_{\rm{rss}, 50}$ [Hz$^{-1/2}$] & {$r_{50}$ [kpc]} \\
\hline
% Data Rows
SFHx & $1.1\times 10^{-9}$ & $267$ & $1.07\times 10^{-22}$ & $7.8$ \\
s25 & $2.7\times 10^{-8}$ & $1132$ & $2.06\times 10^{-22}$ & $4.9$ \\
D15 & $8.9\times 10^{-9}$ & $1102$ & $2.83\times 10^{-22}$ & $2.1$ \\
mesa20\_pert & $9.4\times 10^{-10}$ & $1103$ & $2.15\times 10^{-22}$ & $0.9$ \\
s18 & $1.6\times 10^{-8}$ & $818$ & $2.01\times 10^{-22}$ & $5.3$ \\
\hline
\end{tabular}
\caption{Estimating the source distance $r_{50}$ corresponding to 50\% detection efficiency. The columns from left to right list: (i) the CCSN models, (ii) the fixed total energy emitted by the source $E_{\rm GW}$, (iii) the peak frequency $f_{\rm peak}$ of the GW signal, (iv) the $h_{\rm{rss}, 50}$ derived from Figure~\ref{fig:hrss50}, and (v) the corresponding $r_{50}$ calculated using Equation~\ref{eq:hrss_vs_e}, assuming that $f_0 \approx f_{\rm peak}$.}
\label{table:hrss_dist}
\end{table}

The source distance estimates in Table~\ref{table:hrss_dist} are broadly consistent with those from the O3 targeted CCSN search, as reported in Table III of Ref.~\cite{CCSN_optical_O3}. Two possible reasons for the discrepancy in the distance estimates are: (i) our analysis adopts a stricter detection benchmark of FAR $\leq 1 \pyr$, whereas Ref.~\cite{CCSN_optical_O3} tolerates higher FARs corresponding to the loudest event associated with each CCSN; and (ii) our estimates are derived from an all-sky search, whereas Ref.~\cite{CCSN_optical_O3} is based on targeted searches. Since Ref.~\cite{CCSN_optical_O3} uses a less stringent detection benchmark, one would expect its $r_{50}$ estimates to exceed those in Table~\ref{table:hrss_dist}. This, however, is not always true; specifically, the $r_{50}$ for s25 and s18 in Table~\ref{table:hrss_dist} are further than all corresponding estimates in Ref.~\cite{CCSN_optical_O3}. This comparison suggests that targeted searches do not necessarily provide better sensitivity than all-sky searches, justifying our use of an all-sky search to minimize bias in our results toward any particular sky location.

\subsection{Results and discussion}
\label{sec:ccsn_results}

We inject approximately 350 waveforms per CCSN model at the corresponding $h_{\rm{rss}, 50}$ (as per Figure \ref{fig:hrss50}), into the same 9.5-day segment of O3a data that produced the background measurements in Figure \ref{fig:CCSN_background}. We perform the full-band cWB analysis on the injections to identify eligible candidates with FAR $\leq1\pyr$ for the LF follow-up analysis. The lowest fraction of candidates (71\% of 384 injections) is obtained for model s25, while the highest (80\% of 378 injections) is for mesa20\_pert. For each model, we arbitrarily select 175 of the eligible candidates, to ensure that all CCSN models have equal-sized datasets for a fair comparison. 

\begin{figure}[t]
     \centering
     \includegraphics[width=0.95\linewidth]{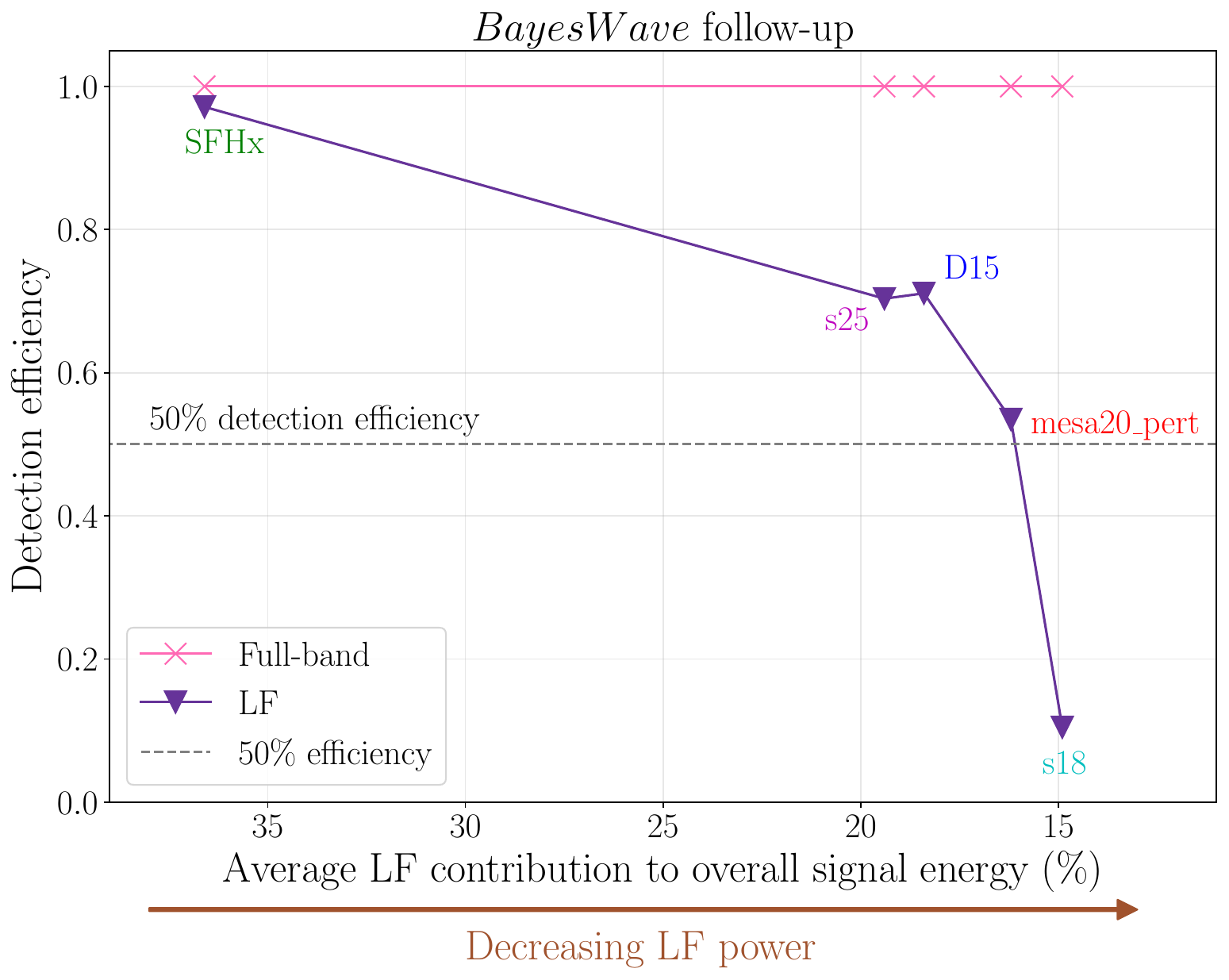}
     \caption{Detection efficiency for events with FAR $\leq 1 \pyr$ for the five CCSN models, based on the \textit{BayesWave} follow-up analyses. The purple triangles and pink crosses show the efficiencies for the LF and full-band analyses respectively. The horizontal axis shows the average LF contribution to the overall signal energy, as reported in Section~\ref{sec:CCSN_models}. CCSN model names are displayed next to their corresponding LF data points to aid interpretation. The gray dashed line indicates 50\% detection efficiency, for reference.}
     \label{fig:CCSN_detEff}
 \end{figure}

The selected cWB candidates are followed-up using the \textit{BayesWave} LF and full-band analyses. For each CCSN model and analysis, we calculate the detection efficiency, defined as the number of candidates recovered with $\bsg \geq 0$ by \textit{BayesWave}, divided by the total number of candidates (175). Figure~\ref{fig:CCSN_detEff} presents the detection efficiency of the five CCSN models as a function of their average LF contribution to the total signal energy. As per the dedicated-frequency follow-up criteria, the detection efficiency for the cWB full-band analysis must be unity. Therefore, it is unsurprising that the \textit{BayesWave} full-band analysis, i.e. the pink crosses in Figure~\ref{fig:CCSN_detEff}, yields the same results. This observation also suggests that the detection efficiency of the full-band analysis is independent of the signal’s LF content. In contrast, the detection efficiencies of the \textit{BayesWave} LF follow-up analysis are non-trivial. We find that the LF detection efficiency generally decreases from left to right across the plot, i.e. as the LF power in the CCSN model decreases. We note, in particular, that the s18 model, which exhibit minimal SASI and has the lowest LF power ($14.9\%$ of the total power), has the lowest detection efficiency at 0.10. In contrast, the detection efficiency for SFHx, with 36.6\% LF power and strong SASI emission, amounts to 0.97. That is, the detectability of SFHx with the LF follow-up is highly likely. However, for s25, D15 and mesa20\_pert, which have ${\sim}10\%$ less LF power than SFHx but still exhibit SASI emissions, the detection efficiency ranges from 0.5 to 0.75. That is, CCSN models with moderate SASI-related LF emission have a fair chance of being overlooked in LF follow-up analyses. In Appendix~\ref{app:overlap_CCSN}, we analyze the reconstruction accuracy for CCSN waveforms, confirming that successful detections in both full-band and LF analyses are accurately reconstructed, whereas non-detections are poorly reconstructed.

\begin{figure}[t]
     \centering
     \includegraphics[width=\linewidth]{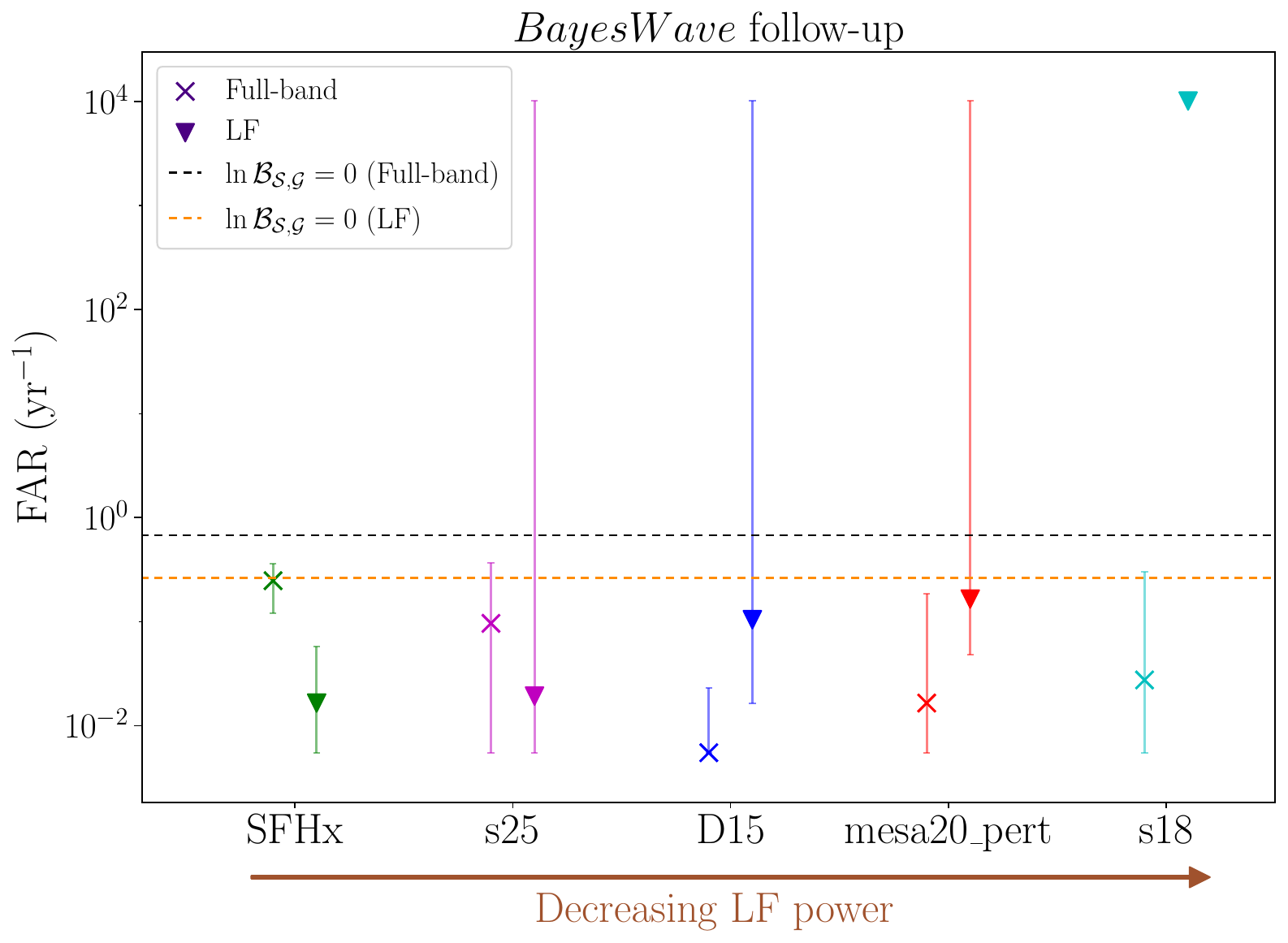}
     \caption{False alarm rate (FAR) of the CCSN injections produced by the \textit{BayesWave} full-band and LF analyses. The crosses (triangles) show the median FAR of the 175 injections for each CCSN model, obtained using the full-band (LF) analysis. The error bars indicate the interquartile range (IQR) of the FARs. Non-detections, as defined by Equation~\ref{eq:BW_nontriggers}, are assigned the maximum FAR ($1\times10^4\pyr$) of the cWB background. The LF error bars for the s18 model are invisible because all injections within the IQR are non-detections. The black (orange) horizontal dashed line at $\mathrm{FAR}=0.68\pyr$ ($0.26\pyr$) indicate the \textit{BayesWave} detection threshold for the full-band (LF) analysis, below which an event is considered a detection. These thresholds correspond to $\bsg=0$ and are derived from the bottom panel of Figure~\ref{fig:CCSN_background}. The CCSN models are displayed on the horizontal axis in order of decreasing LF power, equally spaced from left to right.}
     \label{fig:CCSN_FAR}
 \end{figure}

Detection efficiency can only be evaluated for injected signals, where both the number of injected and recovered signals are known. For detections in real data, the true population of signals is unknown, so the fraction of detected and missed signals cannot be determined. For this reason, detectability is typically quantified by the FAR. Accordingly, we present a FAR-based analysis of the CCSN injections to demonstrate how LF analysis results are interpreted for detections in real data. In Figure~\ref{fig:CCSN_FAR}, the crosses (triangles) indicate the median FAR of the 175 injections per model, recovered by the \textit{BayesWave} full-band (LF) analysis. The error bars bracket the interquartile ranges (IQRs), i.e. the middle 50\% of the FARs. CCSN injections that satisfy Equation~\ref{eq:BW_nontriggers} are not successfully detected by \textit{BayesWave}. However, these events cannot be excluded from the FAR analysis, as it would skew the IQRs and medians. Therefore, in Figure~\ref{fig:CCSN_FAR}, the non-detections are assigned the maximum FAR=$1\times10^4\pyr$ derived from the cWB background, to reflect minimal astrophysical significance. The black (orange) horizontal dashed line represents the FAR threshold, below which an event is considered a successful detection for the full-band (LF) analysis; the FAR threshold corresponds to $\bsg=0$. For all five CCSN models, the median FARs (cross symbols) and the IQRs of the full-band analysis fall below the horizontal black dashed line. That is, $\geq50\%$ of the waveforms per CCSN model satisfy the full-band detection threshold. This observation is consistent with the unit detection efficiency for the full-band analysis in Figure~\ref{fig:CCSN_detEff}. The LF-analysis median FAR (triangle symbols), on the other hand, increases from left to right. This suggests that CCSN models with less LF power are generally recovered with higher FAR by the LF analyses, leading to a lower corresponding detection efficiency, as noted in Figure~\ref{fig:CCSN_detEff}. We also note that the LF FARs for SFHx and s25 are lower than their corresponding full-band FARs, while the opposite is true for the remaining three models. However, the LF and full-band FARs are not directly comparable, as the LF follow-up analyzes only a subset of the full-band signal. That is, since the two analyses probe different spectral bands of the signal, the resulting FARs pertain to fundamentally different detection domains. Overall, Figures~\ref{fig:CCSN_detEff} and \ref{fig:CCSN_FAR} have similar implications for the LF follow-up performance, but Figure~\ref{fig:CCSN_FAR} presents the results in terms of a measurable quantity for detections in real data.

In summary, we find that the \textit{BayesWave} LF follow-up analyses detect CCSN models with stronger LF content at lower FAR. This result suggests that the LF analysis can serve as a follow-up tool for confirming the presence of low-frequency content in a GW detection candidate. A successful detection with the LF follow-up analysis is defined as having a FAR below a nominal threshold, i.e. FAR$=1\pyr$ in this study. The presence of low-frequency GW content inferred from such a detection can then be used to constrain the explosion mechanism of the CCSN. The absence of an LF detection, however, does not necessarily imply the absence of low-frequency content, as the analysis may fail to detect low-frequency signals if their power is relatively weak compared to the overall signal. Therefore, we conclude that a successful LF follow-up detection is useful for constraining the explosion model of a CCSN candidate, but the lack of a detection remains inconclusive. We reiterate that the LF analysis cannot be used to infer the detection significance of the CCSN candidate as a whole, since it analyzes only a subset of the full signal.

\section{High-frequency analysis: loudest event during SN~2019fcn}
\label{sec:HF_SN2019fcn}

In this section, we demonstrate another application of the dedicated-frequency framework: to follow-up a potential CCSN GW burst candidate with the HF analysis. We note in advance that the following illustrative study on the loudest event during SN~2019fcn does not yield astrophysical results. The aim, rather, is to demonstrate by way of a concrete example how to apply the HF analysis to astrophysically significant candidates produced by future GW burst searches. %it serves merely as a demonstrative example for the potential applications of HF analysis, should more suitable detection candidates arise in future GW burst searches. 

\subsection{Selection of a candidate}
Ref. \cite{CCSN_optical_O3} presents cWB search results for GW bursts associated with optically observed CCSNe during O3. The CCSNe are selected based on (i) their optically inferred distances and (ii) whether there is sufficient GW data to generate a few years of time-shifted background for significance assessment. For each CCSN event, the cWB search is conducted over a methodically selected time interval, otherwise known as the on-source window (OSW), which is expected to contain the GW signal up to a specified probability (see Refs.~\cite{CCSN_optical_O1_O2} and \cite{CCSN_optical_O3} for details). OSWs typically last between several hours and a few days, depending on factors such as the cadence of electromagnetic observations and the time interval between core collapse and shock breakout. Although no detections were reported from the searches, the OSW (i.e. non-time-shifted) analysis for each CCSN using cWB produces a list of triggers. The trigger with the lowest FAR is called the `loudest event' and is considered the most plausible GW candidate for the associated CCSN. It is therefore interesting to apply the dedicated-frequency follow-up to one of the loudest events, to study whether the follow-ups can improve the detection significance in practical observing scenarios. 

\begin{figure}[t!]
    \centering
    \includegraphics[width=\linewidth]{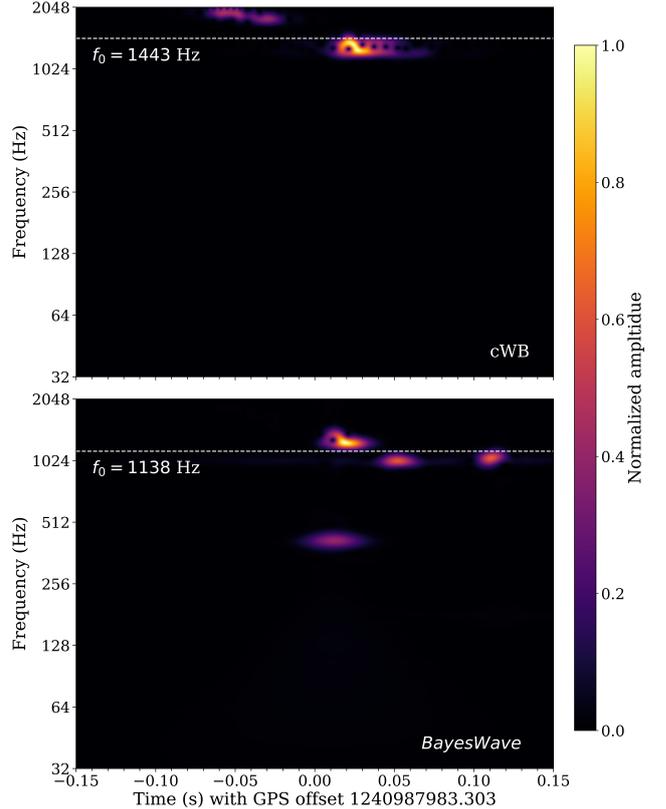}
    \caption{Whitened CWT time-frequency spectrogram of the loudest event during SN~2019fcn. The top panel shows the cWB full-band reconstruction. The horizontal axis indicates the time relative to the central GPS time of the trigger. The color bars show the normalized amplitude, defined the same way as in Figure \ref{fig:CCSN_spectrograms}. The horizontal line indicates $f_0=1443 \Hz$. The bottom panel shows the same as the top panel but for \textit{BayesWave}, with $f_0=1138 \Hz$. }
    \label{fig:SN2019fcn_spectrogram}
\end{figure}

In order to choose an illustrative follow-up candidate, we follow up all of the loudest cWB triggers in Ref.~\cite{CCSN_optical_O3} using the full-band \textit{BayesWave} analysis. Valid candidates are those with positive $\ln \mathcal{B}_{\mathcal{S}, \mathcal{N}}$ and $\ln \mathcal{B}_{\mathcal{S}, \mathcal{G}}$. Among four valid candidates, the loudest event during SN~2019fcn has the highest $\ln \mathcal{B}_{\mathcal{S}, \mathcal{N}}$ (14.7) and the second highest $\ln \mathcal{B}_{\mathcal{S}, \mathcal{G}}$ (7.5). It also has the lowest cWB FAR ($22 \, \pyr$), according to Table II in Ref.~\cite{CCSN_optical_O3}; the FAR is evaluated using a background measured with time-shifted data from the 4.54-day OSW of SN~2019fcn. Altogether, this trigger is the best available O3 burst candidate for a dedicated-frequency follow-up, according to cWB and \textit{BayesWave}. The cWB and \textit{BayesWave} full-band analysis reconstructions of SN~2019fcn's loudest event, displayed in Figure \ref{fig:SN2019fcn_spectrogram}, show that the signal predominantly comprises high-frequency power with central frequency $f_0\sim 1 \kHz$. With minimal low-frequency emission, this trigger is well-suited for the HF follow-up, which focuses on the band where the signal is expected and suppresses noise from irrelevant frequencies to enhance overall detectability. We reiterate that the trigger is not a realistic candidate due to its high ($\geq 1 \pyr$) FAR. Therefore the results of the HF follow-up do not have any astrophysical implications. The aim is simply to demonstrate the methodology.

Figure \ref{fig:SN2019fcn_spectrogram} shows that the cWB reconstruction picks up signal power at ${\sim} 2 \kHz$ which is absent from the \textit{BayesWave} reconstruction. In contrast, the \textit{BayesWave} reconstruction picks up signal power at ${\sim} 0.5 \kHz$, which cWB does not. This observation suggests that cWB and \textit{BayesWave} are sensitive to different parts of the spectrum, but with only one event reconstruction, we are unable to reach a definitive conclusion. Analyzing more events presents an interesting avenue for future studies.

\subsection{Background measurements}
\label{sec:background_SN2019fcn}

The cWB background of SN~2019fcn is measured in Ref.~\cite{CCSN_optical_O3}. Unlike in Section~\ref{sec:LF_CCSN}, the background measurement and analysis of SN~2019fcn in Ref.~\cite{CCSN_optical_O3} is conducted without XGBoost and uses an older version of cWB, because the updated methods were not yet available. We refer the reader to Section IIIB of Ref.~\cite{CCSN_optical_O3} for the full description of the ranking statistic $\etaC$ used in this section.

We aim primarily to follow-up the SN~2019fcn trigger using the \textit{BayesWave} HF analysis, but for completeness we also follow-up using the \textit{BayesWave} full-band analysis. Hence we measure two backgrounds. As before, the \textit{BayesWave} background measurement follows up cWB background triggers above a nominal threshold. Here, the follow-up threshold for the full and HF \textit{BayesWave} background measurements is $\etaC=6.7$, i.e. the value at which the cWB full-band analysis recovers the loudest event during SN~2019fcn, as reported in Ref.~\cite{CCSN_optical_O3}. 

\begin{figure}[t!]
    \centering
    \includegraphics[width=\linewidth]{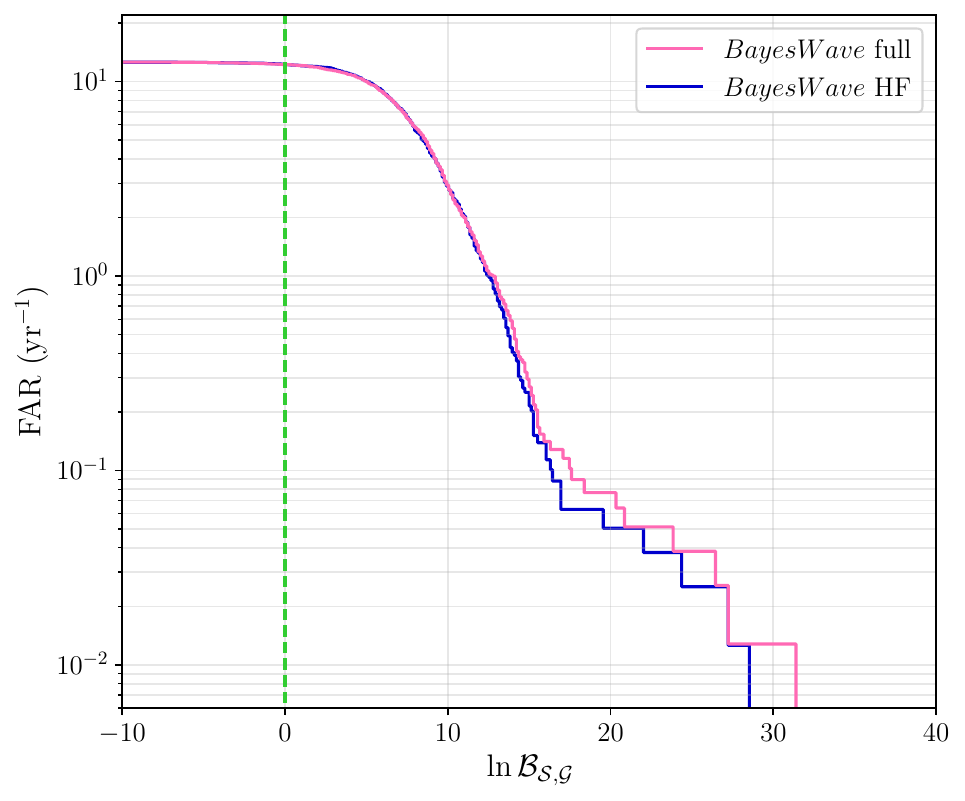}
    \caption{SN~2019fcn background measured using the \textit{BayesWave} full-band (pink curve) and HF (blue curve) follow-up analyses. The vertical green line at $\bsg=0$ indicates the \textit{BayesWave} detection threshold.}
    \label{fig:SN2019fcn_background}
\end{figure}

The pink and blue curves in Figure~\ref{fig:SN2019fcn_background} show the background measured by \textit{BayesWave} using the full-band and HF analyses respectively. Although the HF background is lower than the full-band background, the discrepancy is smaller than that observed between the LF and full-band backgrounds in Figure~\ref{fig:CCSN_background}. This makes sense because the frequency range of the HF analysis (256$-$2048 Hz) overlaps more with the full-band analysis (32$-$2048 Hz) than the LF analysis (32$-$256 Hz). That is, the SNR of the HF triggers should be similar to the full-band analysis triggers, and the number of wavelets $N$ used for reconstruction should likewise be similar, as should $\bsg$. The discrepancy could also partially be due to the different cWB versions used to measure the backgrounds for SN~2019fcn and Figure~\ref{fig:CCSN_background}.

We use the backgrounds in Figure~\ref{fig:SN2019fcn_background} to evaluate in Section~\ref{sec:FAR_SN2019fcn} the \textit{BayesWave} full-band and HF FARs for the loudest event during SN~2019fcn.

\subsection{FAR analysis}
\label{sec:FAR_SN2019fcn}

Table~\ref{table:SN2019fcn} summarizes the detection statistics, $\etaC$ and $\bsg$, recovered by the cWB and \textit{BayesWave} analyses respectively, for the loudest event during SN~2019fcn. The cWB full-band analysis results are taken directly from Ref.~\cite{CCSN_optical_O3}. The \textit{BayesWave} FARs are evaluated by comparing the full-band and HF $\bsg$ values to the corresponding background measurements in Figure~\ref{fig:SN2019fcn_background}. First, we note that the \textit{BayesWave} full-band analysis achieves a lower FAR ($6.4\pyr$) than cWB ($22.1\pyr$). This demonstrates how the hierarchical pipeline enhances detection significance, as previously reported in Ref.~\cite{Kanner_2016}. Second, the HF follow-up further reduces the FAR to $4.9\pyr$. This is because the HF follow-up yields higher $\bsg=8.6$, cf. $\bsg=7.5$ in the full-band analysis. According to Figure~\ref{fig:SN2019fcn_background}, higher $\bsg$ results in lower FAR. The increase in $\bsg$ with the HF follow-up is justified as follows. Since the SN~2019fcn trigger has minimal power below 256 Hz, the HF and full-band signal reconstructions are similar, i.e. they use approximately the same number of wavelets (median $N=9$). The signal evidence depends on the volume of the parameter space, $V$ and the subvolume that the model occupies, $\Delta V$~\cite{BayesWave2}; a higher signal evidence is achieved when $\Delta V/V$ is larger. If the HF signal reconstruction is similar to the full-band reconstruction, then $\Delta V_{\text{full}}\approx\Delta V_{\text{HF}}$. However, the HF analysis is restricted to a narrower bandwidth than the full-band analysis, so it follows that $V_{\text{HF}}<V_{\text{full}}$. As a result, the HF analysis yields higher signal evidence, and consequently higher $\bsg$. 
% $\Delta V_{\text{HF}}/V_{\text{HF}}>\Delta V_{\text{full}}/V_{\text{full}}$, 

\begin{table}[t]
\setlength{\tabcolsep}{0pt}
\rowcolors{4}{white}{white}
\renewcommand{\arraystretch}{1.4} 
\begin{tabular}{
    p{1.85cm}
    >{\centering\arraybackslash}p{0.8cm}
    >{\centering\arraybackslash}p{1cm}
    >{\centering\arraybackslash}p{1.3cm}
    >{\centering\arraybackslash}p{0.2cm}
    >{\centering\arraybackslash}p{0.8cm}
    >{\centering\arraybackslash}p{1cm}
    >{\centering\arraybackslash}p{1.3cm}
}
\hline
{ } & \multicolumn{3}{c}{Full-band} && \multicolumn{3}{c}{HF} \\
\cmidrule[1pt]{2-4} \cmidrule[1pt]{6-8}
Pipeline & $\etaC$ & $\bsg$ & FAR ($\pyr$) && $\etaC$ & $\bsg$ & FAR ($\pyr$) \cr 
\hline
% Data Rows
cWB & 6.7 & - & 22.1 && - & - & - \\
\textit{BayesWave} & - & 7.5 & 6.4 && - & 8.6 & 4.9 \\
\hline
\end{tabular}
\caption{Follow-up analysis output for the loudest event during SN~2019fcn. $\eta_c$ and $\bsg$ are the detection statistics produced by cWB and \textit{BayesWave} respectively. The cWB and \textit{BayesWave} FARs are estimated independently using the corresponding backgrounds of the full-band and HF analyses; see Section~\ref{sec:background_SN2019fcn}.}
\label{table:SN2019fcn}
\end{table}

To summarize, Table~\ref{table:SN2019fcn} demonstrates that (i) the successive application of cWB and \textit{BayesWave} reduces the FAR by a factor of $3.4$, and (ii) the HF \textit{BayesWave} follow-up, limited to $256 \Hz \leq f \leq 2048 \Hz$, further reduces the FAR for the loudest event during SN~2019fcn by a factor of $1.3$. 

We reiterate that the results in Table~\ref{table:SN2019fcn} have no astrophysical significance, because the loudest event during SN~2019fcn is not a real GW burst candidate. The goal in this paper is to demonstrate concretely how to apply HF follow-up to one representative event. This method is suitable for following up future detection candidates with marginal significance (e.g. FAR$\sim1\pyr$) and minimal low-frequency ($f \leq 256 \Hz$) power. One can also adapt the method by adjusting the LF-HF boundary to match other emission mechanisms.

\section{Conclusions}
\label{sec:conclusion}

GW signals from CCSNe contain spectral signatures which reflect the physical mechanisms that occur within the progenitor star immediately prior to its explosion. In particular, low-frequency gravitational wave signatures ($f\lesssim 250 \Hz$), can be used to detect hydrodynamical instabilities, such as the SASI and neutrino-driven convection, which are thought to play a crucial role in driving CCSN explosions. In this work, we introduce the dedicated-frequency framework, a versatile and multifaceted follow-up tool for detecting and characterizing spectral signatures of GW burst candidates. The framework uses a hierarchical detection pipeline comprising two minimally-modeled burst analysis algorithms: cWB, to identify eligible candidates based on their astrophysical significance, and \textit{BayesWave}, to follow-up eligible burst candidates using bandpass analyses. We demonstrate two distinct applications of the dedicated-frequency framework: (i) to identify GW signatures associated with the SASI and neutrino-driven convection in CCSNe using a low-frequency (LF) analysis, limited to the range $32\Hz \leq f \leq 256\Hz$; and (ii) to enhance detectability of the loudest event from SN 2019fcn using a high-frequency (HF) analysis, limited to the range $256\Hz \leq f \leq 2048\Hz$.

The LF study uses GW waveforms from five different CCSN models with typical (non-rotating, with solar metallicity) progenitors: SFHx, s25, D15, mesa20\_pert and s18. These models range from the highest to the lowest LF power, respectively. The waveforms are injected into real O3a data, and the distribution of background noise triggers in the data, i.e. the false-alarm rate (FAR), is measured using a time-shift analysis. The backgrounds for the cWB full-band analysis, \textit{BayesWave} full-band analysis and \textit{BayesWave} LF analysis are evaluated separately to allow for an independent assessment of detection significance for each algorithm and analysis band. To qualify for the LF follow-up in this study, a CCSN injection must satisfy FAR $\leq 1\pyr$ according to the full-band cWB analysis. Hence, the waveforms for each CCSN model are injected at amplitudes corresponding to the 50\% detection efficiency ($h_{\rm{rss}, 50}$) at FAR $\leq 1\pyr$. The value of $h_{\rm{rss}, 50}$ is determined empirically using the full-band cWB analysis. For each CCSN model, 175 injections are chosen arbitrarily from the full list of successful (FAR $\leq 1\pyr$) detection candidates to conduct the LF and full-band \textit{BayesWave} follow-ups. Figure \ref{fig:CCSN_detEff} shows that the LF detection efficiency increases, as the LF power increases. This is because the LF FAR reduces with increasing LF power, as shown Figure \ref{fig:CCSN_FAR}. The study shows that detecting CCSN models like s25, D15, and mesa20\_pert with the LF analysis is not guaranteed, even if they exhibit SASI-related LF emission. That is, an unsuccessful detection with the LF follow-up does not imply an absence of LF emission. However, the converse holds true: a successful detection using the LF follow-up indicates the presence of LF emission and can therefore be used to constrain the CCSN explosion models for real detection candidates.

 To demonstrate another application of the dedicated-frequency framework, we perform a HF follow-up analyses of the loudest event during SN 2019fcn~\cite{CCSN_optical_O3}. The selected trigger is not a real detection candidate because its cWB FAR does not satisfy the LVK detection benchmark ($\mathrm{FAR}\leq 0.01\pyr$); it serves purely to demonstrate the method in practice, using the best available GW burst candidate from O3. The follow-up is conducted using the \textit{BayesWave} full-band and HF analyses, with the noise background for each analysis measured separately. We find that the full-band \textit{BayesWave} follow-up reduces the FAR to $6.4 \pyr$, down from the initial cWB full-band analysis FAR of $22.1 \pyr$.  This finding reinforces the result from Ref.~\cite{Kanner_2016}, that the successive application of cWB and \textit{BayesWave} improves detection confidence. Since the SN 2019fcn trigger has minimal power in the range $f\leq 256 \Hz$, limiting the HF follow-up analyses range to $f\geq 256 \Hz$ reduces false-alarm triggers in the detector backgrounds. The HF \textit{BayesWave} follow-up further reduces the FAR from $6.4 \pyr$ to $4.6 \pyr$. Altogether we find that the HF follow-up improves the detection of burst triggers with minimal low-frequency power. We also note that the \textit{BayesWave} full-band analysis reconstruction of the SN~2019fcn trigger reveals non-negligible power at $f\sim0.5\kHz$, a feature not observed in the corresponding cWB reconstruction. Conversely, cWB detects power at $f\sim2\kHz$, yet \textit{BayesWave} does not. This suggest that cWB and \textit{BayesWave} may be sensitive to different frequency ranges. While our results do not provide sufficient evidence to confirm this claim, we recommend exploring the topic further in future work.

In conclusion, the \textit{BayesWave} LF follow-ups within the dedicated-frequency framework facilitate interpretation of explosion mechanisms through the detection of low-frequency GW emissions in CCSN detection candidates. The HF analysis of the loudest event during SN 2019fcn demonstrates that, in principle, the detection significance of events with minimal low-frequency power can be enhanced by applying follow-up analysis focused exclusively on the high-frequency components, where the signal power is predominantly concentrated. Although not demonstrated in this paper, a similar benefit is expected for LF follow-ups of signals dominated by low-frequency power.

In this paper, cWB is not used for dedicated-frequency analyses for the reasons discussed in Section~\ref{sec:df_workflow}. However, it is an interesting avenue for future work to develop a framework which allows cWB to infer frequency-specific content of GW bursts independently of \textit{BayesWave}, e.g. by analyzing subsets of the full-band reconstructions, instead of bandpass analyses. The dedicated-frequency framework is a versatile GW burst follow-up tool, which can improve detection confidence and characterization of the signal within a specific frequency range. The dedicated-frequency follow-ups can be tailored to any user-defined frequency range, and may therefore be tuned in the future to detect other types of burst signals with frequency-specific GW signatures, e.g. binary neutron-star post-merger remnants~\cite{Kiuchi_2009, Clark_2016} (${\gtrsim} 1$ kHz) and eccentric binary black-holes~\cite{Patterson_2024}.

\section*{Acknowledgements}

This material is based upon work supported by NSF’s LIGO Laboratory which is a major facility fully funded by the National Science Foundation (NSF). Parts of this research were conducted by the Australian Research Council Centre of Excellence for Gravitational Wave Discovery (OzGrav), through project numbers CE170100004 and CE230100016. The authors are grateful for computational resources provided by the LIGO Laboratory and supported by National Science Foundation Grants PHY-0757058 and PHY-0823459. Y.~S.~C.~L. is supported by a Melbourne Research Scholarship. M.~S. acknowledges Polish National Science Centre Grant No. UMO-2023/49/B/ST9/02777 and the Polish National Agency for Academic Exchange within Polish Returns Programme Grant No. BPN/PPO/2023/1/00019.  M.~M. gratefully acknowledges support from the NSF through grant PHY-2409714.

The authors thank Chad Henshaw for the codebase that was used to produce the CWT time-frequency spectrograms presented in this paper. The authors also thank Anthony Mezzacappa, Jade Powell and Yanyan Zheng for their helpful comments.

\appendix
\section{CCSNe waveform reconstruction accuracy} 
\label{app:overlap_CCSN}

The detection efficiency of the \textit{BayesWave} follow-up analyses presented in Figure~\ref{fig:CCSN_detEff} does not necessarily reflect the accuracy of reconstruction. For example, \textit{BayesWave} may register an event that satisfies the detection threshold, but the recovered signal could deviate significantly from the true (injected) signal. Therefore, it is necessary to assess whether \textit{BayesWave} accurately reconstructs the CCSN waveforms used in this study, to ensure reliability of the detection efficiency and FAR measurements in Figures~\ref{fig:CCSN_detEff} and \ref{fig:CCSN_FAR} respectively.

A standard metric for assessing reconstruction accuracy is the \textit{overlap}~\cite{GWTC2}:
\begin{equation}
\mathcal{O} = \frac{\expval{\tilde{h}_{\rm inj} \mid \tilde{h}_{\rm rec}}}{\sqrt{\expval{\tilde{h}_{\rm inj} \mid \tilde{h}_{\rm inj}}\expval{\tilde{h}_{\rm rec} \mid \tilde{h}_{\rm rec}}}}, 
\label{eq:CCSN_overlap}
\end{equation}  
which quantifies the agreement between the inject ($\tilde{h}_{\rm inj}$) and recovered ($\tilde{h}_{\rm inj}$) waveforms in the frequency domain. In Equation~\ref{eq:CCSN_overlap}, $\expval{ \cdot \mid \cdot }$ denote the noise-weighted inner product, defined as:
\begin{equation}
\expval{\tilde{h}_a \mid \tilde{h}_b } = 4 \Re \int_0^{\infty} df \frac{\tilde{h}_a (f) \tilde{h}_b^* (f)}{S_n(f)}, 
\label{eq:CCSN_innerprod}
\end{equation}
where $\tilde{h}^*$ denotes the complex conjugate of $\tilde{h}$ and $S_n(f)=S_n(-f)$ is the one-sided noise PSD of a given detector. By definition,  $\mathcal{O}$ is bounded between -1 and 1;  $\mathcal{O}=1$ ($-1$) indicates perfect (anti-)correlation between the injected and recovered waveforms, and $\mathcal{O}=0$ indicates no correlation. For a network with $\mathcal{I} \geq 2$ detectors, the overall network overlap $\mathcal{O}_{\rm net}$ is calculated by replacing the inner products in Equation~\ref{eq:CCSN_overlap} with their netork sum, viz.
\begin{equation}
\expval{\tilde{h}_{\rm inj} \mid \tilde{h}_{\rm rec}} = \sum_{i=1}^\mathcal{I} \expval{\tilde{h}_{\rm inj}^i \mid \tilde{h}_{\rm rec}^i}, 
\end{equation}
and similarly for $\expval{\tilde{h}_{\rm inj} \mid \tilde{h}_{\rm inj}}$ and $\expval{\tilde{h}_{\rm rec} \mid \tilde{h}_{\rm rec}}$.

\begin{figure*}[t]
    \centering
    \includegraphics[width=\linewidth]{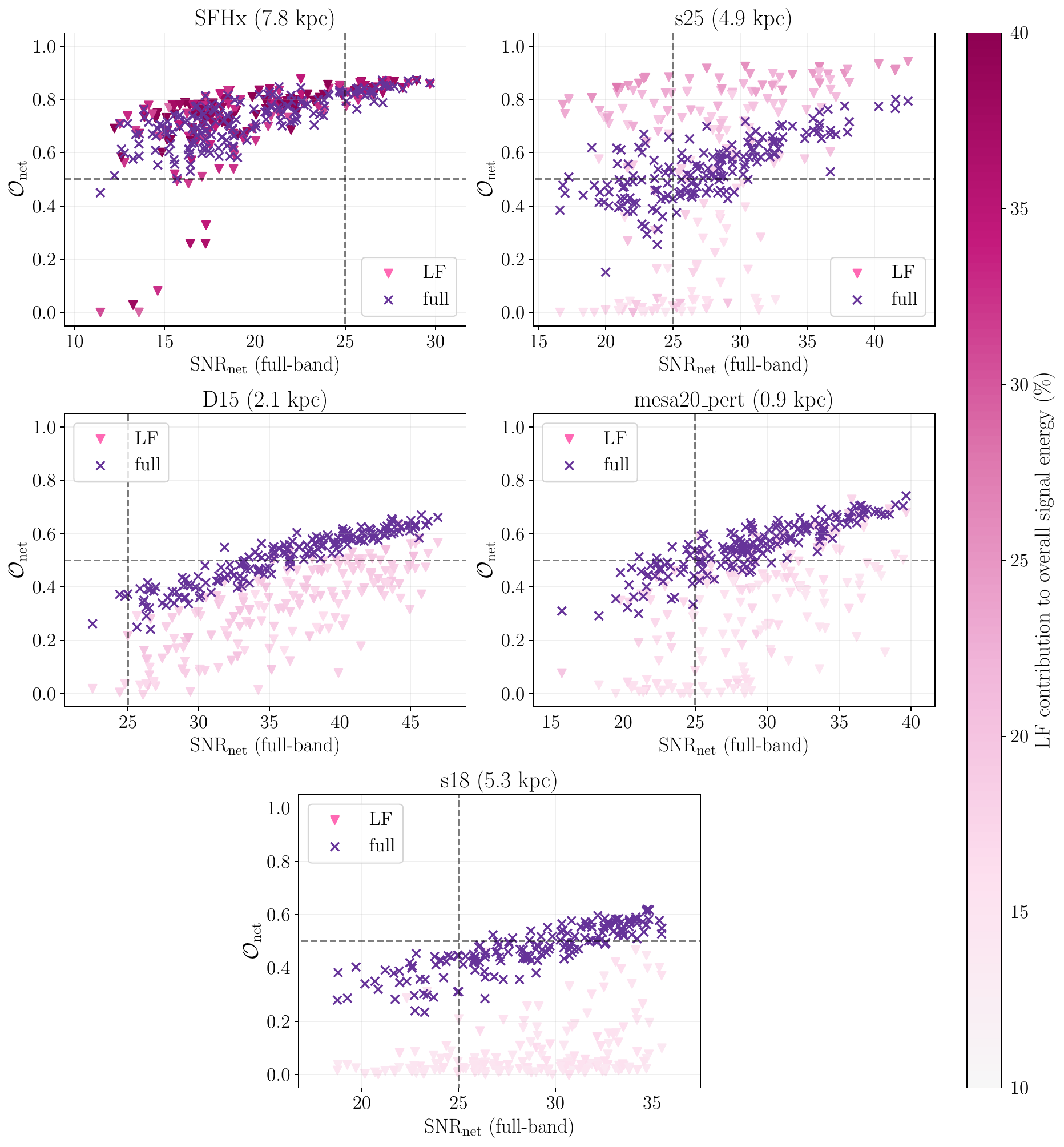}
    \caption{\textit{BayesWave}'s CCSNe waveform reconstruction accuracy. Each panel shows the network overlap $\mathcal{O}_{\rm net}$ versus the injected network SNR (SNR$_{\rm net}$) of the full-band signal for a given CCSN model, with model names indicated at the top and the corresponding $r_{50}$ values in parentheses for reference. Purple crosses and pink triangles at the same SNR$_{\rm net}$ correspond to the same CCSN signal, analyzed in the full- and LF-bands respectively. The color bar applies only to the LF (pink) data points and indicates the percentage LF contribution to the overall signal energy. The color bar for the full-band (purple) data points is omitted because the variation in LF contribution within a given model has minimal impact on the full-band $\mathcal{O}_{\rm net}$. The impact of LF contribution on $\mathcal{O}_{\rm net}$ across different models is assessed by comparing the distribution of purple data points in each panel relative to $\mathcal{O}{\rm net} = 0.5$, as indicated by the horizontal dashed lines. Since the range of SNR$_{\rm net}$ varies between models, a vertical line indicating  SNR$_{\rm net}=25$ is shown in each panel to facilitate comparison.}
    \label{fig:BWoverlap}
\end{figure*}

Figure~\ref{fig:BWoverlap} shows the $\mathcal{O}_{\rm net}$ versus the network SNR (SNR$_{\rm net}$) of the full-band signal. We begin by discussing the full-band $\mathcal{O}_{\rm net}$ (purple crosses), as accurate reconstruction of the full signal is critical for assessing whether it has been properly detected. We observe the expected trend that $\mathcal{O}_{\rm net}$ increases with the full-band injected SNR$_{\rm net}$ within each model, and this trend is consistent across all five models. However, this trend does not necessarily hold between models; for instance, the SFHx waveforms are generally recovered with higher full-band $\mathcal{O}_{\rm net}$ values than s18, despite having lower overall $\mathrm{SNR}_{\rm net}$. This suggests that $\mathcal{O}_{\rm net}$ depends not only on $\mathrm{SNR}_{\rm net}$ but also on waveform morphology, specifically its spectral composition. The horizontal dashed lines at $\mathcal{O}_{\rm net} = 0.5$ in each panel of Figure~\ref{fig:BWoverlap} are used to compare the overall $\mathcal{O}_{\rm net}$ distributions across the CCSN models. SFHx shows the largest number of full-band (purple) data points above $\mathcal{O}_{\rm net} = 0.5$; s25, D15, and mesa20\_pert exhibit roughly equal numbers of data points above and below this threshold; and s18 shows more data points below $\mathcal{O}_{\rm net} = 0.5$. Quantitatively, the full-band $\mathcal{O}_{\rm net}$ values, averaged over the 175 waveforms for each model, are 0.72, 0.54, 0.51, 0.55, and 0.47 for SFHx, s25, D15, mesa20\_pert, and s18 respectively. Notably, SFHx, which has the highest LF power, is reconstructed most accurately among the five models, whereas s18, with the lowest LF power, is reconstructed least accurately. This is because the LF band ($32 \Hz \leq f \leq 256 \Hz$) lies within the most sensitive bands of the LIGO detectors (see Figure 2 of Ref.~\cite{GWTC2}); the sensitivity decreases for $f\gtrsim300 \Hz$. Consequently, models like s18 with waveforms primarily composed of high-frequency ($f\gtrsim 300 \Hz$) features are reconstructed less accurately due to the reduced detector sensitivity at these frequencies. Nevertheless, all five CCSN models injected at their corresponding $h_{\rm rss, 50}$ are recovered with $\mathcal{O}_{\rm net} \approx 0.50$ on average in the full band. Since CCSN waveforms are highly complex, it is unlikely that they can be reconstructed with the same precision as the more systematic compact binary signals ($\mathcal{O}_{\rm net}{\sim}0.8$ at SNR$_{\rm net}{=}20$). Accordingly, the fiducial threshold for acceptable CCSN reconstruction is defined as $\mathcal{O}_{\rm net} \geq 0.2$ in Ref.~\cite{Szczepanczyk_2022}, and all CCSN injections in this study fall well within this acceptable range. This result confirms that the CCSN injections are accurately reconstructed in the full-band analysis, thereby validating the 100\% detection efficiency reported in Figure~\ref{fig:CCSN_detEff}.

We now discuss the reconstruction accuracy of the LF analysis, represented by the pink triangles in Figure~\ref{fig:BWoverlap}. $\mathcal{O}_{\rm net}$ for the LF analysis is computed by restricting the integral in Equation~\ref{eq:CCSN_innerprod} to the range $32 \Hz \leq f \leq 256 \Hz$; and the data points are colored according to the percentage of LF contribution to the overall signal energy, as shown by the color bar. As with the full-band analysis, we observe that the LF $\mathcal{O}_{\rm net}$ generally increases with the SNR$_{\rm net}$ for each model. That is, louder signals achieve more accurate LF reconstructions. However, the s25 model deviates from this trend: disparities in $\mathcal{O}_{\rm net}$ are observed among signals with similar SNR$_{\rm net}$. These disparities arise from variations in LF power, as indicated by the color bar. Specifically, given two s25 waveforms with the same $\mathrm{SNR}_{\rm net}$, the one with stronger LF content (darker colored data point) has a higher LF $\mathcal{O}_{\rm net}$, meaning it is reconstructed more accurately by the LF analysis. A similar trend is observed when comparing $\mathcal{O}_{\rm net}$ across the five models: SFHx, which has the highest LF power among the five models, yields the highest LF $\mathcal{O}_{\rm net}$ overall, averaging 0.73; followed by s25 at 0.54, mesa20\_pert at 0.33, D15 at 0.32, and s18, with the lowest LF power, averaging 0.09. These results align closely with the LF detection efficiencies shown in Figure~\ref{fig:CCSN_detEff}, with SFHx nearing unity and s18 nearing zero.

To summarize, the full-band \textit{BayesWave} analysis reconstructs the CCSN waveforms described in Section~\ref{sec:ccsn_injections} reasonably well, with $\mathcal{O}_{\rm net} \gtrsim 0.5$. A comparable level of accuracy is achieved by the LF analysis when the waveform contains sufficient LF power (e.g. SFHx); however, for waveforms dominated by high-frequency content (e.g. s18), the LF reconstruction performs poorly. Overall, waveforms with higher LF power are reconstructed more accurately by both analyses, as the sensitivity of the LIGO detectors peaks in the LF band. Altogether, these results confirm the reliability of the successful detections claimed in Figure~\ref{fig:CCSN_detEff} (and hence Figure~\ref{fig:CCSN_FAR}) by demonstrating that they achieve acceptable reconstruction accuracy.

\bibliography{citations}

%apsrev4-2.bst 2019-01-14 (MD) hand-edited version of apsrev4-1.bst
%Control: key (0)
%Control: author (8) initials jnrlst
%Control: editor formatted (1) identically to author
%Control: production of article title (0) allowed
%Control: page (0) single
%Control: year (1) truncated
%Control: production of eprint (0) enabled
\begin{thebibliography}{82}%
\makeatletter
\providecommand \@ifxundefined [1]{%
 \@ifx{#1\undefined}
}%
\providecommand \@ifnum [1]{%
 \ifnum #1\expandafter \@firstoftwo
 \else \expandafter \@secondoftwo
 \fi
}%
\providecommand \@ifx [1]{%
 \ifx #1\expandafter \@firstoftwo
 \else \expandafter \@secondoftwo
 \fi
}%
\providecommand \natexlab [1]{#1}%
\providecommand \enquote  [1]{``#1''}%
\providecommand \bibnamefont  [1]{#1}%
\providecommand \bibfnamefont [1]{#1}%
\providecommand \citenamefont [1]{#1}%
\providecommand \href@noop [0]{\@secondoftwo}%
\providecommand \href [0]{\begingroup \@sanitize@url \@href}%
\providecommand \@href[1]{\@@startlink{#1}\@@href}%
\providecommand \@@href[1]{\endgroup#1\@@endlink}%
\providecommand \@sanitize@url [0]{\catcode `\\12\catcode `\$12\catcode `\&12\catcode `\#12\catcode `\^12\catcode `\_12\catcode `\%12\relax}%
\providecommand \@@startlink[1]{}%
\providecommand \@@endlink[0]{}%
\providecommand \url  [0]{\begingroup\@sanitize@url \@url }%
\providecommand \@url [1]{\endgroup\@href {#1}{\urlprefix }}%
\providecommand \urlprefix  [0]{URL }%
\providecommand \Eprint [0]{\href }%
\providecommand \doibase [0]{https://doi.org/}%
\providecommand \selectlanguage [0]{\@gobble}%
\providecommand \bibinfo  [0]{\@secondoftwo}%
\providecommand \bibfield  [0]{\@secondoftwo}%
\providecommand \translation [1]{[#1]}%
\providecommand \BibitemOpen [0]{}%
\providecommand \bibitemStop [0]{}%
\providecommand \bibitemNoStop [0]{.\EOS\space}%
\providecommand \EOS [0]{\spacefactor3000\relax}%
\providecommand \BibitemShut  [1]{\csname bibitem#1\endcsname}%
\let\auto@bib@innerbib\@empty
%</preamble>
\bibitem [{\citenamefont {{Woosley}}\ \emph {et~al.}(2002)\citenamefont {{Woosley}}, \citenamefont {{Heger}},\ and\ \citenamefont {{Weaver}}}]{Woosley_2002}%
  \BibitemOpen
  \bibfield  {author} {\bibinfo {author} {\bibfnamefont {S.~E.}\ \bibnamefont {{Woosley}}}, \bibinfo {author} {\bibfnamefont {A.}~\bibnamefont {{Heger}}},\ and\ \bibinfo {author} {\bibfnamefont {T.~A.}\ \bibnamefont {{Weaver}}},\ }\bibfield  {title} {\bibinfo {title} {{The evolution and explosion of massive stars}},\ }\href {https://doi.org/10.1103/RevModPhys.74.1015} {\bibfield  {journal} {\bibinfo  {journal} {Reviews of Modern Physics}\ }\textbf {\bibinfo {volume} {74}},\ \bibinfo {pages} {1015} (\bibinfo {year} {2002})}\BibitemShut {NoStop}%
\bibitem [{\citenamefont {{Janka}}(2012)}]{Janka_2012}%
  \BibitemOpen
  \bibfield  {author} {\bibinfo {author} {\bibfnamefont {H.-T.}\ \bibnamefont {{Janka}}},\ }\bibfield  {title} {\bibinfo {title} {{Explosion Mechanisms of Core-Collapse Supernovae}},\ }\href {https://doi.org/10.1146/annurev-nucl-102711-094901} {\bibfield  {journal} {\bibinfo  {journal} {Annual Review of Nuclear and Particle Science}\ }\textbf {\bibinfo {volume} {62}},\ \bibinfo {pages} {407} (\bibinfo {year} {2012})},\ \Eprint {https://arxiv.org/abs/1206.2503} {arXiv:1206.2503 [astro-ph.SR]} \BibitemShut {NoStop}%
\bibitem [{\citenamefont {{Burrows}}\ \emph {et~al.}(1995)\citenamefont {{Burrows}}, \citenamefont {{Hayes}},\ and\ \citenamefont {{Fryxell}}}]{Burrows_1995}%
  \BibitemOpen
  \bibfield  {author} {\bibinfo {author} {\bibfnamefont {A.}~\bibnamefont {{Burrows}}}, \bibinfo {author} {\bibfnamefont {J.}~\bibnamefont {{Hayes}}},\ and\ \bibinfo {author} {\bibfnamefont {B.~A.}\ \bibnamefont {{Fryxell}}},\ }\bibfield  {title} {\bibinfo {title} {{On the Nature of Core-Collapse Supernova Explosions}},\ }\href {https://doi.org/10.1086/176188} {\bibfield  {journal} {\bibinfo  {journal} {\apj}\ }\textbf {\bibinfo {volume} {450}},\ \bibinfo {pages} {830} (\bibinfo {year} {1995})},\ \Eprint {https://arxiv.org/abs/astro-ph/9506061} {arXiv:astro-ph/9506061 [astro-ph]} \BibitemShut {NoStop}%
\bibitem [{\citenamefont {{Woosley}}\ and\ \citenamefont {{Janka}}(2005)}]{Woosley_2005}%
  \BibitemOpen
  \bibfield  {author} {\bibinfo {author} {\bibfnamefont {S.}~\bibnamefont {{Woosley}}}\ and\ \bibinfo {author} {\bibfnamefont {H.-T.}\ \bibnamefont {{Janka}}},\ }\bibfield  {title} {\bibinfo {title} {{The physics of core-collapse supernovae}},\ }\href {https://doi.org/10.1038/nphys172} {\bibfield  {journal} {\bibinfo  {journal} {Nature Physics}\ }\textbf {\bibinfo {volume} {1}},\ \bibinfo {pages} {147} (\bibinfo {year} {2005})},\ \Eprint {https://arxiv.org/abs/astro-ph/0601261} {arXiv:astro-ph/0601261 [astro-ph]} \BibitemShut {NoStop}%
\bibitem [{\citenamefont {{Kotake}}\ \emph {et~al.}(2006)\citenamefont {{Kotake}}, \citenamefont {{Sato}},\ and\ \citenamefont {{Takahashi}}}]{Kotake_2006}%
  \BibitemOpen
  \bibfield  {author} {\bibinfo {author} {\bibfnamefont {K.}~\bibnamefont {{Kotake}}}, \bibinfo {author} {\bibfnamefont {K.}~\bibnamefont {{Sato}}},\ and\ \bibinfo {author} {\bibfnamefont {K.}~\bibnamefont {{Takahashi}}},\ }\bibfield  {title} {\bibinfo {title} {{Explosion mechanism, neutrino burst and gravitational wave in core-collapse supernovae}},\ }\href {https://doi.org/10.1088/0034-4885/69/4/R03} {\bibfield  {journal} {\bibinfo  {journal} {Reports on Progress in Physics}\ }\textbf {\bibinfo {volume} {69}},\ \bibinfo {pages} {971} (\bibinfo {year} {2006})},\ \Eprint {https://arxiv.org/abs/astro-ph/0509456} {arXiv:astro-ph/0509456 [astro-ph]} \BibitemShut {NoStop}%
\bibitem [{\citenamefont {Aasi}\ \emph {et~al.}(2015)\citenamefont {Aasi} \emph {et~al.}}]{LIGOmain}%
  \BibitemOpen
  \bibfield  {author} {\bibinfo {author} {\bibfnamefont {J.}~\bibnamefont {Aasi}} \emph {et~al.} (\bibinfo {collaboration} {LIGO Scientific Collaboration}),\ }\bibfield  {title} {\bibinfo {title} {{Advanced LIGO}},\ }\href {https://doi.org/10.1088/0264-9381/32/7/074001} {\bibfield  {journal} {\bibinfo  {journal} {Classical and Quantum Gravity}\ }\textbf {\bibinfo {volume} {32}},\ \bibinfo {pages} {074001} (\bibinfo {year} {2015})},\ \Eprint {https://arxiv.org/abs/1411.4547} {arXiv:1411.4547 [gr-qc]} \BibitemShut {NoStop}%
\bibitem [{\citenamefont {Acernese}\ \emph {et~al.}(2015)\citenamefont {Acernese} \emph {et~al.}}]{VIRGOmain}%
  \BibitemOpen
  \bibfield  {author} {\bibinfo {author} {\bibfnamefont {F.}~\bibnamefont {Acernese}} \emph {et~al.} (\bibinfo {collaboration} {Virgo Collaboration}),\ }\bibfield  {title} {\bibinfo {title} {{Advanced Virgo: a second-generation interferometric gravitational wave detector}},\ }\href {https://doi.org/10.1088/0264-9381/32/2/024001} {\bibfield  {journal} {\bibinfo  {journal} {Classical and Quantum Gravity}\ }\textbf {\bibinfo {volume} {32}},\ \bibinfo {pages} {024001} (\bibinfo {year} {2015})},\ \Eprint {https://arxiv.org/abs/1408.3978} {arXiv:1408.3978 [gr-qc]} \BibitemShut {NoStop}%
\bibitem [{\citenamefont {{Bethe}}\ and\ \citenamefont {{Wilson}}(1985)}]{Bethe_1985}%
  \BibitemOpen
  \bibfield  {author} {\bibinfo {author} {\bibfnamefont {H.~A.}\ \bibnamefont {{Bethe}}}\ and\ \bibinfo {author} {\bibfnamefont {J.~R.}\ \bibnamefont {{Wilson}}},\ }\bibfield  {title} {\bibinfo {title} {{Revival of a stalled supernova shock by neutrino heating}},\ }\href {https://doi.org/10.1086/163343} {\bibfield  {journal} {\bibinfo  {journal} {\apj}\ }\textbf {\bibinfo {volume} {295}},\ \bibinfo {pages} {14} (\bibinfo {year} {1985})}\BibitemShut {NoStop}%
\bibitem [{\citenamefont {{Janka}}(2017)}]{Janka_2017}%
  \BibitemOpen
  \bibfield  {author} {\bibinfo {author} {\bibfnamefont {H.-T.}\ \bibnamefont {{Janka}}},\ }\bibfield  {title} {\bibinfo {title} {{Neutrino-Driven Explosions}},\ }in\ \href {https://doi.org/10.1007/978-3-319-21846-5_109} {\emph {\bibinfo {booktitle} {Handbook of Supernovae}}},\ \bibinfo {editor} {edited by\ \bibinfo {editor} {\bibfnamefont {A.~W.}\ \bibnamefont {{Alsabti}}}\ and\ \bibinfo {editor} {\bibfnamefont {P.}~\bibnamefont {{Murdin}}}}\ (\bibinfo {year} {2017})\ p.\ \bibinfo {pages} {1095}\BibitemShut {NoStop}%
\bibitem [{\citenamefont {{M{\"u}ller}}(2020)}]{Muller_2020}%
  \BibitemOpen
  \bibfield  {author} {\bibinfo {author} {\bibfnamefont {B.}~\bibnamefont {{M{\"u}ller}}},\ }\bibfield  {title} {\bibinfo {title} {{Hydrodynamics of core-collapse supernovae and their progenitors}},\ }\href {https://doi.org/10.1007/s41115-020-0008-5} {\bibfield  {journal} {\bibinfo  {journal} {Living Reviews in Computational Astrophysics}\ }\textbf {\bibinfo {volume} {6}},\ \bibinfo {eid} {3} (\bibinfo {year} {2020})},\ \Eprint {https://arxiv.org/abs/2006.05083} {arXiv:2006.05083 [astro-ph.SR]} \BibitemShut {NoStop}%
\bibitem [{\citenamefont {{Abdikamalov}}\ \emph {et~al.}(2020)\citenamefont {{Abdikamalov}}, \citenamefont {{Pagliaroli}},\ and\ \citenamefont {{Radice}}}]{Abdikamalov_2020}%
  \BibitemOpen
  \bibfield  {author} {\bibinfo {author} {\bibfnamefont {E.}~\bibnamefont {{Abdikamalov}}}, \bibinfo {author} {\bibfnamefont {G.}~\bibnamefont {{Pagliaroli}}},\ and\ \bibinfo {author} {\bibfnamefont {D.}~\bibnamefont {{Radice}}},\ }\bibfield  {title} {\bibinfo {title} {{Gravitational Waves from Core-Collapse Supernovae}},\ }\href {https://doi.org/10.48550/arXiv.2010.04356} {\bibfield  {journal} {\bibinfo  {journal} {arXiv e-prints}\ ,\ \bibinfo {eid} {arXiv:2010.04356}} (\bibinfo {year} {2020})},\ \Eprint {https://arxiv.org/abs/2010.04356} {arXiv:2010.04356 [astro-ph.SR]} \BibitemShut {NoStop}%
\bibitem [{\citenamefont {Bethe}(1990)}]{Bethe_1990}%
  \BibitemOpen
  \bibfield  {author} {\bibinfo {author} {\bibfnamefont {H.~A.}\ \bibnamefont {Bethe}},\ }\bibfield  {title} {\bibinfo {title} {Supernova mechanisms},\ }\href {https://doi.org/10.1103/RevModPhys.62.801} {\bibfield  {journal} {\bibinfo  {journal} {Rev. Mod. Phys.}\ }\textbf {\bibinfo {volume} {62}},\ \bibinfo {pages} {801} (\bibinfo {year} {1990})}\BibitemShut {NoStop}%
\bibitem [{\citenamefont {{Hirata}}\ \emph {et~al.}(1987)\citenamefont {{Hirata}}, \citenamefont {{Kajita}}, \citenamefont {{Koshiba}}, \citenamefont {{Nakahata}}, \citenamefont {{Oyama}}, \citenamefont {{Sato}}, \citenamefont {{Suzuki}}, \citenamefont {{Takita}}, \citenamefont {{Totsuka}}, \citenamefont {{Kifune}},\ and\ \citenamefont {et~al.}}]{SN1987A_1}%
  \BibitemOpen
  \bibfield  {author} {\bibinfo {author} {\bibfnamefont {K.}~\bibnamefont {{Hirata}}}, \bibinfo {author} {\bibfnamefont {T.}~\bibnamefont {{Kajita}}}, \bibinfo {author} {\bibfnamefont {M.}~\bibnamefont {{Koshiba}}}, \bibinfo {author} {\bibfnamefont {M.}~\bibnamefont {{Nakahata}}}, \bibinfo {author} {\bibfnamefont {Y.}~\bibnamefont {{Oyama}}}, \bibinfo {author} {\bibfnamefont {N.}~\bibnamefont {{Sato}}}, \bibinfo {author} {\bibfnamefont {A.}~\bibnamefont {{Suzuki}}}, \bibinfo {author} {\bibfnamefont {M.}~\bibnamefont {{Takita}}}, \bibinfo {author} {\bibfnamefont {Y.}~\bibnamefont {{Totsuka}}}, \bibinfo {author} {\bibfnamefont {T.}~\bibnamefont {{Kifune}}},\ and\ \bibinfo {author} {\bibnamefont {et~al.}},\ }\bibfield  {title} {\bibinfo {title} {{Observation of a neutrino burst from the supernova SN1987A}},\ }\href {https://doi.org/10.1103/PhysRevLett.58.1490} {\bibfield  {journal} {\bibinfo  {journal} {\prl}\ }\textbf {\bibinfo {volume} {58}},\ \bibinfo {pages} {1490} (\bibinfo {year} {1987})}\BibitemShut
  {NoStop}%
\bibitem [{\citenamefont {{Bionta}}\ \emph {et~al.}(1987)\citenamefont {{Bionta}}, \citenamefont {{Blewitt}}, \citenamefont {{Bratton}}, \citenamefont {{Casper}}, \citenamefont {{Ciocio}}, \citenamefont {{Claus}}, \citenamefont {{Cortez}}, \citenamefont {{Crouch}}, \citenamefont {{Dye}}, \citenamefont {{Errede}},\ and\ \citenamefont {et~al.}}]{SN1987A_2}%
  \BibitemOpen
  \bibfield  {author} {\bibinfo {author} {\bibfnamefont {R.~M.}\ \bibnamefont {{Bionta}}}, \bibinfo {author} {\bibfnamefont {G.}~\bibnamefont {{Blewitt}}}, \bibinfo {author} {\bibfnamefont {C.~B.}\ \bibnamefont {{Bratton}}}, \bibinfo {author} {\bibfnamefont {D.}~\bibnamefont {{Casper}}}, \bibinfo {author} {\bibfnamefont {A.}~\bibnamefont {{Ciocio}}}, \bibinfo {author} {\bibfnamefont {R.}~\bibnamefont {{Claus}}}, \bibinfo {author} {\bibfnamefont {B.}~\bibnamefont {{Cortez}}}, \bibinfo {author} {\bibfnamefont {M.}~\bibnamefont {{Crouch}}}, \bibinfo {author} {\bibfnamefont {S.~T.}\ \bibnamefont {{Dye}}}, \bibinfo {author} {\bibfnamefont {S.}~\bibnamefont {{Errede}}},\ and\ \bibinfo {author} {\bibnamefont {et~al.}},\ }\bibfield  {title} {\bibinfo {title} {{Observation of a neutrino burst in coincidence with supernova 1987A in the Large Magellanic Cloud}},\ }\href {https://doi.org/10.1103/PhysRevLett.58.1494} {\bibfield  {journal} {\bibinfo  {journal} {\prl}\ }\textbf {\bibinfo {volume} {58}},\ \bibinfo {pages}
  {1494} (\bibinfo {year} {1987})}\BibitemShut {NoStop}%
\bibitem [{\citenamefont {{Alexeyev}}\ \emph {et~al.}(1988)\citenamefont {{Alexeyev}}, \citenamefont {{Alexeyeva}}, \citenamefont {{Krivosheina}},\ and\ \citenamefont {{Volchenko}}}]{SN1987A_3}%
  \BibitemOpen
  \bibfield  {author} {\bibinfo {author} {\bibfnamefont {E.~N.}\ \bibnamefont {{Alexeyev}}}, \bibinfo {author} {\bibfnamefont {L.~N.}\ \bibnamefont {{Alexeyeva}}}, \bibinfo {author} {\bibfnamefont {I.~V.}\ \bibnamefont {{Krivosheina}}},\ and\ \bibinfo {author} {\bibfnamefont {V.~I.}\ \bibnamefont {{Volchenko}}},\ }\bibfield  {title} {\bibinfo {title} {{Detection of the neutrino signal from SN 1987A in the LMC using the INR Baksan underground scintillation telescope}},\ }\href {https://doi.org/10.1016/0370-2693(88)91651-6} {\bibfield  {journal} {\bibinfo  {journal} {Physics Letters B}\ }\textbf {\bibinfo {volume} {205}},\ \bibinfo {pages} {209} (\bibinfo {year} {1988})}\BibitemShut {NoStop}%
\bibitem [{\citenamefont {{Burrows}}\ and\ \citenamefont {{Lattimer}}(1986)}]{Burrows_1986}%
  \BibitemOpen
  \bibfield  {author} {\bibinfo {author} {\bibfnamefont {A.}~\bibnamefont {{Burrows}}}\ and\ \bibinfo {author} {\bibfnamefont {J.~M.}\ \bibnamefont {{Lattimer}}},\ }\bibfield  {title} {\bibinfo {title} {{The Birth of Neutron Stars}},\ }\href {https://doi.org/10.1086/164405} {\bibfield  {journal} {\bibinfo  {journal} {\apj}\ }\textbf {\bibinfo {volume} {307}},\ \bibinfo {pages} {178} (\bibinfo {year} {1986})}\BibitemShut {NoStop}%
\bibitem [{\citenamefont {{Woosley}}(1988)}]{Woosley_SN1987A}%
  \BibitemOpen
  \bibfield  {author} {\bibinfo {author} {\bibfnamefont {S.~E.}\ \bibnamefont {{Woosley}}},\ }\bibfield  {title} {\bibinfo {title} {{SN 1987A: After the Peak}},\ }\href {https://doi.org/10.1086/166468} {\bibfield  {journal} {\bibinfo  {journal} {\apj}\ }\textbf {\bibinfo {volume} {330}},\ \bibinfo {pages} {218} (\bibinfo {year} {1988})}\BibitemShut {NoStop}%
\bibitem [{\citenamefont {{Ott}}(2009)}]{Ott_2009}%
  \BibitemOpen
  \bibfield  {author} {\bibinfo {author} {\bibfnamefont {C.~D.}\ \bibnamefont {{Ott}}},\ }\bibfield  {title} {\bibinfo {title} {{The gravitational-wave signature of core-collapse supernovae}},\ }\href {https://doi.org/10.1088/0264-9381/26/6/063001} {\bibfield  {journal} {\bibinfo  {journal} {Classical and Quantum Gravity}\ }\textbf {\bibinfo {volume} {26}},\ \bibinfo {eid} {063001} (\bibinfo {year} {2009})},\ \Eprint {https://arxiv.org/abs/0809.0695} {arXiv:0809.0695 [astro-ph]} \BibitemShut {NoStop}%
\bibitem [{\citenamefont {Abbott}\ \emph {et~al.}(2019{\natexlab{a}})\citenamefont {Abbott} \emph {et~al.}}]{GWTC1}%
  \BibitemOpen
  \bibfield  {author} {\bibinfo {author} {\bibfnamefont {B.~P.}\ \bibnamefont {Abbott}} \emph {et~al.} (\bibinfo {collaboration} {LIGO Scientific, Virgo}),\ }\bibfield  {title} {\bibinfo {title} {{GWTC-1: A Gravitational-Wave Transient Catalog of Compact Binary Mergers Observed by LIGO and Virgo during the First and Second Observing Runs}},\ }\href {https://doi.org/10.1103/PhysRevX.9.031040} {\bibfield  {journal} {\bibinfo  {journal} {Phys. Rev. X}\ }\textbf {\bibinfo {volume} {9}},\ \bibinfo {pages} {031040} (\bibinfo {year} {2019}{\natexlab{a}})},\ \Eprint {https://arxiv.org/abs/1811.12907} {arXiv:1811.12907 [astro-ph.HE]} \BibitemShut {NoStop}%
\bibitem [{\citenamefont {Abbott}\ \emph {et~al.}(2021{\natexlab{a}})\citenamefont {Abbott} \emph {et~al.}}]{GWTC2}%
  \BibitemOpen
  \bibfield  {author} {\bibinfo {author} {\bibfnamefont {R.}~\bibnamefont {Abbott}} \emph {et~al.} (\bibinfo {collaboration} {LIGO Scientific, Virgo}),\ }\bibfield  {title} {\bibinfo {title} {{GWTC-2: Compact Binary Coalescences Observed by LIGO and Virgo During the First Half of the Third Observing Run}},\ }\href {https://doi.org/10.1103/PhysRevX.11.021053} {\bibfield  {journal} {\bibinfo  {journal} {Physical Review X}\ }\textbf {\bibinfo {volume} {11}},\ \bibinfo {pages} {021053} (\bibinfo {year} {2021}{\natexlab{a}})},\ \Eprint {https://arxiv.org/abs/2010.14527} {arXiv:2010.14527 [gr-qc]} \BibitemShut {NoStop}%
\bibitem [{\citenamefont {{Abbott}}\ \emph {et~al.}(2024)\citenamefont {{Abbott}}, \citenamefont {{Abbott}}, \citenamefont {{Acernese}}, \citenamefont {{Ackley}}, \citenamefont {{Adams}}, \citenamefont {{Adhikari}}, \citenamefont {{Adhikari}}, \citenamefont {{Adya}}, \citenamefont {{Affeldt}}, \citenamefont {{Agarwal}},\ and\ \citenamefont {et~al.}}]{GWTC2.1}%
  \BibitemOpen
  \bibfield  {author} {\bibinfo {author} {\bibfnamefont {R.}~\bibnamefont {{Abbott}}}, \bibinfo {author} {\bibfnamefont {T.~D.}\ \bibnamefont {{Abbott}}}, \bibinfo {author} {\bibfnamefont {F.}~\bibnamefont {{Acernese}}}, \bibinfo {author} {\bibfnamefont {K.}~\bibnamefont {{Ackley}}}, \bibinfo {author} {\bibfnamefont {C.}~\bibnamefont {{Adams}}}, \bibinfo {author} {\bibfnamefont {N.}~\bibnamefont {{Adhikari}}}, \bibinfo {author} {\bibfnamefont {R.~X.}\ \bibnamefont {{Adhikari}}}, \bibinfo {author} {\bibfnamefont {V.~B.}\ \bibnamefont {{Adya}}}, \bibinfo {author} {\bibfnamefont {C.}~\bibnamefont {{Affeldt}}}, \bibinfo {author} {\bibfnamefont {D.}~\bibnamefont {{Agarwal}}},\ and\ \bibinfo {author} {\bibnamefont {et~al.}},\ }\bibfield  {title} {\bibinfo {title} {{GWTC-2.1: Deep extended catalog of compact binary coalescences observed by LIGO and Virgo during the first half of the third observing run}},\ }\href {https://doi.org/10.1103/PhysRevD.109.022001} {\bibfield  {journal} {\bibinfo  {journal} {\prd}\
  }\textbf {\bibinfo {volume} {109}},\ \bibinfo {eid} {022001} (\bibinfo {year} {2024})},\ \Eprint {https://arxiv.org/abs/2108.01045} {arXiv:2108.01045 [gr-qc]} \BibitemShut {NoStop}%
\bibitem [{\citenamefont {Abbott}\ \emph {et~al.}(2023{\natexlab{a}})\citenamefont {Abbott} \emph {et~al.}}]{GWTC3}%
  \BibitemOpen
  \bibfield  {author} {\bibinfo {author} {\bibfnamefont {R.}~\bibnamefont {Abbott}} \emph {et~al.} (\bibinfo {collaboration} {LIGO Scientific, Virgo, and KAGRA}),\ }\bibfield  {title} {\bibinfo {title} {{GWTC-3: Compact Binary Coalescences Observed by LIGO and Virgo during the Second Part of the Third Observing Run}},\ }\href {https://doi.org/10.1103/PhysRevX.13.041039} {\bibfield  {journal} {\bibinfo  {journal} {Phys. Rev. X}\ }\textbf {\bibinfo {volume} {13}},\ \bibinfo {pages} {041039} (\bibinfo {year} {2023}{\natexlab{a}})},\ \Eprint {https://arxiv.org/abs/2111.03606} {arXiv:2111.03606 [gr-qc]} \BibitemShut {NoStop}%
\bibitem [{\citenamefont {{Blondin}}\ \emph {et~al.}(2003)\citenamefont {{Blondin}}, \citenamefont {{Mezzacappa}},\ and\ \citenamefont {{DeMarino}}}]{Blondin_2002}%
  \BibitemOpen
  \bibfield  {author} {\bibinfo {author} {\bibfnamefont {J.~M.}\ \bibnamefont {{Blondin}}}, \bibinfo {author} {\bibfnamefont {A.}~\bibnamefont {{Mezzacappa}}},\ and\ \bibinfo {author} {\bibfnamefont {C.}~\bibnamefont {{DeMarino}}},\ }\bibfield  {title} {\bibinfo {title} {{Stability of Standing Accretion Shocks, with an Eye toward Core-Collapse Supernovae}},\ }\href {https://doi.org/10.1086/345812} {\bibfield  {journal} {\bibinfo  {journal} {\apj}\ }\textbf {\bibinfo {volume} {584}},\ \bibinfo {pages} {971} (\bibinfo {year} {2003})},\ \Eprint {https://arxiv.org/abs/astro-ph/0210634} {arXiv:astro-ph/0210634 [astro-ph]} \BibitemShut {NoStop}%
\bibitem [{\citenamefont {Szczepa\'{n}czyk}\ and\ \citenamefont {Zanolin}(2022)}]{Szczepanczyk_2022}%
  \BibitemOpen
  \bibfield  {author} {\bibinfo {author} {\bibfnamefont {M.}~\bibnamefont {Szczepa\'{n}czyk}}\ and\ \bibinfo {author} {\bibfnamefont {M.}~\bibnamefont {Zanolin}},\ }\bibfield  {title} {\bibinfo {title} {{Gravitational Waves from a Core-Collapse Supernova: Perspectives with Detectors in the Late 2020s and Early 2030s}},\ }\href {https://doi.org/10.3390/galaxies10030070} {\bibfield  {journal} {\bibinfo  {journal} {Galaxies}\ }\textbf {\bibinfo {volume} {10}},\ \bibinfo {pages} {70} (\bibinfo {year} {2022})}\BibitemShut {NoStop}%
\bibitem [{\citenamefont {{Kuroda}}\ \emph {et~al.}(2016)\citenamefont {{Kuroda}}, \citenamefont {{Kotake}},\ and\ \citenamefont {{Takiwaki}}}]{SFHx}%
  \BibitemOpen
  \bibfield  {author} {\bibinfo {author} {\bibfnamefont {T.}~\bibnamefont {{Kuroda}}}, \bibinfo {author} {\bibfnamefont {K.}~\bibnamefont {{Kotake}}},\ and\ \bibinfo {author} {\bibfnamefont {T.}~\bibnamefont {{Takiwaki}}},\ }\bibfield  {title} {\bibinfo {title} {{A New Gravitational-wave Signature from Standing Accretion Shock Instability in Supernovae}},\ }\href {https://doi.org/10.3847/2041-8205/829/1/L14} {\bibfield  {journal} {\bibinfo  {journal} {\apjl}\ }\textbf {\bibinfo {volume} {829}},\ \bibinfo {eid} {L14} (\bibinfo {year} {2016})},\ \Eprint {https://arxiv.org/abs/1605.09215} {arXiv:1605.09215 [astro-ph.HE]} \BibitemShut {NoStop}%
\bibitem [{\citenamefont {{O'Connor}}\ and\ \citenamefont {{Couch}}(2018)}]{mesa20_pert}%
  \BibitemOpen
  \bibfield  {author} {\bibinfo {author} {\bibfnamefont {E.~P.}\ \bibnamefont {{O'Connor}}}\ and\ \bibinfo {author} {\bibfnamefont {S.~M.}\ \bibnamefont {{Couch}}},\ }\bibfield  {title} {\bibinfo {title} {{Exploring Fundamentally Three-dimensional Phenomena in High-fidelity Simulations of Core-collapse Supernovae}},\ }\href {https://doi.org/10.3847/1538-4357/aadcf7} {\bibfield  {journal} {\bibinfo  {journal} {\apj}\ }\textbf {\bibinfo {volume} {865}},\ \bibinfo {eid} {81} (\bibinfo {year} {2018})},\ \Eprint {https://arxiv.org/abs/1807.07579} {arXiv:1807.07579 [astro-ph.HE]} \BibitemShut {NoStop}%
\bibitem [{\citenamefont {{Buikema}}\ \emph {et~al.}(2020)\citenamefont {{Buikema}} \emph {et~al.}}]{Buikema_2020}%
  \BibitemOpen
  \bibfield  {author} {\bibinfo {author} {\bibfnamefont {A.}~\bibnamefont {{Buikema}}} \emph {et~al.} (\bibinfo {collaboration} {LIGO Instrument Science}),\ }\bibfield  {title} {\bibinfo {title} {{Sensitivity and performance of the Advanced LIGO detectors in the third observing run}},\ }\href {https://doi.org/10.1103/PhysRevD.102.062003} {\bibfield  {journal} {\bibinfo  {journal} {\prd}\ }\textbf {\bibinfo {volume} {102}},\ \bibinfo {eid} {062003} (\bibinfo {year} {2020})},\ \Eprint {https://arxiv.org/abs/2008.01301} {arXiv:2008.01301 [astro-ph.IM]} \BibitemShut {NoStop}%
\bibitem [{\citenamefont {Abbott}\ \emph {et~al.}(2021{\natexlab{b}})\citenamefont {Abbott} \emph {et~al.}}]{allskyO3}%
  \BibitemOpen
  \bibfield  {author} {\bibinfo {author} {\bibfnamefont {R.}~\bibnamefont {Abbott}} \emph {et~al.} (\bibinfo {collaboration} {LIGO Scientific, Virgo and KAGRA Collaboration}),\ }\bibfield  {title} {\bibinfo {title} {{All-sky search for short gravitational-wave bursts in the third Advanced LIGO and Advanced Virgo run}},\ }\href {https://doi.org/10.1103/PhysRevD.104.122004} {\bibfield  {journal} {\bibinfo  {journal} {\prd}\ }\textbf {\bibinfo {volume} {104}},\ \bibinfo {pages} {122004} (\bibinfo {year} {2021}{\natexlab{b}})},\ \Eprint {https://arxiv.org/abs/2107.03701} {arXiv:2107.03701 [gr-qc]} \BibitemShut {NoStop}%
\bibitem [{\citenamefont {Glanzer}\ \emph {et~al.}(2023)\citenamefont {Glanzer} \emph {et~al.}}]{GravitySpy_O3}%
  \BibitemOpen
  \bibfield  {author} {\bibinfo {author} {\bibfnamefont {J.}~\bibnamefont {Glanzer}} \emph {et~al.},\ }\bibfield  {title} {\bibinfo {title} {{Data quality up to the third observing run of advanced LIGO: Gravity Spy glitch classifications}},\ }\href {https://doi.org/10.1088/1361-6382/acb633} {\bibfield  {journal} {\bibinfo  {journal} {Class. Quant. Grav.}\ }\textbf {\bibinfo {volume} {40}},\ \bibinfo {pages} {065004} (\bibinfo {year} {2023})},\ \Eprint {https://arxiv.org/abs/2208.12849} {arXiv:2208.12849 [gr-qc]} \BibitemShut {NoStop}%
\bibitem [{\citenamefont {{Klimenko}}\ \emph {et~al.}(2008)\citenamefont {{Klimenko}}, \citenamefont {{Yakushin}}, \citenamefont {{Mercer}},\ and\ \citenamefont {{Mitselmakher}}}]{cWB1}%
  \BibitemOpen
  \bibfield  {author} {\bibinfo {author} {\bibfnamefont {S.}~\bibnamefont {{Klimenko}}}, \bibinfo {author} {\bibfnamefont {I.}~\bibnamefont {{Yakushin}}}, \bibinfo {author} {\bibfnamefont {A.}~\bibnamefont {{Mercer}}},\ and\ \bibinfo {author} {\bibfnamefont {G.}~\bibnamefont {{Mitselmakher}}},\ }\bibfield  {title} {\bibinfo {title} {{A coherent method for detection of gravitational wave bursts}},\ }\href {https://doi.org/10.1088/0264-9381/25/11/114029} {\bibfield  {journal} {\bibinfo  {journal} {Classical and Quantum Gravity}\ }\textbf {\bibinfo {volume} {25}},\ \bibinfo {eid} {114029} (\bibinfo {year} {2008})},\ \Eprint {https://arxiv.org/abs/0802.3232} {arXiv:0802.3232 [gr-qc]} \BibitemShut {NoStop}%
\bibitem [{\citenamefont {{Klimenko}}\ \emph {et~al.}(2016)\citenamefont {{Klimenko}}, \citenamefont {{Vedovato}}, \citenamefont {{Drago}}, \citenamefont {{Salemi}}, \citenamefont {{Tiwari}}, \citenamefont {{Prodi}}, \citenamefont {{Lazzaro}}, \citenamefont {{Ackley}}, \citenamefont {{Tiwari}}, \citenamefont {{Da Silva}},\ and\ \citenamefont {{Mitselmakher}}}]{cWB2}%
  \BibitemOpen
  \bibfield  {author} {\bibinfo {author} {\bibfnamefont {S.}~\bibnamefont {{Klimenko}}}, \bibinfo {author} {\bibfnamefont {G.}~\bibnamefont {{Vedovato}}}, \bibinfo {author} {\bibfnamefont {M.}~\bibnamefont {{Drago}}}, \bibinfo {author} {\bibfnamefont {F.}~\bibnamefont {{Salemi}}}, \bibinfo {author} {\bibfnamefont {V.}~\bibnamefont {{Tiwari}}}, \bibinfo {author} {\bibfnamefont {G.~A.}\ \bibnamefont {{Prodi}}}, \bibinfo {author} {\bibfnamefont {C.}~\bibnamefont {{Lazzaro}}}, \bibinfo {author} {\bibfnamefont {K.}~\bibnamefont {{Ackley}}}, \bibinfo {author} {\bibfnamefont {S.}~\bibnamefont {{Tiwari}}}, \bibinfo {author} {\bibfnamefont {C.~F.}\ \bibnamefont {{Da Silva}}},\ and\ \bibinfo {author} {\bibfnamefont {G.}~\bibnamefont {{Mitselmakher}}},\ }\bibfield  {title} {\bibinfo {title} {{Method for detection and reconstruction of gravitational wave transients with networks of advanced detectors}},\ }\href {https://doi.org/10.1103/PhysRevD.93.042004} {\bibfield  {journal} {\bibinfo  {journal} {\prd}\ }\textbf
  {\bibinfo {volume} {93}},\ \bibinfo {eid} {042004} (\bibinfo {year} {2016})},\ \Eprint {https://arxiv.org/abs/1511.05999} {arXiv:1511.05999 [gr-qc]} \BibitemShut {NoStop}%
\bibitem [{\citenamefont {{Drago}}\ \emph {et~al.}(2021)\citenamefont {{Drago}}, \citenamefont {{Klimenko}}, \citenamefont {{Lazzaro}}, \citenamefont {{Milotti}}, \citenamefont {{Mitselmakher}}, \citenamefont {{Necula}}, \citenamefont {{O'Brian}}, \citenamefont {{Prodi}}, \citenamefont {{Salemi}}, \citenamefont {{Szczepanczyk}}, \citenamefont {{Tiwari}}, \citenamefont {{Tiwari}}, \citenamefont {{Gayathri}}, \citenamefont {{Vedovato}},\ and\ \citenamefont {{Yakushin}}}]{cWB3}%
  \BibitemOpen
  \bibfield  {author} {\bibinfo {author} {\bibfnamefont {M.}~\bibnamefont {{Drago}}}, \bibinfo {author} {\bibfnamefont {S.}~\bibnamefont {{Klimenko}}}, \bibinfo {author} {\bibfnamefont {C.}~\bibnamefont {{Lazzaro}}}, \bibinfo {author} {\bibfnamefont {E.}~\bibnamefont {{Milotti}}}, \bibinfo {author} {\bibfnamefont {G.}~\bibnamefont {{Mitselmakher}}}, \bibinfo {author} {\bibfnamefont {V.}~\bibnamefont {{Necula}}}, \bibinfo {author} {\bibfnamefont {B.}~\bibnamefont {{O'Brian}}}, \bibinfo {author} {\bibfnamefont {G.~A.}\ \bibnamefont {{Prodi}}}, \bibinfo {author} {\bibfnamefont {F.}~\bibnamefont {{Salemi}}}, \bibinfo {author} {\bibfnamefont {M.}~\bibnamefont {{Szczepanczyk}}}, \bibinfo {author} {\bibfnamefont {S.}~\bibnamefont {{Tiwari}}}, \bibinfo {author} {\bibfnamefont {V.}~\bibnamefont {{Tiwari}}}, \bibinfo {author} {\bibfnamefont {V.}~\bibnamefont {{Gayathri}}}, \bibinfo {author} {\bibfnamefont {G.}~\bibnamefont {{Vedovato}}},\ and\ \bibinfo {author} {\bibfnamefont {I.}~\bibnamefont {{Yakushin}}},\
  }\bibfield  {title} {\bibinfo {title} {{coherent WaveBurst, a pipeline for unmodeled gravitational-wave data analysis}},\ }\href {https://doi.org/10.1016/j.softx.2021.100678} {\bibfield  {journal} {\bibinfo  {journal} {SoftwareX}\ }\textbf {\bibinfo {volume} {14}},\ \bibinfo {eid} {100678} (\bibinfo {year} {2021})}\BibitemShut {NoStop}%
\bibitem [{\citenamefont {Klimenko}\ \emph {et~al.}(2021)\citenamefont {Klimenko}, \citenamefont {Vedovato}, \citenamefont {Necula}, \citenamefont {Salemi}, \citenamefont {Drago}, \citenamefont {Chassande-Mottin}, \citenamefont {Tiwari}, \citenamefont {Lazzaro}, \citenamefont {O'Brian}, \citenamefont {Szczepanczyk}, \citenamefont {Tiwari},\ and\ \citenamefont {Gayathri}}]{cWB_library}%
  \BibitemOpen
  \bibfield  {author} {\bibinfo {author} {\bibfnamefont {S.}~\bibnamefont {Klimenko}}, \bibinfo {author} {\bibfnamefont {G.}~\bibnamefont {Vedovato}}, \bibinfo {author} {\bibfnamefont {V.}~\bibnamefont {Necula}}, \bibinfo {author} {\bibfnamefont {F.}~\bibnamefont {Salemi}}, \bibinfo {author} {\bibfnamefont {M.}~\bibnamefont {Drago}}, \bibinfo {author} {\bibfnamefont {E.}~\bibnamefont {Chassande-Mottin}}, \bibinfo {author} {\bibfnamefont {V.}~\bibnamefont {Tiwari}}, \bibinfo {author} {\bibfnamefont {C.}~\bibnamefont {Lazzaro}}, \bibinfo {author} {\bibfnamefont {B.}~\bibnamefont {O'Brian}}, \bibinfo {author} {\bibfnamefont {M.}~\bibnamefont {Szczepanczyk}}, \bibinfo {author} {\bibfnamefont {S.}~\bibnamefont {Tiwari}},\ and\ \bibinfo {author} {\bibfnamefont {V.}~\bibnamefont {Gayathri}},\ }\href {https://doi.org/10.5281/zenodo.4419902} {\bibinfo {title} {cwb pipeline library: 6.4.0}} (\bibinfo {year} {2021})\BibitemShut {NoStop}%
\bibitem [{\citenamefont {{Mishra}}\ \emph {et~al.}(2025)\citenamefont {{Mishra}}, \citenamefont {{Bhaumik}}, \citenamefont {{Gayathri}}, \citenamefont {{Szczepa{\'n}czyk}}, \citenamefont {{Bartos}},\ and\ \citenamefont {{Klimenko}}}]{Mishra_2024}%
  \BibitemOpen
  \bibfield  {author} {\bibinfo {author} {\bibfnamefont {T.}~\bibnamefont {{Mishra}}}, \bibinfo {author} {\bibfnamefont {S.}~\bibnamefont {{Bhaumik}}}, \bibinfo {author} {\bibfnamefont {V.}~\bibnamefont {{Gayathri}}}, \bibinfo {author} {\bibfnamefont {M.~J.}\ \bibnamefont {{Szczepa{\'n}czyk}}}, \bibinfo {author} {\bibfnamefont {I.}~\bibnamefont {{Bartos}}},\ and\ \bibinfo {author} {\bibfnamefont {S.}~\bibnamefont {{Klimenko}}},\ }\bibfield  {title} {\bibinfo {title} {{Gravitational waves detected by a burst search in LIGO/Virgo's third observing run}},\ }\href {https://doi.org/10.1103/PhysRevD.111.023054} {\bibfield  {journal} {\bibinfo  {journal} {\prd}\ }\textbf {\bibinfo {volume} {111}},\ \bibinfo {eid} {023054} (\bibinfo {year} {2025})},\ \Eprint {https://arxiv.org/abs/2410.15191} {arXiv:2410.15191 [astro-ph.HE]} \BibitemShut {NoStop}%
\bibitem [{\citenamefont {{Cornish}}\ and\ \citenamefont {{Littenberg}}(2015)}]{BayesWave}%
  \BibitemOpen
  \bibfield  {author} {\bibinfo {author} {\bibfnamefont {N.~J.}\ \bibnamefont {{Cornish}}}\ and\ \bibinfo {author} {\bibfnamefont {T.~B.}\ \bibnamefont {{Littenberg}}},\ }\bibfield  {title} {\bibinfo {title} {{Bayeswave: Bayesian inference for gravitational wave bursts and instrument glitches}},\ }\href {https://doi.org/10.1088/0264-9381/32/13/135012} {\bibfield  {journal} {\bibinfo  {journal} {Classical and Quantum Gravity}\ }\textbf {\bibinfo {volume} {32}},\ \bibinfo {eid} {135012} (\bibinfo {year} {2015})},\ \Eprint {https://arxiv.org/abs/1410.3835} {arXiv:1410.3835 [gr-qc]} \BibitemShut {NoStop}%
\bibitem [{\citenamefont {Littenberg}\ \emph {et~al.}(2016)\citenamefont {Littenberg}, \citenamefont {Kanner}, \citenamefont {Cornish},\ and\ \citenamefont {Millhouse}}]{BayesWave2}%
  \BibitemOpen
  \bibfield  {author} {\bibinfo {author} {\bibfnamefont {T.~B.}\ \bibnamefont {Littenberg}}, \bibinfo {author} {\bibfnamefont {J.~B.}\ \bibnamefont {Kanner}}, \bibinfo {author} {\bibfnamefont {N.~J.}\ \bibnamefont {Cornish}},\ and\ \bibinfo {author} {\bibfnamefont {M.}~\bibnamefont {Millhouse}},\ }\bibfield  {title} {\bibinfo {title} {Enabling high confidence detections of gravitational-wave bursts},\ }\href {https://doi.org/10.1103/PhysRevD.94.044050} {\bibfield  {journal} {\bibinfo  {journal} {Phys. Rev. D}\ }\textbf {\bibinfo {volume} {94}},\ \bibinfo {pages} {044050} (\bibinfo {year} {2016})}\BibitemShut {NoStop}%
\bibitem [{\citenamefont {Cornish}\ \emph {et~al.}(2021)\citenamefont {Cornish}, \citenamefont {Littenberg}, \citenamefont {B\'ecsy}, \citenamefont {Chatziioannou}, \citenamefont {Clark}, \citenamefont {Ghonge},\ and\ \citenamefont {Millhouse}}]{BayesWave3}%
  \BibitemOpen
  \bibfield  {author} {\bibinfo {author} {\bibfnamefont {N.~J.}\ \bibnamefont {Cornish}}, \bibinfo {author} {\bibfnamefont {T.~B.}\ \bibnamefont {Littenberg}}, \bibinfo {author} {\bibfnamefont {B.}~\bibnamefont {B\'ecsy}}, \bibinfo {author} {\bibfnamefont {K.}~\bibnamefont {Chatziioannou}}, \bibinfo {author} {\bibfnamefont {J.~A.}\ \bibnamefont {Clark}}, \bibinfo {author} {\bibfnamefont {S.}~\bibnamefont {Ghonge}},\ and\ \bibinfo {author} {\bibfnamefont {M.}~\bibnamefont {Millhouse}},\ }\bibfield  {title} {\bibinfo {title} {Bayeswave analysis pipeline in the era of gravitational wave observations},\ }\href {https://doi.org/10.1103/PhysRevD.103.044006} {\bibfield  {journal} {\bibinfo  {journal} {Phys. Rev. D}\ }\textbf {\bibinfo {volume} {103}},\ \bibinfo {pages} {044006} (\bibinfo {year} {2021})}\BibitemShut {NoStop}%
\bibitem [{\citenamefont {{Szczepa{\'n}czyk}}\ \emph {et~al.}(2021)\citenamefont {{Szczepa{\'n}czyk}}, \citenamefont {{Antelis}}, \citenamefont {{Benjamin}}, \citenamefont {{Cavagli{\`a}}}, \citenamefont {{Gondek-Rosi{\'n}ska}}, \citenamefont {{Hansen}}, \citenamefont {{Klimenko}}, \citenamefont {{Morales}}, \citenamefont {{Moreno}}, \citenamefont {{Mukherjee}},\ and\ \citenamefont {et~al.}}]{Szczepanczyk_2021}%
  \BibitemOpen
  \bibfield  {author} {\bibinfo {author} {\bibfnamefont {M.~J.}\ \bibnamefont {{Szczepa{\'n}czyk}}}, \bibinfo {author} {\bibfnamefont {J.~M.}\ \bibnamefont {{Antelis}}}, \bibinfo {author} {\bibfnamefont {M.}~\bibnamefont {{Benjamin}}}, \bibinfo {author} {\bibfnamefont {M.}~\bibnamefont {{Cavagli{\`a}}}}, \bibinfo {author} {\bibfnamefont {D.}~\bibnamefont {{Gondek-Rosi{\'n}ska}}}, \bibinfo {author} {\bibfnamefont {T.}~\bibnamefont {{Hansen}}}, \bibinfo {author} {\bibfnamefont {S.}~\bibnamefont {{Klimenko}}}, \bibinfo {author} {\bibfnamefont {M.~D.}\ \bibnamefont {{Morales}}}, \bibinfo {author} {\bibfnamefont {C.}~\bibnamefont {{Moreno}}}, \bibinfo {author} {\bibfnamefont {S.}~\bibnamefont {{Mukherjee}}},\ and\ \bibinfo {author} {\bibnamefont {et~al.}},\ }\bibfield  {title} {\bibinfo {title} {{Detecting and reconstructing gravitational waves from the next galactic core-collapse supernova in the advanced detector era}},\ }\href {https://doi.org/10.1103/PhysRevD.104.102002} {\bibfield  {journal} {\bibinfo
  {journal} {\prd}\ }\textbf {\bibinfo {volume} {104}},\ \bibinfo {eid} {102002} (\bibinfo {year} {2021})},\ \Eprint {https://arxiv.org/abs/2104.06462} {arXiv:2104.06462 [astro-ph.HE]} \BibitemShut {NoStop}%
\bibitem [{\citenamefont {Raza}\ \emph {et~al.}(2022)\citenamefont {Raza}, \citenamefont {McIver}, \citenamefont {D\'alya},\ and\ \citenamefont {Raffai}}]{Raza_2022}%
  \BibitemOpen
  \bibfield  {author} {\bibinfo {author} {\bibfnamefont {N.}~\bibnamefont {Raza}}, \bibinfo {author} {\bibfnamefont {J.}~\bibnamefont {McIver}}, \bibinfo {author} {\bibfnamefont {G.}~\bibnamefont {D\'alya}},\ and\ \bibinfo {author} {\bibfnamefont {P.}~\bibnamefont {Raffai}},\ }\bibfield  {title} {\bibinfo {title} {{Prospects for reconstructing the gravitational-wave signals from core-collapse supernovae with Advanced LIGO-Virgo and the BayesWave algorithm}},\ }\href {https://doi.org/10.1103/PhysRevD.106.063014} {\bibfield  {journal} {\bibinfo  {journal} {Phys. Rev. D}\ }\textbf {\bibinfo {volume} {106}},\ \bibinfo {pages} {063014} (\bibinfo {year} {2022})},\ \Eprint {https://arxiv.org/abs/2203.08960} {arXiv:2203.08960 [astro-ph.HE]} \BibitemShut {NoStop}%
\bibitem [{\citenamefont {Kanner}\ \emph {et~al.}(2016)\citenamefont {Kanner}, \citenamefont {Littenberg}, \citenamefont {Cornish}, \citenamefont {Millhouse}, \citenamefont {Xhakaj}, \citenamefont {Salemi}, \citenamefont {Drago}, \citenamefont {Vedovato},\ and\ \citenamefont {Klimenko}}]{Kanner_2016}%
  \BibitemOpen
  \bibfield  {author} {\bibinfo {author} {\bibfnamefont {J.~B.}\ \bibnamefont {Kanner}}, \bibinfo {author} {\bibfnamefont {T.~B.}\ \bibnamefont {Littenberg}}, \bibinfo {author} {\bibfnamefont {N.}~\bibnamefont {Cornish}}, \bibinfo {author} {\bibfnamefont {M.}~\bibnamefont {Millhouse}}, \bibinfo {author} {\bibfnamefont {E.}~\bibnamefont {Xhakaj}}, \bibinfo {author} {\bibfnamefont {F.}~\bibnamefont {Salemi}}, \bibinfo {author} {\bibfnamefont {M.}~\bibnamefont {Drago}}, \bibinfo {author} {\bibfnamefont {G.}~\bibnamefont {Vedovato}},\ and\ \bibinfo {author} {\bibfnamefont {S.}~\bibnamefont {Klimenko}},\ }\bibfield  {title} {\bibinfo {title} {Leveraging waveform complexity for confident detection of gravitational waves},\ }\href {https://doi.org/10.1103/PhysRevD.93.022002} {\bibfield  {journal} {\bibinfo  {journal} {Phys. Rev. D}\ }\textbf {\bibinfo {volume} {93}},\ \bibinfo {pages} {022002} (\bibinfo {year} {2016})}\BibitemShut {NoStop}%
\bibitem [{\citenamefont {Abbott}\ \emph {et~al.}(2017)\citenamefont {Abbott} \emph {et~al.}}]{allskyO1}%
  \BibitemOpen
  \bibfield  {author} {\bibinfo {author} {\bibfnamefont {B.~P.}\ \bibnamefont {Abbott}} \emph {et~al.} (\bibinfo {collaboration} {LIGO Scientific and Virgo Collaboration}),\ }\bibfield  {title} {\bibinfo {title} {{All-sky search for short gravitational-wave bursts in the first Advanced LIGO run}},\ }\href {https://doi.org/10.1103/PhysRevD.95.042003} {\bibfield  {journal} {\bibinfo  {journal} {\prd}\ }\textbf {\bibinfo {volume} {95}},\ \bibinfo {pages} {042003} (\bibinfo {year} {2017})},\ \Eprint {https://arxiv.org/abs/1611.02972} {arXiv:1611.02972 [gr-qc]} \BibitemShut {NoStop}%
\bibitem [{\citenamefont {Abbott}\ \emph {et~al.}(2019{\natexlab{b}})\citenamefont {Abbott} \emph {et~al.}}]{allskyO2}%
  \BibitemOpen
  \bibfield  {author} {\bibinfo {author} {\bibfnamefont {B.~P.}\ \bibnamefont {Abbott}} \emph {et~al.} (\bibinfo {collaboration} {LIGO Scientific and Virgo Collaboration}),\ }\bibfield  {title} {\bibinfo {title} {{All-Sky Search for Short Gravitational-Wave Bursts in the Second Advanced LIGO and Advanced Virgo Run}},\ }\href {https://doi.org/10.1103/PhysRevD.100.024017} {\bibfield  {journal} {\bibinfo  {journal} {\prd}\ }\textbf {\bibinfo {volume} {100}},\ \bibinfo {pages} {024017} (\bibinfo {year} {2019}{\natexlab{b}})},\ \Eprint {https://arxiv.org/abs/1905.03457} {arXiv:1905.03457 [gr-qc]} \BibitemShut {NoStop}%
\bibitem [{\citenamefont {Abbott}\ \emph {et~al.}(2020)\citenamefont {Abbott} \emph {et~al.}}]{CCSN_optical_O1_O2}%
  \BibitemOpen
  \bibfield  {author} {\bibinfo {author} {\bibfnamefont {B.~P.}\ \bibnamefont {Abbott}} \emph {et~al.} (\bibinfo {collaboration} {LIGO Scientific, Virgo}),\ }\bibfield  {title} {\bibinfo {title} {{Optically targeted search for gravitational waves emitted by core-collapse supernovae during the first and second observing runs of advanced LIGO and advanced Virgo}},\ }\href {https://doi.org/10.1103/PhysRevD.101.084002} {\bibfield  {journal} {\bibinfo  {journal} {Phys. Rev. D}\ }\textbf {\bibinfo {volume} {101}},\ \bibinfo {pages} {084002} (\bibinfo {year} {2020})},\ \Eprint {https://arxiv.org/abs/1908.03584} {arXiv:1908.03584 [astro-ph.HE]} \BibitemShut {NoStop}%
\bibitem [{\citenamefont {Szczepa\'nczyk}\ \emph {et~al.}(2024)\citenamefont {Szczepa\'nczyk} \emph {et~al.}}]{CCSN_optical_O3}%
  \BibitemOpen
  \bibfield  {author} {\bibinfo {author} {\bibfnamefont {M.~J.}\ \bibnamefont {Szczepa\'nczyk}} \emph {et~al.},\ }\bibfield  {title} {\bibinfo {title} {{Optically targeted search for gravitational waves emitted by core-collapse supernovae during the third observing run of Advanced LIGO and Advanced Virgo}},\ }\href {https://doi.org/10.1103/PhysRevD.110.042007} {\bibfield  {journal} {\bibinfo  {journal} {Phys. Rev. D}\ }\textbf {\bibinfo {volume} {110}},\ \bibinfo {pages} {042007} (\bibinfo {year} {2024})},\ \Eprint {https://arxiv.org/abs/2305.16146} {arXiv:2305.16146 [astro-ph.HE]} \BibitemShut {NoStop}%
\bibitem [{\citenamefont {{Abac}}\ \emph {et~al.}(2025)\citenamefont {{Abac}}, \citenamefont {{Abbott}}, \citenamefont {{Abouelfettouh}}, \citenamefont {{Acernese}}, \citenamefont {{Ackley}}, \citenamefont {{Adhicary}}, \citenamefont {{Adhikari}}, \citenamefont {{Adhikari}}, \citenamefont {{Adkins}}, \citenamefont {{Agarwal}},\ and\ \citenamefont {et~al.}}]{SN2023ixf}%
  \BibitemOpen
  \bibfield  {author} {\bibinfo {author} {\bibfnamefont {A.~G.}\ \bibnamefont {{Abac}}}, \bibinfo {author} {\bibfnamefont {R.}~\bibnamefont {{Abbott}}}, \bibinfo {author} {\bibfnamefont {I.}~\bibnamefont {{Abouelfettouh}}}, \bibinfo {author} {\bibfnamefont {F.}~\bibnamefont {{Acernese}}}, \bibinfo {author} {\bibfnamefont {K.}~\bibnamefont {{Ackley}}}, \bibinfo {author} {\bibfnamefont {S.}~\bibnamefont {{Adhicary}}}, \bibinfo {author} {\bibfnamefont {N.}~\bibnamefont {{Adhikari}}}, \bibinfo {author} {\bibfnamefont {R.~X.}\ \bibnamefont {{Adhikari}}}, \bibinfo {author} {\bibfnamefont {V.~K.}\ \bibnamefont {{Adkins}}}, \bibinfo {author} {\bibfnamefont {D.}~\bibnamefont {{Agarwal}}},\ and\ \bibinfo {author} {\bibnamefont {et~al.}},\ }\bibfield  {title} {\bibinfo {title} {{Search for Gravitational Waves Emitted from SN 2023ixf}},\ }\href {https://doi.org/10.3847/1538-4357/adc681} {\bibfield  {journal} {\bibinfo  {journal} {\apj}\ }\textbf {\bibinfo {volume} {985}},\ \bibinfo {eid} {183} (\bibinfo {year} {2025})},\
  \Eprint {https://arxiv.org/abs/2410.16565} {arXiv:2410.16565 [astro-ph.HE]} \BibitemShut {NoStop}%
\bibitem [{\citenamefont {Abbott}\ \emph {et~al.}(2023{\natexlab{b}})\citenamefont {Abbott} \emph {et~al.}}]{OpenDataII}%
  \BibitemOpen
  \bibfield  {author} {\bibinfo {author} {\bibfnamefont {R.}~\bibnamefont {Abbott}} \emph {et~al.} (\bibinfo {collaboration} {LIGO Scientific, Virgo, and KAGRA}),\ }\bibfield  {title} {\bibinfo {title} {{Open Data from the Third Observing Run of LIGO, Virgo, KAGRA, and GEO}},\ }\href {https://doi.org/10.3847/1538-4365/acdc9f} {\bibfield  {journal} {\bibinfo  {journal} {Astrophys. J. Suppl.}\ }\textbf {\bibinfo {volume} {267}},\ \bibinfo {pages} {29} (\bibinfo {year} {2023}{\natexlab{b}})},\ \Eprint {https://arxiv.org/abs/2302.03676} {arXiv:2302.03676 [gr-qc]} \BibitemShut {NoStop}%
\bibitem [{\citenamefont {{Bethe}}\ \emph {et~al.}(1979)\citenamefont {{Bethe}}, \citenamefont {{Brown}}, \citenamefont {{Applegate}},\ and\ \citenamefont {{Lattimer}}}]{Bethe_1979}%
  \BibitemOpen
  \bibfield  {author} {\bibinfo {author} {\bibfnamefont {H.~A.}\ \bibnamefont {{Bethe}}}, \bibinfo {author} {\bibfnamefont {G.~E.}\ \bibnamefont {{Brown}}}, \bibinfo {author} {\bibfnamefont {J.}~\bibnamefont {{Applegate}}},\ and\ \bibinfo {author} {\bibfnamefont {J.~M.}\ \bibnamefont {{Lattimer}}},\ }\bibfield  {title} {\bibinfo {title} {{Equation of state in the gravitational collapse of stars}},\ }\href {https://doi.org/10.1016/0375-9474(79)90596-7} {\bibfield  {journal} {\bibinfo  {journal} {Nuclear Physics A}\ }\textbf {\bibinfo {volume} {324}},\ \bibinfo {pages} {487} (\bibinfo {year} {1979})}\BibitemShut {NoStop}%
\bibitem [{\citenamefont {{Baron}}\ \emph {et~al.}(1985)\citenamefont {{Baron}}, \citenamefont {{Cooperstein}},\ and\ \citenamefont {{Kahana}}}]{Baron_1985}%
  \BibitemOpen
  \bibfield  {author} {\bibinfo {author} {\bibfnamefont {E.}~\bibnamefont {{Baron}}}, \bibinfo {author} {\bibfnamefont {J.}~\bibnamefont {{Cooperstein}}},\ and\ \bibinfo {author} {\bibfnamefont {S.}~\bibnamefont {{Kahana}}},\ }\bibfield  {title} {\bibinfo {title} {{Type II supernovae in 12M$_{cirdot}$ and 15M$_{cirdot}$ stars: The equation of state and general relativity}},\ }\href {https://doi.org/10.1103/PhysRevLett.55.126} {\bibfield  {journal} {\bibinfo  {journal} {\prl}\ }\textbf {\bibinfo {volume} {55}},\ \bibinfo {pages} {126} (\bibinfo {year} {1985})}\BibitemShut {NoStop}%
\bibitem [{\citenamefont {{Colgate}}\ and\ \citenamefont {{White}}(1966)}]{Colgate_1966}%
  \BibitemOpen
  \bibfield  {author} {\bibinfo {author} {\bibfnamefont {S.~A.}\ \bibnamefont {{Colgate}}}\ and\ \bibinfo {author} {\bibfnamefont {R.~H.}\ \bibnamefont {{White}}},\ }\bibfield  {title} {\bibinfo {title} {{The Hydrodynamic Behavior of Supernovae Explosions}},\ }\href {https://doi.org/10.1086/148549} {\bibfield  {journal} {\bibinfo  {journal} {\apj}\ }\textbf {\bibinfo {volume} {143}},\ \bibinfo {pages} {626} (\bibinfo {year} {1966})}\BibitemShut {NoStop}%
\bibitem [{\citenamefont {Burrows}(2013)}]{Burrows_2013}%
  \BibitemOpen
  \bibfield  {author} {\bibinfo {author} {\bibfnamefont {A.}~\bibnamefont {Burrows}},\ }\bibfield  {title} {\bibinfo {title} {Colloquium: Perspectives on core-collapse supernova theory},\ }\href {https://doi.org/10.1103/RevModPhys.85.245} {\bibfield  {journal} {\bibinfo  {journal} {Rev. Mod. Phys.}\ }\textbf {\bibinfo {volume} {85}},\ \bibinfo {pages} {245} (\bibinfo {year} {2013})}\BibitemShut {NoStop}%
\bibitem [{\citenamefont {{Burrows}}\ and\ \citenamefont {{Vartanyan}}(2021)}]{Burrows_2021}%
  \BibitemOpen
  \bibfield  {author} {\bibinfo {author} {\bibfnamefont {A.}~\bibnamefont {{Burrows}}}\ and\ \bibinfo {author} {\bibfnamefont {D.}~\bibnamefont {{Vartanyan}}},\ }\bibfield  {title} {\bibinfo {title} {{Core-collapse supernova explosion theory}},\ }\href {https://doi.org/10.1038/s41586-020-03059-w} {\bibfield  {journal} {\bibinfo  {journal} {\nat}\ }\textbf {\bibinfo {volume} {589}},\ \bibinfo {pages} {29} (\bibinfo {year} {2021})},\ \Eprint {https://arxiv.org/abs/2009.14157} {arXiv:2009.14157 [astro-ph.SR]} \BibitemShut {NoStop}%
\bibitem [{\citenamefont {{Heger}}\ \emph {et~al.}(2005)\citenamefont {{Heger}}, \citenamefont {{Woosley}},\ and\ \citenamefont {{Spruit}}}]{Heger_2005}%
  \BibitemOpen
  \bibfield  {author} {\bibinfo {author} {\bibfnamefont {A.}~\bibnamefont {{Heger}}}, \bibinfo {author} {\bibfnamefont {S.~E.}\ \bibnamefont {{Woosley}}},\ and\ \bibinfo {author} {\bibfnamefont {H.~C.}\ \bibnamefont {{Spruit}}},\ }\bibfield  {title} {\bibinfo {title} {{Presupernova Evolution of Differentially Rotating Massive Stars Including Magnetic Fields}},\ }\href {https://doi.org/10.1086/429868} {\bibfield  {journal} {\bibinfo  {journal} {\apj}\ }\textbf {\bibinfo {volume} {626}},\ \bibinfo {pages} {350} (\bibinfo {year} {2005})},\ \Eprint {https://arxiv.org/abs/astro-ph/0409422} {arXiv:astro-ph/0409422 [astro-ph]} \BibitemShut {NoStop}%
\bibitem [{\citenamefont {{Morozova}}\ \emph {et~al.}(2018)\citenamefont {{Morozova}}, \citenamefont {{Radice}}, \citenamefont {{Burrows}},\ and\ \citenamefont {{Vartanyan}}}]{Morozova_2018}%
  \BibitemOpen
  \bibfield  {author} {\bibinfo {author} {\bibfnamefont {V.}~\bibnamefont {{Morozova}}}, \bibinfo {author} {\bibfnamefont {D.}~\bibnamefont {{Radice}}}, \bibinfo {author} {\bibfnamefont {A.}~\bibnamefont {{Burrows}}},\ and\ \bibinfo {author} {\bibfnamefont {D.}~\bibnamefont {{Vartanyan}}},\ }\bibfield  {title} {\bibinfo {title} {{The Gravitational Wave Signal from Core-collapse Supernovae}},\ }\href {https://doi.org/10.3847/1538-4357/aac5f1} {\bibfield  {journal} {\bibinfo  {journal} {\apj}\ }\textbf {\bibinfo {volume} {861}},\ \bibinfo {eid} {10} (\bibinfo {year} {2018})},\ \Eprint {https://arxiv.org/abs/1801.01914} {arXiv:1801.01914 [astro-ph.HE]} \BibitemShut {NoStop}%
\bibitem [{\citenamefont {{Mezzacappa}}\ and\ \citenamefont {{Zanolin}}(2024)}]{Mezzacappa_2024}%
  \BibitemOpen
  \bibfield  {author} {\bibinfo {author} {\bibfnamefont {A.}~\bibnamefont {{Mezzacappa}}}\ and\ \bibinfo {author} {\bibfnamefont {M.}~\bibnamefont {{Zanolin}}},\ }\bibfield  {title} {\bibinfo {title} {{Gravitational Waves from Neutrino-Driven Core Collapse Supernovae: Predictions, Detection, and Parameter Estimation}},\ }\href {https://doi.org/10.48550/arXiv.2401.11635} {\bibfield  {journal} {\bibinfo  {journal} {arXiv e-prints}\ ,\ \bibinfo {eid} {arXiv:2401.11635}} (\bibinfo {year} {2024})},\ \Eprint {https://arxiv.org/abs/2401.11635} {arXiv:2401.11635 [astro-ph.HE]} \BibitemShut {NoStop}%
\bibitem [{\citenamefont {{Andresen}}\ \emph {et~al.}(2017)\citenamefont {{Andresen}}, \citenamefont {{M{\"u}ller}}, \citenamefont {{M{\"u}ller}},\ and\ \citenamefont {{Janka}}}]{Andresen_2016}%
  \BibitemOpen
  \bibfield  {author} {\bibinfo {author} {\bibfnamefont {H.}~\bibnamefont {{Andresen}}}, \bibinfo {author} {\bibfnamefont {B.}~\bibnamefont {{M{\"u}ller}}}, \bibinfo {author} {\bibfnamefont {E.}~\bibnamefont {{M{\"u}ller}}},\ and\ \bibinfo {author} {\bibfnamefont {H.~T.}\ \bibnamefont {{Janka}}},\ }\bibfield  {title} {\bibinfo {title} {{Gravitational wave signals from 3D neutrino hydrodynamics simulations of core-collapse supernovae}},\ }\href {https://doi.org/10.1093/mnras/stx618} {\bibfield  {journal} {\bibinfo  {journal} {\mnras}\ }\textbf {\bibinfo {volume} {468}},\ \bibinfo {pages} {2032} (\bibinfo {year} {2017})},\ \Eprint {https://arxiv.org/abs/1607.05199} {arXiv:1607.05199 [astro-ph.HE]} \BibitemShut {NoStop}%
\bibitem [{\citenamefont {{Wilson}}\ and\ \citenamefont {{Mayle}}(1988)}]{Wilson_1988}%
  \BibitemOpen
  \bibfield  {author} {\bibinfo {author} {\bibfnamefont {J.~R.}\ \bibnamefont {{Wilson}}}\ and\ \bibinfo {author} {\bibfnamefont {R.~W.}\ \bibnamefont {{Mayle}}},\ }\bibfield  {title} {\bibinfo {title} {{Convection in core collapse supernovae.}},\ }\href {https://doi.org/10.1016/0370-1573(88)90036-1} {\bibfield  {journal} {\bibinfo  {journal} {Physics Reports}\ }\textbf {\bibinfo {volume} {163}},\ \bibinfo {pages} {63} (\bibinfo {year} {1988})}\BibitemShut {NoStop}%
\bibitem [{\citenamefont {{Murphy}}\ \emph {et~al.}(2009)\citenamefont {{Murphy}}, \citenamefont {{Ott}},\ and\ \citenamefont {{Burrows}}}]{Murphy_2009}%
  \BibitemOpen
  \bibfield  {author} {\bibinfo {author} {\bibfnamefont {J.~W.}\ \bibnamefont {{Murphy}}}, \bibinfo {author} {\bibfnamefont {C.~D.}\ \bibnamefont {{Ott}}},\ and\ \bibinfo {author} {\bibfnamefont {A.}~\bibnamefont {{Burrows}}},\ }\bibfield  {title} {\bibinfo {title} {{A Model for Gravitational Wave Emission from Neutrino-Driven Core-Collapse Supernovae}},\ }\href {https://doi.org/10.1088/0004-637X/707/2/1173} {\bibfield  {journal} {\bibinfo  {journal} {\apj}\ }\textbf {\bibinfo {volume} {707}},\ \bibinfo {pages} {1173} (\bibinfo {year} {2009})},\ \Eprint {https://arxiv.org/abs/0907.4762} {arXiv:0907.4762 [astro-ph.SR]} \BibitemShut {NoStop}%
\bibitem [{\citenamefont {{Radice}}\ \emph {et~al.}(2019)\citenamefont {{Radice}}, \citenamefont {{Morozova}}, \citenamefont {{Burrows}}, \citenamefont {{Vartanyan}},\ and\ \citenamefont {{Nagakura}}}]{s25}%
  \BibitemOpen
  \bibfield  {author} {\bibinfo {author} {\bibfnamefont {D.}~\bibnamefont {{Radice}}}, \bibinfo {author} {\bibfnamefont {V.}~\bibnamefont {{Morozova}}}, \bibinfo {author} {\bibfnamefont {A.}~\bibnamefont {{Burrows}}}, \bibinfo {author} {\bibfnamefont {D.}~\bibnamefont {{Vartanyan}}},\ and\ \bibinfo {author} {\bibfnamefont {H.}~\bibnamefont {{Nagakura}}},\ }\bibfield  {title} {\bibinfo {title} {{Characterizing the Gravitational Wave Signal from Core-collapse Supernovae}},\ }\href {https://doi.org/10.3847/2041-8213/ab191a} {\bibfield  {journal} {\bibinfo  {journal} {\apjl}\ }\textbf {\bibinfo {volume} {876}},\ \bibinfo {eid} {L9} (\bibinfo {year} {2019})},\ \Eprint {https://arxiv.org/abs/1812.07703} {arXiv:1812.07703 [astro-ph.HE]} \BibitemShut {NoStop}%
\bibitem [{\citenamefont {{Mezzacappa}}\ \emph {et~al.}(1998)\citenamefont {{Mezzacappa}}, \citenamefont {{Calder}}, \citenamefont {{Bruenn}}, \citenamefont {{Blondin}}, \citenamefont {{Guidry}}, \citenamefont {{Strayer}},\ and\ \citenamefont {{Umar}}}]{Mezzacappa_1998}%
  \BibitemOpen
  \bibfield  {author} {\bibinfo {author} {\bibfnamefont {A.}~\bibnamefont {{Mezzacappa}}}, \bibinfo {author} {\bibfnamefont {A.~C.}\ \bibnamefont {{Calder}}}, \bibinfo {author} {\bibfnamefont {S.~W.}\ \bibnamefont {{Bruenn}}}, \bibinfo {author} {\bibfnamefont {J.~M.}\ \bibnamefont {{Blondin}}}, \bibinfo {author} {\bibfnamefont {M.~W.}\ \bibnamefont {{Guidry}}}, \bibinfo {author} {\bibfnamefont {M.~R.}\ \bibnamefont {{Strayer}}},\ and\ \bibinfo {author} {\bibfnamefont {A.~S.}\ \bibnamefont {{Umar}}},\ }\bibfield  {title} {\bibinfo {title} {{An Investigation of Neutrino-driven Convection and the Core Collapse Supernova Mechanism Using Multigroup Neutrino Transport}},\ }\href {https://doi.org/10.1086/305338} {\bibfield  {journal} {\bibinfo  {journal} {\apj}\ }\textbf {\bibinfo {volume} {495}},\ \bibinfo {pages} {911} (\bibinfo {year} {1998})},\ \Eprint {https://arxiv.org/abs/astro-ph/9709188} {arXiv:astro-ph/9709188 [astro-ph]} \BibitemShut {NoStop}%
\bibitem [{\citenamefont {{Hayama}}\ \emph {et~al.}(2018)\citenamefont {{Hayama}}, \citenamefont {{Kuroda}}, \citenamefont {{Kotake}},\ and\ \citenamefont {{Takiwaki}}}]{Hayama_2018}%
  \BibitemOpen
  \bibfield  {author} {\bibinfo {author} {\bibfnamefont {K.}~\bibnamefont {{Hayama}}}, \bibinfo {author} {\bibfnamefont {T.}~\bibnamefont {{Kuroda}}}, \bibinfo {author} {\bibfnamefont {K.}~\bibnamefont {{Kotake}}},\ and\ \bibinfo {author} {\bibfnamefont {T.}~\bibnamefont {{Takiwaki}}},\ }\bibfield  {title} {\bibinfo {title} {{Circular polarization of gravitational waves from non-rotating supernova cores: a new probe into the pre-explosion hydrodynamics}},\ }\href {https://doi.org/10.1093/mnrasl/sly055} {\bibfield  {journal} {\bibinfo  {journal} {\mnras}\ }\textbf {\bibinfo {volume} {477}},\ \bibinfo {pages} {L96} (\bibinfo {year} {2018})},\ \Eprint {https://arxiv.org/abs/1802.03842} {arXiv:1802.03842 [astro-ph.HE]} \BibitemShut {NoStop}%
\bibitem [{\citenamefont {{Powell}}\ \emph {et~al.}(2021)\citenamefont {{Powell}}, \citenamefont {{M{\"u}ller}},\ and\ \citenamefont {{Heger}}}]{Powell_2021}%
  \BibitemOpen
  \bibfield  {author} {\bibinfo {author} {\bibfnamefont {J.}~\bibnamefont {{Powell}}}, \bibinfo {author} {\bibfnamefont {B.}~\bibnamefont {{M{\"u}ller}}},\ and\ \bibinfo {author} {\bibfnamefont {A.}~\bibnamefont {{Heger}}},\ }\bibfield  {title} {\bibinfo {title} {{The final core collapse of pulsational pair instability supernovae}},\ }\href {https://doi.org/10.1093/mnras/stab614} {\bibfield  {journal} {\bibinfo  {journal} {\mnras}\ }\textbf {\bibinfo {volume} {503}},\ \bibinfo {pages} {2108} (\bibinfo {year} {2021})},\ \Eprint {https://arxiv.org/abs/2101.06889} {arXiv:2101.06889 [astro-ph.HE]} \BibitemShut {NoStop}%
\bibitem [{\citenamefont {{Foglizzo}}\ \emph {et~al.}(2006)\citenamefont {{Foglizzo}}, \citenamefont {{Scheck}},\ and\ \citenamefont {{Janka}}}]{Foglizzo_2006}%
  \BibitemOpen
  \bibfield  {author} {\bibinfo {author} {\bibfnamefont {T.}~\bibnamefont {{Foglizzo}}}, \bibinfo {author} {\bibfnamefont {L.}~\bibnamefont {{Scheck}}},\ and\ \bibinfo {author} {\bibfnamefont {H.~T.}\ \bibnamefont {{Janka}}},\ }\bibfield  {title} {\bibinfo {title} {{Neutrino-driven Convection versus Advection in Core-Collapse Supernovae}},\ }\href {https://doi.org/10.1086/508443} {\bibfield  {journal} {\bibinfo  {journal} {\apj}\ }\textbf {\bibinfo {volume} {652}},\ \bibinfo {pages} {1436} (\bibinfo {year} {2006})},\ \Eprint {https://arxiv.org/abs/astro-ph/0507636} {arXiv:astro-ph/0507636 [astro-ph]} \BibitemShut {NoStop}%
\bibitem [{\citenamefont {{Marek}}\ \emph {et~al.}(2009)\citenamefont {{Marek}}, \citenamefont {{Janka}},\ and\ \citenamefont {{M{\"u}ller}}}]{Marek_2009}%
  \BibitemOpen
  \bibfield  {author} {\bibinfo {author} {\bibfnamefont {A.}~\bibnamefont {{Marek}}}, \bibinfo {author} {\bibfnamefont {H.~T.}\ \bibnamefont {{Janka}}},\ and\ \bibinfo {author} {\bibfnamefont {E.}~\bibnamefont {{M{\"u}ller}}},\ }\bibfield  {title} {\bibinfo {title} {{Equation-of-state dependent features in shock-oscillation modulated neutrino and gravitational-wave signals from supernovae}},\ }\href {https://doi.org/10.1051/0004-6361/200810883} {\bibfield  {journal} {\bibinfo  {journal} {Astronomy and Astrophysics}\ }\textbf {\bibinfo {volume} {496}},\ \bibinfo {pages} {475} (\bibinfo {year} {2009})},\ \Eprint {https://arxiv.org/abs/0808.4136} {arXiv:0808.4136 [astro-ph]} \BibitemShut {NoStop}%
\bibitem [{\citenamefont {{Klimenko}}(2022)}]{Wavescan}%
  \BibitemOpen
  \bibfield  {author} {\bibinfo {author} {\bibfnamefont {S.}~\bibnamefont {{Klimenko}}},\ }\bibfield  {title} {\bibinfo {title} {{Wavescan: multiresolution regression of gravitational-wave data}},\ }\href {https://doi.org/10.48550/arXiv.2201.01096} {\bibfield  {journal} {\bibinfo  {journal} {arXiv e-prints}\ ,\ \bibinfo {eid} {arXiv:2201.01096}} (\bibinfo {year} {2022})},\ \Eprint {https://arxiv.org/abs/2201.01096} {arXiv:2201.01096 [physics.data-an]} \BibitemShut {NoStop}%
\bibitem [{\citenamefont {{Klimenko}}\ \emph {et~al.}(2005)\citenamefont {{Klimenko}}, \citenamefont {{Mohanty}}, \citenamefont {{Rakhmanov}},\ and\ \citenamefont {{Mitselmakher}}}]{cWB4}%
  \BibitemOpen
  \bibfield  {author} {\bibinfo {author} {\bibfnamefont {S.}~\bibnamefont {{Klimenko}}}, \bibinfo {author} {\bibfnamefont {S.}~\bibnamefont {{Mohanty}}}, \bibinfo {author} {\bibfnamefont {M.}~\bibnamefont {{Rakhmanov}}},\ and\ \bibinfo {author} {\bibfnamefont {G.}~\bibnamefont {{Mitselmakher}}},\ }\bibfield  {title} {\bibinfo {title} {{Constraint likelihood analysis for a network of gravitational wave detectors}},\ }\href {https://doi.org/10.1103/PhysRevD.72.122002} {\bibfield  {journal} {\bibinfo  {journal} {\prd}\ }\textbf {\bibinfo {volume} {72}},\ \bibinfo {eid} {122002} (\bibinfo {year} {2005})},\ \Eprint {https://arxiv.org/abs/gr-qc/0508068} {arXiv:gr-qc/0508068 [gr-qc]} \BibitemShut {NoStop}%
\bibitem [{\citenamefont {{Mishra}}\ \emph {et~al.}(2021)\citenamefont {{Mishra}}, \citenamefont {{O'Brien}}, \citenamefont {{Gayathri}}, \citenamefont {{Szczepa{\'n}czyk}}, \citenamefont {{Bhaumik}}, \citenamefont {{Bartos}},\ and\ \citenamefont {{Klimenko}}}]{Mishra_2021}%
  \BibitemOpen
  \bibfield  {author} {\bibinfo {author} {\bibfnamefont {T.}~\bibnamefont {{Mishra}}}, \bibinfo {author} {\bibfnamefont {B.}~\bibnamefont {{O'Brien}}}, \bibinfo {author} {\bibfnamefont {V.}~\bibnamefont {{Gayathri}}}, \bibinfo {author} {\bibfnamefont {M.}~\bibnamefont {{Szczepa{\'n}czyk}}}, \bibinfo {author} {\bibfnamefont {S.}~\bibnamefont {{Bhaumik}}}, \bibinfo {author} {\bibfnamefont {I.}~\bibnamefont {{Bartos}}},\ and\ \bibinfo {author} {\bibfnamefont {S.}~\bibnamefont {{Klimenko}}},\ }\bibfield  {title} {\bibinfo {title} {{Optimization of model independent gravitational wave search for binary black hole mergers using machine learning}},\ }\href {https://doi.org/10.1103/PhysRevD.104.023014} {\bibfield  {journal} {\bibinfo  {journal} {\prd}\ }\textbf {\bibinfo {volume} {104}},\ \bibinfo {eid} {023014} (\bibinfo {year} {2021})},\ \Eprint {https://arxiv.org/abs/2105.04739} {arXiv:2105.04739 [gr-qc]} \BibitemShut {NoStop}%
\bibitem [{\citenamefont {{Mishra}}\ \emph {et~al.}(2022)\citenamefont {{Mishra}}, \citenamefont {{O'Brien}}, \citenamefont {{Szczepa{\'n}czyk}}, \citenamefont {{Vedovato}}, \citenamefont {{Bhaumik}}, \citenamefont {{Gayathri}}, \citenamefont {{Prodi}}, \citenamefont {{Salemi}}, \citenamefont {{Milotti}}, \citenamefont {{Bartos}},\ and\ \citenamefont {et~al.}}]{Mishra_2022}%
  \BibitemOpen
  \bibfield  {author} {\bibinfo {author} {\bibfnamefont {T.}~\bibnamefont {{Mishra}}}, \bibinfo {author} {\bibfnamefont {B.}~\bibnamefont {{O'Brien}}}, \bibinfo {author} {\bibfnamefont {M.}~\bibnamefont {{Szczepa{\'n}czyk}}}, \bibinfo {author} {\bibfnamefont {G.}~\bibnamefont {{Vedovato}}}, \bibinfo {author} {\bibfnamefont {S.}~\bibnamefont {{Bhaumik}}}, \bibinfo {author} {\bibfnamefont {V.}~\bibnamefont {{Gayathri}}}, \bibinfo {author} {\bibfnamefont {G.}~\bibnamefont {{Prodi}}}, \bibinfo {author} {\bibfnamefont {F.}~\bibnamefont {{Salemi}}}, \bibinfo {author} {\bibfnamefont {E.}~\bibnamefont {{Milotti}}}, \bibinfo {author} {\bibfnamefont {I.}~\bibnamefont {{Bartos}}},\ and\ \bibinfo {author} {\bibnamefont {et~al.}},\ }\bibfield  {title} {\bibinfo {title} {{Search for binary black hole mergers in the third observing run of Advanced LIGO-Virgo using coherent WaveBurst enhanced with machine learning}},\ }\href {https://doi.org/10.1103/PhysRevD.105.083018} {\bibfield  {journal} {\bibinfo  {journal} {\prd}\
  }\textbf {\bibinfo {volume} {105}},\ \bibinfo {eid} {083018} (\bibinfo {year} {2022})},\ \Eprint {https://arxiv.org/abs/2201.01495} {arXiv:2201.01495 [gr-qc]} \BibitemShut {NoStop}%
\bibitem [{\citenamefont {Szczepa\'nczyk}\ \emph {et~al.}(2023)\citenamefont {Szczepa\'nczyk} \emph {et~al.}}]{Szczepanczyk2022_XGBoost}%
  \BibitemOpen
  \bibfield  {author} {\bibinfo {author} {\bibfnamefont {M.~J.}\ \bibnamefont {Szczepa\'nczyk}} \emph {et~al.},\ }\bibfield  {title} {\bibinfo {title} {{Search for gravitational-wave bursts in the third Advanced LIGO-Virgo run with coherent WaveBurst enhanced by machine learning}},\ }\href {https://doi.org/10.1103/PhysRevD.107.062002} {\bibfield  {journal} {\bibinfo  {journal} {Phys. Rev. D}\ }\textbf {\bibinfo {volume} {107}},\ \bibinfo {pages} {062002} (\bibinfo {year} {2023})},\ \Eprint {https://arxiv.org/abs/2210.01754} {arXiv:2210.01754 [gr-qc]} \BibitemShut {NoStop}%
\bibitem [{\citenamefont {Littenberg}\ and\ \citenamefont {Cornish}(2009)}]{BW_TI}%
  \BibitemOpen
  \bibfield  {author} {\bibinfo {author} {\bibfnamefont {T.~B.}\ \bibnamefont {Littenberg}}\ and\ \bibinfo {author} {\bibfnamefont {N.~J.}\ \bibnamefont {Cornish}},\ }\bibfield  {title} {\bibinfo {title} {{A Bayesian Approach to the Detection Problem in Gravitational Wave Astronomy}},\ }\href {https://doi.org/10.1103/PhysRevD.80.063007} {\bibfield  {journal} {\bibinfo  {journal} {Phys. Rev. D}\ }\textbf {\bibinfo {volume} {80}},\ \bibinfo {pages} {063007} (\bibinfo {year} {2009})},\ \Eprint {https://arxiv.org/abs/0902.0368} {arXiv:0902.0368 [gr-qc]} \BibitemShut {NoStop}%
\bibitem [{\citenamefont {Lee}\ \emph {et~al.}(2021)\citenamefont {Lee}, \citenamefont {Millhouse},\ and\ \citenamefont {Melatos}}]{Lee_2021}%
  \BibitemOpen
  \bibfield  {author} {\bibinfo {author} {\bibfnamefont {Y.~S.~C.}\ \bibnamefont {Lee}}, \bibinfo {author} {\bibfnamefont {M.}~\bibnamefont {Millhouse}},\ and\ \bibinfo {author} {\bibfnamefont {A.}~\bibnamefont {Melatos}},\ }\bibfield  {title} {\bibinfo {title} {Enhancing the gravitational-wave burst detection confidence in expanded detector networks with the bayeswave pipeline},\ }\href {https://doi.org/10.1103/PhysRevD.103.062002} {\bibfield  {journal} {\bibinfo  {journal} {Phys. Rev. D}\ }\textbf {\bibinfo {volume} {103}},\ \bibinfo {pages} {062002} (\bibinfo {year} {2021})}\BibitemShut {NoStop}%
\bibitem [{\citenamefont {{Steiner}}\ \emph {et~al.}(2013)\citenamefont {{Steiner}}, \citenamefont {{Hempel}},\ and\ \citenamefont {{Fischer}}}]{SFHx_eos}%
  \BibitemOpen
  \bibfield  {author} {\bibinfo {author} {\bibfnamefont {A.~W.}\ \bibnamefont {{Steiner}}}, \bibinfo {author} {\bibfnamefont {M.}~\bibnamefont {{Hempel}}},\ and\ \bibinfo {author} {\bibfnamefont {T.}~\bibnamefont {{Fischer}}},\ }\bibfield  {title} {\bibinfo {title} {{Core-collapse Supernova Equations of State Based on Neutron Star Observations}},\ }\href {https://doi.org/10.1088/0004-637X/774/1/17} {\bibfield  {journal} {\bibinfo  {journal} {\apj}\ }\textbf {\bibinfo {volume} {774}},\ \bibinfo {eid} {17} (\bibinfo {year} {2013})},\ \Eprint {https://arxiv.org/abs/1207.2184} {arXiv:1207.2184 [astro-ph.SR]} \BibitemShut {NoStop}%
\bibitem [{\citenamefont {{Mezzacappa}}\ \emph {et~al.}(2023)\citenamefont {{Mezzacappa}}, \citenamefont {{Marronetti}}, \citenamefont {{Landfield}}, \citenamefont {{Lentz}}, \citenamefont {{Murphy}}, \citenamefont {{Raphael Hix}}, \citenamefont {{Harris}}, \citenamefont {{Bruenn}}, \citenamefont {{Blondin}}, \citenamefont {{Bronson Messer}}, \citenamefont {{Casanova}},\ and\ \citenamefont {{Kronzer}}}]{D15}%
  \BibitemOpen
  \bibfield  {author} {\bibinfo {author} {\bibfnamefont {A.}~\bibnamefont {{Mezzacappa}}}, \bibinfo {author} {\bibfnamefont {P.}~\bibnamefont {{Marronetti}}}, \bibinfo {author} {\bibfnamefont {R.~E.}\ \bibnamefont {{Landfield}}}, \bibinfo {author} {\bibfnamefont {E.~J.}\ \bibnamefont {{Lentz}}}, \bibinfo {author} {\bibfnamefont {R.~D.}\ \bibnamefont {{Murphy}}}, \bibinfo {author} {\bibfnamefont {W.}~\bibnamefont {{Raphael Hix}}}, \bibinfo {author} {\bibfnamefont {J.~A.}\ \bibnamefont {{Harris}}}, \bibinfo {author} {\bibfnamefont {S.~W.}\ \bibnamefont {{Bruenn}}}, \bibinfo {author} {\bibfnamefont {J.~M.}\ \bibnamefont {{Blondin}}}, \bibinfo {author} {\bibfnamefont {O.~E.}\ \bibnamefont {{Bronson Messer}}}, \bibinfo {author} {\bibfnamefont {J.}~\bibnamefont {{Casanova}}},\ and\ \bibinfo {author} {\bibfnamefont {L.~L.}\ \bibnamefont {{Kronzer}}},\ }\bibfield  {title} {\bibinfo {title} {{Core collapse supernova gravitational wave emission for progenitors of 9.6, 15, and 25M{\ensuremath{\odot}}}},\ }\href
  {https://doi.org/10.1103/PhysRevD.107.043008} {\bibfield  {journal} {\bibinfo  {journal} {\prd}\ }\textbf {\bibinfo {volume} {107}},\ \bibinfo {eid} {043008} (\bibinfo {year} {2023})},\ \Eprint {https://arxiv.org/abs/2208.10643} {arXiv:2208.10643 [astro-ph.SR]} \BibitemShut {NoStop}%
\bibitem [{\citenamefont {{Bruenn}}\ \emph {et~al.}(2020)\citenamefont {{Bruenn}}, \citenamefont {{Blondin}}, \citenamefont {{Hix}}, \citenamefont {{Lentz}}, \citenamefont {{Messer}}, \citenamefont {{Mezzacappa}}, \citenamefont {{Endeve}}, \citenamefont {{Harris}}, \citenamefont {{Marronetti}}, \citenamefont {{Budiardja}},\ and\ \citenamefont {et~al.}}]{Chimera}%
  \BibitemOpen
  \bibfield  {author} {\bibinfo {author} {\bibfnamefont {S.~W.}\ \bibnamefont {{Bruenn}}}, \bibinfo {author} {\bibfnamefont {J.~M.}\ \bibnamefont {{Blondin}}}, \bibinfo {author} {\bibfnamefont {W.~R.}\ \bibnamefont {{Hix}}}, \bibinfo {author} {\bibfnamefont {E.~J.}\ \bibnamefont {{Lentz}}}, \bibinfo {author} {\bibfnamefont {O.~E.~B.}\ \bibnamefont {{Messer}}}, \bibinfo {author} {\bibfnamefont {A.}~\bibnamefont {{Mezzacappa}}}, \bibinfo {author} {\bibfnamefont {E.}~\bibnamefont {{Endeve}}}, \bibinfo {author} {\bibfnamefont {J.~A.}\ \bibnamefont {{Harris}}}, \bibinfo {author} {\bibfnamefont {P.}~\bibnamefont {{Marronetti}}}, \bibinfo {author} {\bibfnamefont {R.~D.}\ \bibnamefont {{Budiardja}}},\ and\ \bibinfo {author} {\bibnamefont {et~al.}},\ }\bibfield  {title} {\bibinfo {title} {{CHIMERA: A Massively Parallel Code for Core-collapse Supernova Simulations}},\ }\href {https://doi.org/10.3847/1538-4365/ab7aff} {\bibfield  {journal} {\bibinfo  {journal} {\apjs}\ }\textbf {\bibinfo {volume} {248}},\ \bibinfo {eid}
  {11} (\bibinfo {year} {2020})}\BibitemShut {NoStop}%
\bibitem [{\citenamefont {{Powell}}\ and\ \citenamefont {{M{\"u}ller}}(2019)}]{s18}%
  \BibitemOpen
  \bibfield  {author} {\bibinfo {author} {\bibfnamefont {J.}~\bibnamefont {{Powell}}}\ and\ \bibinfo {author} {\bibfnamefont {B.}~\bibnamefont {{M{\"u}ller}}},\ }\bibfield  {title} {\bibinfo {title} {{Gravitational wave emission from 3D explosion models of core-collapse supernovae with low and normal explosion energies}},\ }\href {https://doi.org/10.1093/mnras/stz1304} {\bibfield  {journal} {\bibinfo  {journal} {\mnras}\ }\textbf {\bibinfo {volume} {487}},\ \bibinfo {pages} {1178} (\bibinfo {year} {2019})},\ \Eprint {https://arxiv.org/abs/1812.05738} {arXiv:1812.05738 [astro-ph.HE]} \BibitemShut {NoStop}%
\bibitem [{\citenamefont {{M{\"u}ller}}\ and\ \citenamefont {{Janka}}(2015)}]{coconut_fmt}%
  \BibitemOpen
  \bibfield  {author} {\bibinfo {author} {\bibfnamefont {B.}~\bibnamefont {{M{\"u}ller}}}\ and\ \bibinfo {author} {\bibfnamefont {H.~T.}\ \bibnamefont {{Janka}}},\ }\bibfield  {title} {\bibinfo {title} {{Non-radial instabilities and progenitor asphericities in core-collapse supernovae}},\ }\href {https://doi.org/10.1093/mnras/stv101} {\bibfield  {journal} {\bibinfo  {journal} {\mnras}\ }\textbf {\bibinfo {volume} {448}},\ \bibinfo {pages} {2141} (\bibinfo {year} {2015})},\ \Eprint {https://arxiv.org/abs/1409.4783} {arXiv:1409.4783 [astro-ph.SR]} \BibitemShut {NoStop}%
\bibitem [{\citenamefont {{Henshaw}}\ \emph {et~al.}(2024)\citenamefont {{Henshaw}}, \citenamefont {{Arogeti}}, \citenamefont {{Heranval}},\ and\ \citenamefont {{Cadonati}}}]{Henshaw_2024}%
  \BibitemOpen
  \bibfield  {author} {\bibinfo {author} {\bibfnamefont {C.}~\bibnamefont {{Henshaw}}}, \bibinfo {author} {\bibfnamefont {M.}~\bibnamefont {{Arogeti}}}, \bibinfo {author} {\bibfnamefont {A.}~\bibnamefont {{Heranval}}},\ and\ \bibinfo {author} {\bibfnamefont {L.}~\bibnamefont {{Cadonati}}},\ }\bibfield  {title} {\bibinfo {title} {{Visualization of frequency structures in gravitational wave signals}},\ }\href {https://doi.org/10.48550/arXiv.2402.16533} {\bibfield  {journal} {\bibinfo  {journal} {arXiv e-prints}\ ,\ \bibinfo {eid} {arXiv:2402.16533}} (\bibinfo {year} {2024})},\ \Eprint {https://arxiv.org/abs/2402.16533} {arXiv:2402.16533 [gr-qc]} \BibitemShut {NoStop}%
\bibitem [{\citenamefont {{W{\k{a}}s}}\ \emph {et~al.}(2010)\citenamefont {{W{\k{a}}s}}, \citenamefont {{Bizouard}}, \citenamefont {{Brisson}}, \citenamefont {{Cavalier}}, \citenamefont {{Davier}}, \citenamefont {{Hello}}, \citenamefont {{Leroy}}, \citenamefont {{Robinet}},\ and\ \citenamefont {{Vavoulidis}}}]{timeslide_2010}%
  \BibitemOpen
  \bibfield  {author} {\bibinfo {author} {\bibfnamefont {M.}~\bibnamefont {{W{\k{a}}s}}}, \bibinfo {author} {\bibfnamefont {M.-A.}\ \bibnamefont {{Bizouard}}}, \bibinfo {author} {\bibfnamefont {V.}~\bibnamefont {{Brisson}}}, \bibinfo {author} {\bibfnamefont {F.}~\bibnamefont {{Cavalier}}}, \bibinfo {author} {\bibfnamefont {M.}~\bibnamefont {{Davier}}}, \bibinfo {author} {\bibfnamefont {P.}~\bibnamefont {{Hello}}}, \bibinfo {author} {\bibfnamefont {N.}~\bibnamefont {{Leroy}}}, \bibinfo {author} {\bibfnamefont {F.}~\bibnamefont {{Robinet}}},\ and\ \bibinfo {author} {\bibfnamefont {M.}~\bibnamefont {{Vavoulidis}}},\ }\bibfield  {title} {\bibinfo {title} {{On the background estimation by time slides in a network of gravitational wave detectors}},\ }\href {https://doi.org/10.1088/0264-9381/27/1/015005} {\bibfield  {journal} {\bibinfo  {journal} {Classical and Quantum Gravity}\ }\textbf {\bibinfo {volume} {27}},\ \bibinfo {eid} {015005} (\bibinfo {year} {2010})},\ \Eprint {https://arxiv.org/abs/0906.2120}
  {arXiv:0906.2120 [gr-qc]} \BibitemShut {NoStop}%
\bibitem [{\citenamefont {Lee}\ \emph {et~al.}(2024)\citenamefont {Lee}, \citenamefont {Millhouse},\ and\ \citenamefont {Melatos}}]{Lee2024}%
  \BibitemOpen
  \bibfield  {author} {\bibinfo {author} {\bibfnamefont {Y.~S.~C.}\ \bibnamefont {Lee}}, \bibinfo {author} {\bibfnamefont {M.}~\bibnamefont {Millhouse}},\ and\ \bibinfo {author} {\bibfnamefont {A.}~\bibnamefont {Melatos}},\ }\bibfield  {title} {\bibinfo {title} {{Impact of noise transients on gravitational-wave burst detection efficiency of the BayesWave pipeline with multidetector networks}},\ }\href {https://doi.org/10.1103/PhysRevD.109.082002} {\bibfield  {journal} {\bibinfo  {journal} {Phys. Rev. D}\ }\textbf {\bibinfo {volume} {109}},\ \bibinfo {pages} {082002} (\bibinfo {year} {2024})},\ \Eprint {https://arxiv.org/abs/2403.16837} {arXiv:2403.16837 [gr-qc]} \BibitemShut {NoStop}%
\bibitem [{\citenamefont {Sutton}(2013)}]{Sutton_2013}%
  \BibitemOpen
  \bibfield  {author} {\bibinfo {author} {\bibfnamefont {P.~J.}\ \bibnamefont {Sutton}},\ }\bibfield  {title} {\bibinfo {title} {{A Rule of Thumb for the Detectability of Gravitational-Wave Bursts}},\ }\href@noop {} {\  (\bibinfo {year} {2013})},\ \Eprint {https://arxiv.org/abs/1304.0210} {arXiv:1304.0210 [gr-qc]} \BibitemShut {NoStop}%
\bibitem [{\citenamefont {{Kiuchi}}\ \emph {et~al.}(2009)\citenamefont {{Kiuchi}}, \citenamefont {{Sekiguchi}}, \citenamefont {{Shibata}},\ and\ \citenamefont {{Taniguchi}}}]{Kiuchi_2009}%
  \BibitemOpen
  \bibfield  {author} {\bibinfo {author} {\bibfnamefont {K.}~\bibnamefont {{Kiuchi}}}, \bibinfo {author} {\bibfnamefont {Y.}~\bibnamefont {{Sekiguchi}}}, \bibinfo {author} {\bibfnamefont {M.}~\bibnamefont {{Shibata}}},\ and\ \bibinfo {author} {\bibfnamefont {K.}~\bibnamefont {{Taniguchi}}},\ }\bibfield  {title} {\bibinfo {title} {{Long-term general relativistic simulation of binary neutron stars collapsing to a black hole}},\ }\href {https://doi.org/10.1103/PhysRevD.80.064037} {\bibfield  {journal} {\bibinfo  {journal} {\prd}\ }\textbf {\bibinfo {volume} {80}},\ \bibinfo {eid} {064037} (\bibinfo {year} {2009})},\ \Eprint {https://arxiv.org/abs/0904.4551} {arXiv:0904.4551 [gr-qc]} \BibitemShut {NoStop}%
\bibitem [{\citenamefont {{Clark}}\ \emph {et~al.}(2016)\citenamefont {{Clark}}, \citenamefont {{Bauswein}}, \citenamefont {{Stergioulas}},\ and\ \citenamefont {{Shoemaker}}}]{Clark_2016}%
  \BibitemOpen
  \bibfield  {author} {\bibinfo {author} {\bibfnamefont {J.~A.}\ \bibnamefont {{Clark}}}, \bibinfo {author} {\bibfnamefont {A.}~\bibnamefont {{Bauswein}}}, \bibinfo {author} {\bibfnamefont {N.}~\bibnamefont {{Stergioulas}}},\ and\ \bibinfo {author} {\bibfnamefont {D.}~\bibnamefont {{Shoemaker}}},\ }\bibfield  {title} {\bibinfo {title} {{Observing gravitational waves from the post-merger phase of binary neutron star coalescence}},\ }\href {https://doi.org/10.1088/0264-9381/33/8/085003} {\bibfield  {journal} {\bibinfo  {journal} {Classical and Quantum Gravity}\ }\textbf {\bibinfo {volume} {33}},\ \bibinfo {eid} {085003} (\bibinfo {year} {2016})},\ \Eprint {https://arxiv.org/abs/1509.08522} {arXiv:1509.08522 [astro-ph.HE]} \BibitemShut {NoStop}%
\bibitem [{\citenamefont {{Patterson}}\ \emph {et~al.}(2024)\citenamefont {{Patterson}}, \citenamefont {{Tomson}},\ and\ \citenamefont {{Fairhurst}}}]{Patterson_2024}%
  \BibitemOpen
  \bibfield  {author} {\bibinfo {author} {\bibfnamefont {B.~G.}\ \bibnamefont {{Patterson}}}, \bibinfo {author} {\bibfnamefont {S.~M.}\ \bibnamefont {{Tomson}}},\ and\ \bibinfo {author} {\bibfnamefont {S.}~\bibnamefont {{Fairhurst}}},\ }\bibfield  {title} {\bibinfo {title} {{Identifying Eccentricity in Binary Black Hole mergers using a Harmonic Decomposition of the Gravitational Waveform}},\ }\href {https://doi.org/10.48550/arXiv.2411.04187} {\bibfield  {journal} {\bibinfo  {journal} {arXiv e-prints}\ ,\ \bibinfo {eid} {arXiv:2411.04187}} (\bibinfo {year} {2024})},\ \Eprint {https://arxiv.org/abs/2411.04187} {arXiv:2411.04187 [gr-qc]} \BibitemShut {NoStop}%
\end{thebibliography}%

\end{document}